\documentclass[sigconf, nonacm]{acmart}
\usepackage{enumitem}
\usepackage{graphicx}
\usepackage{subcaption}
\usepackage{balance} 
\usepackage{booktabs}
\usepackage{algorithm, algpseudocode}
\usepackage{setspace}
\usepackage{footnote}
\usepackage{xspace}
\usepackage{enumitem}
\usepackage{amsmath}
\usepackage{bbm}
\usepackage{multirow}

\makesavenoteenv{tabular}
\makesavenoteenv{table}

\newcommand\vldbdoi{XX.XX/XXX.XX}
\newcommand\vldbpages{XXX-XXX}
\newcommand\vldbvolume{14}
\newcommand\vldbissue{1}
\newcommand\vldbyear{2021}
\newcommand\vldbauthors{\authors}
\newcommand\vldbtitle{\shorttitle} 
\newcommand\vldbavailabilityurl{}
\newcommand\vldbpagestyle{plain} 

\newcommand{\system}{\textsc{Kamino}\xspace}

\newcommand{\eat}[1]{}

\newtheorem{theorem}{Theorem}

\newtheorem{lemma}{Lemma}

\newtheorem{definition}{Definition}

\newcounter{exmpc}
\newcommand{\example}[1]{
\refstepcounter{exmpc}
{\vspace*{3pt}\noindent\underline{Example \theexmpc}: #1\qed}
}

\newcommand{\stitle}[1]{\smallskip \noindent{\bf #1}}
\newcommand{\squishlist}{
   \begin{list}{$\bullet$}
    {
      \setlength{\itemsep}{0pt}
      \setlength{\parsep}{3pt}
      \setlength{\topsep}{3pt}
      \setlength{\partopsep}{0pt}
      \setlength{\leftmargin}{1.5em}
      \setlength{\labelwidth}{1em}
      \setlength{\labelsep}{0.5em} } }

\newcommand{\squishend}{
    \end{list}  }
    
\DeclareMathOperator*{\argmax}{arg\,max}

\newcommand{\norm}[1]{\left\lVert#1\right\rVert}

\newcommand{\revise}[1]{#1}

\newif\ifpaper
\paperfalse   

\begin{document}
\title{\system: Constraint-Aware Differentially Private Data Synthesis}

\author{Chang Ge\enspace\enspace\enspace Shubhankar Mohapatra\enspace\enspace\enspace Xi He\enspace\enspace\enspace Ihab F. Ilyas}
\affiliation{%
\institution{University of Waterloo}
}
\affiliation{\{c4ge, s3mohapatra, xihe, ilyas\}@uwaterloo.ca}

\gdef\authors{Chang Ge, Shubhankar Mohapatra, Xi He, and Ihab F. Ilyas}

\begin{abstract}

Organizations are increasingly relying on data to support decisions.
When data contains private and sensitive information, 
the data owner often desires to publish a synthetic database instance that is similarly useful as the true data, while ensuring the privacy of individual data records.
Existing differentially private data synthesis methods aim to generate useful data based on applications, 
but they fail in keeping one of the most fundamental data properties of the structured data --- 
the underlying correlations and dependencies among tuples and attributes (i.e., the structure of the data).
This structure is often expressed as integrity and schema constraints, or with a probabilistic generative process.
As a result, the synthesized data is not useful for any downstream tasks that require this structure to be preserved.

This work presents \system, a data synthesis system to ensure differential privacy and to preserve the structure and correlations present in the original dataset. 
\system takes as input of a database instance, along with its schema (including integrity constraints), 
and produces a synthetic database instance with differential privacy and structure preservation guarantees.
We empirically show that while preserving the structure of the data, 
\system achieves comparable and even better usefulness in applications of training classification models and answering marginal queries than the state-of-the-art methods of differentially private data synthesis.
\end{abstract}

\maketitle

\ifpaper
\pagestyle{\vldbpagestyle}
\begingroup\small\noindent\raggedright\textbf{PVLDB Reference Format:}\\
\vldbauthors. \vldbtitle. PVLDB, \vldbvolume(\vldbissue): \vldbpages, \vldbyear.\\
\href{https://doi.org/\vldbdoi}{doi:\vldbdoi}
\endgroup
\begingroup
\renewcommand\thefootnote{}\footnote{\noindent
This work is licensed under the Creative Commons BY-NC-ND 4.0 International License. Visit \url{https://creativecommons.org/licenses/by-nc-nd/4.0/} to view a copy of this license. For any use beyond those covered by this license, obtain permission by emailing \href{mailto:info@vldb.org}{info@vldb.org}. Copyright is held by the owner/author(s). Publication rights licensed to the VLDB Endowment. \\
\raggedright Proceedings of the VLDB Endowment, Vol. \vldbvolume, No. \vldbissue\ %
ISSN 2150-8097. \\
\href{https://doi.org/\vldbdoi}{doi:\vldbdoi} \\
}\addtocounter{footnote}{-1}\endgroup

\ifdefempty{\vldbavailabilityurl}{}{
\vspace{.3cm}
\begingroup\small\noindent\raggedright\textbf{PVLDB Artifact Availability:}\\
The source code, data, and/or other artifacts have been made available at \url{\vldbavailabilityurl}.
\endgroup
}
\else
\pagestyle{\vldbpagestyle}
\fi

\section{Introduction}
\label{sec:intro}

Organizations have been extensively relying on personal data to support a growing spectrum of businesses, from music recommendations to life-saving coronavirus research~\cite{wang-etal-2020-cord}.
This type of data is often structured and carries sensitive information about individuals.
Reckless data sharing for data-driven applications and research causes great privacy concerns~\cite{user_concern, ibm_report} and penalties~\cite{GDPR}.
Differential privacy (DP)~\cite{DBLP:conf/icalp/Dwork06} has emerged as a standard data privacy guarantee by government agencies~\cite{DBLP:conf/kdd/Abowd18, Hawes2020Implementing} and companies~\cite{Erlingsson14Rappor, Greenberg16Apples, DBLP:journals/pvldb/JohnsonNS18}.
Informally, the output of a data sharing process that satisfies DP has a similar distribution whether an individual's data is used for the computation or not.
Hence, the output cannot be used to infer much about any individual's data and therefore is considered ``private''.

Differential privacy is often achieved via randomization, such as injecting controlled noise into the input data~\cite{rand_response} based on the required privacy level, and hence there is a trade-off between privacy and the utility of this data to downstream applications.
One approach often followed in prior work focuses on the optimization of this trade-off for a given application (e.g., releasing  statistics~\cite{DBLP:conf/kdd/Abowd18, onthemap}, building prediction models~\cite{DBLP:conf/iclr/PapernotSMRTE18, DBLP:conf/ccs/AbadiCGMMT016}, answering SQL queries~\cite{DBLP:conf/sigmod/0002HIM19, DBLP:journals/pvldb/JohnsonNS18, DBLP:journals/pvldb/KotsogiannisTHF19, DBLP:conf/sigmod/McSherry09}). 
For example, APEx~\cite{DBLP:conf/sigmod/0002HIM19} is designed for data exploration; 
for each query, APEx searches the best differentially private algorithm that can answer the query accurately with the minimum privacy cost.
This line of work allows the fine-tuning of an algorithm for the optimal trade-off between the privacy cost and the accuracy of the given application, but the released output may not be useful for other applications.
Running a new application on the same dataset usually requires additional privacy cost.

An attractive alternative approach is to publish a differentially private synthetic database instance with a set of desired properties such as similar value distributions or dependency structure, 
with the hope that it has the same utility or is as useful as the original dataset to a large class of downstream applications that require those properties. 
For example, the US Census Bureau released differentially private census data, and it has been shown useful to keep similar home-workplace distribution as the true data to populate the mapping application~\cite{onthemap}.
Privately releasing synthetic data avoids designing separate mechanism for each target application, and the privacy cost is incurred only once for all supported applications due to the post-processing property of differential privacy~\cite{DBLP:journals/fttcs/DworkR14}.

\subsection{Problems with Current DP Data Synthesis}\label{sec:intro:problems}
For applications that consume structured data with predefined schema in a SQL database, it is important for the synthetic data to keep \emph{the structure of the data} --- the underlying correlations and dependencies among tuples and attributes.
This structure is often expressed as integrity and schema constraints, 
such as functional dependencies between attributes or key constraints between tables.
Otherwise, the synthesized data is not useful for any downstream tasks that require this structure to be preserved.

In general, generating differentially private synthetic data based on true data faces fundamental challenges.
Take answering statistical queries as an example application.
Prior work~\cite{DBLP:conf/stoc/BlumLR08, DBLP:conf/stoc/DworkNRRV09, DBLP:conf/tcc/UllmanV11, DBLP:conf/icml/GaboardiAHRW14} have shown that the running time for sampling a synthetic dataset that is accurate for answering a large family of statistics (e.g., all $\alpha$-way marginals) grows exponentially in the dimension of the data.
On the other hand, an efficient private data generation algorithm fails to offer the same level of accuracy guarantees to all the queries.
Existing practical methods (e.g.,~\cite{DBLP:journals/corr/abs-1812-02274, DBLP:conf/kdd/ChenXZX15, DBLP:conf/iclr/JordonYS19a, DBLP:conf/sigmod/ZhangCPSX14, DBLP:conf/pods/BarakCDKMT07}) therefore choose to privately learn only a subset of queries or correlations to model the true data and then sample database instances based on the learned information.
However, the structure of the data is not explicitly captured by these methods and thus are poorly preserved in the synthetic data. 
In particular, all these methods assume tuples in the database instance are independent and identically distributed (i.i.d.), and sample each tuple independently.
The output database instance has a significant number of violations to the structure constraints in the truth. 

\example{\label{example:post_clean}
Consider the Adult dataset~\cite{Dua:2019} consisting of 15 attributes with denial constraints~\cite{DBLP:books/acm/IlyasC19}, such as `two tuples with the same education category cannot have different education numbers', and `tuples with higher capital gain cannot have lower capital loss'. 
There is no single violation of these constraints in the true data, but the synthetic data generated by the state-of-the-arts including PrivBayes~\cite{DBLP:conf/sigmod/ZhangCPSX14}, PATE-GAN~\cite{ DBLP:conf/iclr/JordonYS19a}, and DP-VAE~\cite{ DBLP:journals/corr/abs-1812-02274} have up to 32\% of the tuple pairs failing these constraints (Table~\ref{tab:dc_vio}).

\begin{figure}[t]
    \centering
    \begin{subfigure}[h]{0.24\textwidth}%
        \includegraphics[width=\textwidth]{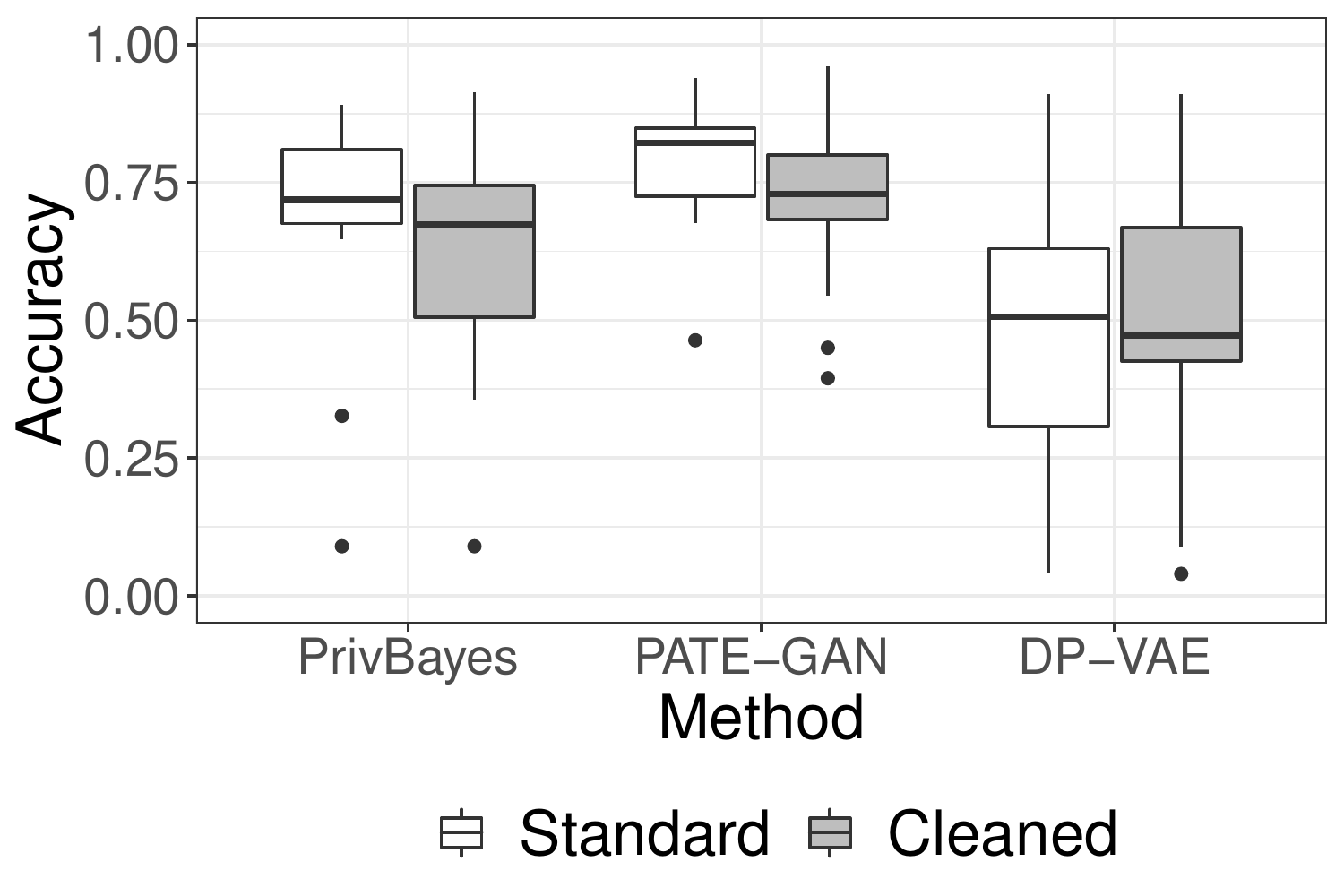}%
        \vspace{-.75em}
        \caption{Accuracy}
        \label{fig:post_clean:accuracy}
    \end{subfigure}%
    \begin{subfigure}[h]{0.24\textwidth}%
        \includegraphics[width=\textwidth]{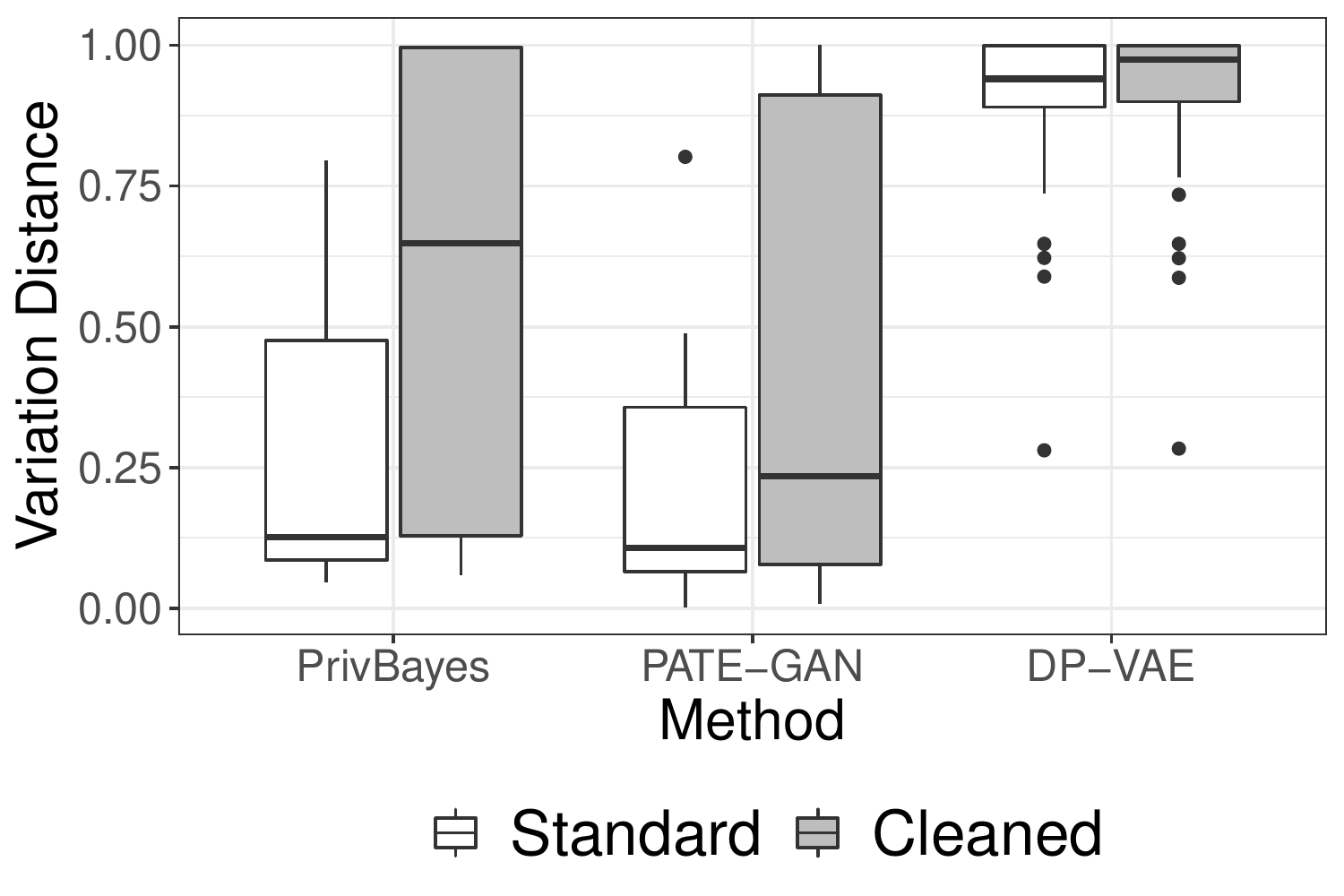}%
        \vspace{-.75em}
        \caption{2-way marginal}
        \label{fig:post_clean:2}
    \end{subfigure}
    \vspace{-.75em}
     \caption{
     A synthetic Adult data using PrivBayes, PATE-GAN and DP-VAE satisfying ($\epsilon=1, \delta=10^{-6}$)-DP, with and without fixing the integrity violations (labeled as `cleaned' and `standard', respectively). 
     Each point in Figure~\ref{fig:post_clean:accuracy} represents the testing accuracy for one target attribute.
     Each point in Figure~\ref{fig:post_clean:2} represents the total variation distance between the true and synthetic Adult.
     More details are in \S~\ref{sec:evaluation}.       
     }
    \label{fig:post_clean}
\end{figure}

However, na\"ivly repairing the incorrect structure constraints in the synthetic data can compromise the usefulness.
We applied state-of-the-art data cleaning method~\cite{DBLP:journals/pvldb/RekatsinasCIR17} to fix the violations in the synthetic data generated by the aforementioned three methods. Then we evaluated their usefulness in training classification models and building 2-way marginals. Figure~\ref{fig:post_clean} shows that the repaired synthetic data (labeled as `cleaned') have lower classification quality (i.e., smaller accuracy score) and poorer marginals (i.e., larger distance) compared to the synthetic data with violations (labeled as `standard'). 
Though the repaired synthetic data managed to comply with structure constraints, they become less useful for training models and releasing marginal statistics. 
}

\subsection{Constraint-Aware DP Data Synthesis}
\revise{To solve the aforementioned problems,} we are motivated to design an end-to-end synthetic data generator that preserves both the structure of the data and the privacy of individual data records.
In this work, we consider an important class of structure constraints, the \emph{denial constraints} (DCs)~\cite{DBLP:books/acm/IlyasC19}, and 
we present \system\footnote{Kamino was the planet in Star Wars, renowned for the technology of clone armies.}, a system for constraint-aware differentially private data synthesis. 

Our solution is built on top of the probabilistic database framework~\cite{DBLP:series/synthesis/2011Suciu, DBLP:conf/icdt/SaIKRR19}, which models a probability distribution over ordinary databases and incorporates the denial constrains as parametric factors.
Database instances that share similar structural and statistical correlations with the true data are modeled to have similar probabilities. 
We first privately learn a parametric model of the probabilistic database,
and then sample a database instance from the model as a post-processing step.
To make it more efficient, we decompose the joint probability of a database instance into a chain of conditional probabilities, and privately estimate tuple distribution using tuple embedding~\cite{aimnet} and the attention mechanism~\cite{DBLP:journals/corr/BahdanauCB14} for mixed data types (categorical and numerical).
As we explicitly consider additional correlation structures compared to prior work, 
\system can incur more performance cost and utility cost for other applications given the same level of privacy constraint.
Our empirical evaluation shows that the performance overhead and accuracy payoff are negligible.
We also show that while preserving DCs, 
\system produces synthetic data that have comparable and even better quality for classification applications and marginal queries than the state-of-the-art methods on differentially private data synthesis.

We highlight the main contributions of this work as follows:
\squishlist
\item We believe this is the first work to consider denial constraints in differentially private data synthesis, which are important properties for structured data. We use probabilistic database framework to incorporate DCs and attribute correlations.
\item We develop an efficient learning and sampling algorithm for \system by decomposing the probabilistic database model into a chain of submodels, based on the given constraints (\S~\ref{sec:ps} \& \S~\ref{sec:solution}). 
\item We design a private learning algorithm in \system to learn the weights of given DCs to allow interpreting them in the model as soft constraints (\S~\ref{sec:weight_learning}).
\item We build the prototype for \system, the first end-to-end system for differentially private data generation with DCs, and apply advanced privacy composition techniques to obtain a tight end-to-end privacy bound (\S~\ref{sec:analysis}).
\squishend
We evaluate \system over real-world datasets and show that the synthetic data have similar violations to the given DCs as in the true data, and they also achieve the best or close to best data usefulness in the marginal queries (variation distance) and the learning tasks (accuracy and F1), compared to the state-of-the-art methods (\S~\ref{sec:evaluation}).

\section{Preliminaries}
\label{sec:preliminaries}
We consider a relational database schema of a single relation $R = \{A_1, \cdots, A_k\}$ with $k$ attributes.
Let $D$ be a database instance of this schema $R$ and consist of $n$ tuples $\{t_1, \cdots, t_n\}$.
Each tuple $t_i \in D$ has an implicit identifier $i$, and $t_i[A_j]$ denotes the value taken by the tuple $t_i$ for attribute $A_j$ from its domain $\mathcal{D}(A_j)$.
Index 1 refers to the first element in a list/array.

\subsection{Denial Constraints}\label{sec:preliminaries:dc}

Denial constraints (DCs)~\cite{DBLP:books/acm/IlyasC19} are used in practice by domain experts to specify the structure of the data, such as functional dependency (FD)~\cite{DBLP:journals/cj/HuhtalaKPT99} and conditional FD~\cite{DBLP:journals/tkde/FanGLX11}.
In case of missing DCs, recent work have designed algorithms to automatically discover DCs from the database instance~\cite{DBLP:journals/pvldb/ChuIP13, DBLP:journals/pvldb/BleifussKN17}. 

We express a DC as a first-order formula in the form of $\phi: \forall t_i, t_j, \cdots \in D, \neg (P_1 \land \cdots \land P_m)$.
Each predict $P_i$ is of the form $(v_1\; o\; v_2)$ or $(v_1\; o\; c)$,
where $v_1, v_2 \in t_x[A]$, $x\in \{i,j, \cdots\}$, $A\in R$, $o\in \{=, \neq, >,\geq, <, \leq \}$, and $c$ is a constant.
We will omit universal quantifiers $\forall t_i,t_j,\ldots$ hereafter for simplicity.

\example{\label{example:dc}
Consider a database instance $D$ with schema $R=\{age, edu\_num,edu, cap\_gain, cap\_loss \}$, and three DCs:

\noindent$\phi_1$: 
$\neg (t_i[edu]=t_j[edu] \land t_i[edu\_num] \neq t_j[edu\_num])$

\noindent$\phi_2$: 
$\neg (t_i[cap\_gain]>t_j[cap\_gain] \land t_i[cap\_loss] < t_j[cap\_loss])$

\noindent$\phi_3$: 
$\neg (t_i[age] < 10 \land t_i[cap\_gain] > 1M)$

The first DC $\phi_1$ expresses an FD $edu \rightarrow edu\_num$.
It states that for any two tuples with same $edu$, their $edu\_num$ must be the same too. 
The second DC $\phi_2$ states that for any two tuples, if one's $cap\_gain$ is greater than the other's, its $cap\_loss$ cannot be smaller.
The third DC $\phi_3$ is a unary DC that enforces every tuple with $age$ less than 10 cannot have $cap\_gain$ more than 1 million.
}

A DC states that all the predicts cannot be true at the same time, otherwise, a violation occurs. 
We use $V(\phi, D)$ to represent the set of tuples (for unary DCs) or tuple groups (for non-unary DCs) that violates DC $\phi$ in a database instance $D$.
we refer to DC $\phi$ as a hard DC if no violations are allowed (i.e., $V(\phi, D)=\emptyset$), or a soft DC if a database instance can have violations.
Note that the set of DC violations expands monotonicity with respect to the size of a database instance, that is for a subset instance $\hat{D} \subset D$, $V(\phi, \hat{D}) \subset V(\phi, D)$.
We also use $\mathcal{A}_\phi$ to represent the set of attributes that participate in the DC $\phi$. For example, $\mathcal{A}_{\phi_1} = \{edu, edu\_num \}$.

\subsection{Probabilistic Database}\label{sec:preliminaries:pdb}
\revise{The probabilistic database framework~\cite{DBLP:conf/icdt/SaIKRR19} has been used in practice~\cite{aimnet, DBLP:journals/pvldb/RekatsinasCIR17} to model observed data that do not fully comply with a given set of DCs. }
Intuitively, a database instance with few violations is more likely. 
Given a set of DCs $\Phi$ and their weights $\{w_\phi \mid \phi \in \Phi\}$,
the probability of an instance $D$ is defined as follows:
\begin{equation}\label{equ:pdb}
\Pr(D) \propto \prod_{t \in D} \Pr(t) \times \exp(-\sum_{\phi \in \Phi} w_\phi \times |V(\phi, D)|)
\end{equation}
where $\prod_{t \in D} \Pr(t)$ models a tuple-independent probabilistic database~\cite{DBLP:series/synthesis/2011Suciu, DBLP:conf/icdt/SaIKRR19},
wherein each tuple independently comes from a probability distribution over tuples,
and $|V(\phi, D)|$ is the size of violations of DC $\phi$ on $D$.
Each DC $\phi$ is associated with a weight $w_\phi$ and each violation of $\phi$ contributes a factor of $\exp(-w_\phi)$ to the probability of a random database instance $D$.
This model captures both hard and soft DCs.
For hard DCs, we set weights to be infinitely large, then a database instance with any violations has a small probability.
For soft DCs, having more violations decreases its probability.

To learn a probabilistic database, one needs to learn the probability of tuples $\Pr(t)$ as well as the weights of DCs $w_{\phi}$.
The goal is to find the set of parameters $\{\Pr(t)$, $w_\phi\}$ that maximizes the product of the likelihoods of all the training database samples~\cite{DBLP:conf/icdt/SaIKRR19}.
\revise{
The observed data will be used to learn the parameters in the model. 
We assume the distribution does not change.
}

\subsection{Tuple Embedding}\label{sec:preliminaries:embedding}

In this work, we express the tuple probability as the product of a chain of conditional probabilities:
\begin{equation}\label{equ:tupleprob}
\Pr(t) = \Pr(t[A_1]) \prod_{j=2}^k \Pr(t[A_j] \mid t[A_1, \cdots, A_{j-1}])
\end{equation}
Each conditional probability is learned as a discriminative model based on \emph{tuple embedding}~\cite{aimnet} and \emph{attention mechanism}~\cite{DBLP:journals/corr/BahdanauCB14}. 
Similar to word embedding that models words in vectors of real numbers~\cite{DBLP:conf/nips/MikolovSCCD13}, tuple embedding has been applied to model tuples by encoding tuples into the space of real numbers~\cite{aimnet, DBLP:journals/pvldb/EbraheemTJOT18}.

\ifpaper

\else
Consider the discriminative model used in AimNet~\cite{aimnet} that predicts the value of \emph{target} attribute $A_j$ based on the values of a set of \emph{context} attributes $\{A_1,\ldots,A_{j-1}\}$.
AimNet transforms each attribute value into a vector embedding with fixed dimension $d$.
For an attribute with continuous values $\vec{x} \in \mathbb{R}^{d'}$, where $d'<d$, 
AimNet first standardizes each dimension to zero mean and unit
variance,
and then apply a linear layer followed by a non-linear ReLU layer to
obtain a non-linear transformation of the input:
$\vec{z} = B\omega(A\vec{\underline{x}} + \vec{c} ) + \vec{d}$, 
where A, B, $\vec{c}$, $\vec{d}$ are learned parameters and $\omega$ is a ReLU.
For each attribute with discrete values, AimNet associates it with a learnable lookup table mapping embeddings to domain values. 

AimNet relies on the attention mechanism~\cite{DBLP:journals/corr/BahdanauCB14} to learn structural dependencies between different attributes of the input data and uses the attention weights to combine the representations of inputs into an vector representation (the context vector) for the target attribute. 
To predict a target attribute value,
it learns the transformation from context vector back to a value in the domain of the target attribute. 
The output of a discriminative model is the learned representation of all the attributes, and a list of prediction probabilities for all values of a target attribute with the discrete domain, or the regression parameters (mean and std) of a Gaussian distribution for a target attribute with a continuous domain.
\fi

\subsection{Differential Privacy}\label{sec:preliminaries:dp}

Differential privacy (DP)~\cite{DworkMNS06,DBLP:journals/fttcs/DworkR14} is used as our measure of privacy.

\begin{definition}[Differential Privacy (DP)~\cite{DBLP:journals/fttcs/DworkR14}]
A randomized algorithm $\mathcal{M}$
achieves ($\epsilon,\delta$)-DP if for all $\mathcal{S} \subseteq$ Range($\mathcal{M}$) and for any two database instances $D,D' \in \mathcal{D}$ that differ only in one tuple:
$$\Pr[\mathcal{M}(D) \in \mathcal{S}]\le e^\epsilon \Pr[\mathcal{M}(D') \in \mathcal{S}] + \delta.$$
\end{definition}

The privacy cost is measured by the parameters ($\epsilon,\delta$). The smaller the privacy parameters are, the stronger the privacy offers.  
Complex DP algorithms can be built from the basic algorithms following two important properties of differential privacy:
1) Post-processing~\cite{DBLP:conf/eurocrypt/DworkKMMN06} states that for any function $g$ defined over the output of the mechanism $\mathcal{M}$, if $\mathcal{M}$ satisfies ($\epsilon,\delta$)-DP, so does $g(\mathcal{M})$;
2) Composability~\cite{DBLP:conf/icalp/Dwork06} states that if $\mathcal{M}_1$, $\mathcal{M}_2$, $\cdots$, $\mathcal{M}_k$ satisfy ($\epsilon_1,\delta_1$)-, ($\epsilon_1,\delta_1$)-, $\cdots$, ($\epsilon_k,\delta_k$)-DP, 
then a mechanism sequentially applying $\mathcal{M}_1$, $\mathcal{M}_2$ to $\mathcal{M}_k$ satisfies ($\sum_{i=1}^k\epsilon_i, \sum_{i=1}^k\delta_i$)-DP.

Gaussian mechanism~\cite{DBLP:journals/fttcs/DworkR14} is a widely used DP algorithm. 
Given a function $f:\mathcal{D} \rightarrow \mathbb{R}^d$, the Gaussian mechanism adds noise sampled from a Gaussian distribution $\mathcal{N}(0,S_f^2\sigma^2)$ to each component of the query output, where $\sigma$ is the noise scale and $S_f$
is the $L_2$ sensitivity of function $f$, 
which is defined as
$S_f = \max_{D,D'\text{differ in a row}} ||f(D) - f(D')||_2$. 
For $\epsilon\in (0,1)$, if $\sigma \geq \sqrt{2\ln(1.25/\delta)}/\epsilon$, 
then the Gaussian mechanism satisfies $(\epsilon,\delta)$-DP.

Gaussian mechanism has been applied to answer counting queries~\cite{DBLP:journals/vldb/LiMHMR15}. 
It has also been used in differentially private stochastic gradient descent (DPSGD)~\cite{WM10, BST14, song2013stochastic, DBLP:conf/ccs/AbadiCGMMT016}.
The gradients of SGD are the random variables to which the noise is added. As there is no a priori bound of the gradient, the sensitivity $S_f$ is set by clipping the maximum $L_2$ norm of the gradient to a user-defined parameter $C$.

\ifpaper
\else
\stitle{Semantics of DP.}
Prior work~\cite{DBLP:journals/jpc/DworkN10,DBLP:conf/kdd/GantaKS08,relationshipprivacy09,coupledworld13,DBLP:journals/tods/KiferM14,DBLP:conf/sigmod/HeMD14,DBLP:journals/jpc/Kasiviswanathan14} 
offer semantic interpretations of DP with respect to adversarial knowledge: \emph{informally, regardless of external knowledge, an adversary with access to the sanitized database draws the same conclusions whether or not one's data is included in the original database}. Ganta et al.~\cite{DBLP:conf/kdd/GantaKS08} formalize
the notion of ``external knowledge'', and of ``drawing conclusions'' 
respectively via 
(i) a \emph{prior} probability distribution $b[\cdot]$ on the database domain $\mathcal{D}$ of size $n$; 
and (ii) the corresponding \emph{posterior} probability distribution: given a transcript $t$ outputted by mechanism $\mathcal{M}$, the adversary updates his belief about the database $D$ using Baye's rule to obtain a posterior
$\hat{b}[D|t] = \frac{\Pr[\mathcal{M}(D)=t]b[D]}{\sum_{D'}\Pr[\mathcal{M}(D')=t]b[D']}$. 
Consider the hypothetical scenario where
person $i$'s data is not used, denoted by $\mathcal{M}(D_{-i})$, given a transcript $t$, the updated posterior belief about the database $D$ is defined as  $\hat{b}_i[D|t]=\frac{\Pr[\mathcal{M}(D_{-i})=t]b[D]}{\sum_{D'}\Pr[\mathcal{M}(D'_{-i})=t]b[D']}$.
A DP mechanism $\mathcal{M}$ with proper privacy parameters (sufficiently small $\delta$) can prevent the adversary from drawing different conclusions about whether or not person $i$'s data was used, i.e., the statistical difference between $\hat{b}[\cdot|t]$ and  $\hat{b}_i[\cdot|t]$ is small for $D$ drawn from $b[\cdot]$ and $t$ drawn from $\mathcal{M}(D)$ with a high probability~\cite{DBLP:conf/kdd/GantaKS08,DBLP:journals/jpc/Kasiviswanathan14}. 
This posterior-to-posterior comparison applies to arbitrary prior of adversaries, unlike prior-to-posterior approaches~\cite{DBLP:journals/jpc/DworkN10,DBLP:journals/tods/KiferM14}. 

\fi

\section{\system Overview}
\label{sec:ps}
To solve the shortcomings of the current differentially private data synthesis approaches mentioned in \S~\ref{sec:intro:problems}, 
we state our problem definition and provide a high-level description of our approach.

\subsection{Problem Statement}
\label{sec:problem}

Given a private database instance $D^*$ with schema and domain, 
a set of denial constraints $\Phi$ with information about their hardness, and a differential privacy budget $(\epsilon,\delta)$, 
we would like to design a process $P$ that generates a useful synthetic database instance $D^\prime$ as $D^*$ (e.g., the same statistics and attribute correlations) while meeting two additional requirements:
\begin{enumerate}[leftmargin=*, label=R\arabic*.]
\item (Data Consistency) 
We consider data consistency with respect to the set of denial constraints $\Phi$ from the input:  
for each DC $\phi \in \Phi$, $D^*$ and $D^\prime$ have a similar number of violations, 
i.e., $|V(\phi, D^\prime)| \approx |V(\phi, D^*)|$.
\item (Privacy Guarantee) The process $P$ that outputs $D^\prime$ achieves ($\epsilon,\delta$)-differential privacy: 
for any set of output instances $\mathbb{D}$ outputted by $P$, $\Pr(P(D_1) \in \mathbb{D}) \leq e^\epsilon\Pr(P(D_2) \in \mathbb{D} ) + \delta$, for any two neighboring $D_1$ and $D_2$ differing in one record.
\end{enumerate}
DC constraints $\Phi$ are public in our problem and can be modeled as part of the adversary's prior. 
This subsumes the special case when $\Phi$ are not public to the adversary. 
\ifpaper
\revise{More discussion on the semantics of DP can be found in our full paper~\cite{full_paper}}.
\else
The semantic privacy results by Ganta et al.~\cite{DBLP:conf/kdd/GantaKS08,DBLP:journals/jpc/Kasiviswanathan14} (\S~\ref{sec:preliminaries:dp}) are applicable to our problem and prior work on DP data synthesis, and hence this work will focus on the design of a DP mechanism. We will leave the mechanism design for stronger semantic privacy guarantees to future work.
\fi

\ifpaper

\else
Na\"ive solutions fail either R1 or R2, or both.
If the privacy guarantee (R2) is not mandatory, then releasing the private database instance $D^*$ would suffice for data consistency (R1).
On the contrary, if privacy is the only goal, then outputting an empty dataset will suffice.
As shown in Example~\ref{example:post_clean}, 
existing practical work on private data synthesis~\cite{DBLP:journals/corr/abs-1812-02274, DBLP:conf/iclr/JordonYS19a, DBLP:conf/sigmod/ZhangCPSX14} do not preserve DCs. 
A post-cleaning step 
can improve the data consistency, but 
it is at the cost of the usefulness of the synthetic for other applications (Figure~\ref{fig:post_clean}). 
To the best of our knowledge, there is no existing work to address both the data consistency and privacy guarantee in the synthesis process.
\fi

\subsection{Methodology Overview}
\label{sec:overview}
Recall from \S~\ref{sec:preliminaries:pdb}, the probabilistic database model is a parametric model to describe the probability of instances.
We adopt the probabilistic database model to represent databases with denial constraints. 
There are two main steps: 
(i) privately learn the unknown parameters in the probabilistic database model with samples from the true data; 
(ii) sample a database instance based on the learned probabilistic database model. 
However, both steps are challenging.
First, it is well known that finding the analytical solution of the parameters of a probabilistic database without privacy concerns is \#P-complete~\cite{DBLP:series/synthesis/2011Suciu, DBLP:journals/ml/RichardsonD06}, and approximate methods such as gradient descent may not converge to a global optimum~\cite{DBLP:conf/icdt/SaIKRR19}, due to the large sampling space of tuples (cross product of all attributes' domain sizes) and of instances (exponential to the number of possible tuples).
Second, prior work~\cite{DBLP:conf/stoc/DworkNRRV09, DBLP:conf/stoc/BlumLR08, DBLP:conf/tcc/UllmanV11, DBLP:conf/icml/GaboardiAHRW14} show that there is no efficient DP algorithm that can generate a database, which maintains accurate answers for an exponential family of learning concepts (e.g. the set of parameters in the probabilistic database model).

To tackle both the efficiency and the privacy challenge, we factorize the probability distribution of a database instance into a set of conditional probabilities given a subset of tuples and attributes, and learn them accordingly.  
\revise{
We sample an instance based on the learned conditional probabilities. 
}

\stitle{Probabilistic database decomposition.}
We express the probability distribution of a database instance in Eqn.~\eqref{equ:pdb} into a chain of conditional probabilities based on two sequences (i) a sequence of tuple ids; and (ii) a sequence of attributes. 

First, given a sequence of tuple ids $(1,2,\ldots,n)$ in $D$, for any DC $\phi$, the set of its violations in $D$, i.e., $V(\phi,D)$, can be iteratively computed by adding new violations introduced by tuple $t_i$ with respect to its prefix tuples $D_{:i} = [t_1, t_2, \cdots, t_{i-1}]$ (with $D_{:1} = \emptyset$) from $D$, for $i=1,\ldots,n$. Let $V(\phi, t_i \mid D_{:i})$
denote the set of new violations caused by tuple $t_i$ with respect to $D_{:i}$. Then we have 
\begin{align}
|V(\phi, D)| 
&= |V(\phi, t_1)| + |V(\phi, t_2 \mid D_{:2})| + \cdots + |V(\phi, t_n \mid D_{:n})| \nonumber \\
&=\sum_{i=1}^n |V(\phi, t_i \mid D_{:i})|
\label{equ:vio}
\end{align}
This allows us to decompose Eqn.~\eqref{equ:pdb} as
\begin{align}
\Pr(D) &\propto \left( \prod_{i=1}^n \Pr(t_i) \right) \times \exp\left(-\sum_{\phi \in \Phi} w_\phi \sum_{i=1}^n |V(\phi, t_i \mid D_{:i})|\right) \nonumber \\
&= \prod_{i=1}^n \left[ \Pr(t_i) \times \exp \left( -\sum_{\phi \in \Phi} w_\phi \times |V(\phi, t_i \mid D_{:i})| \right) \right]
\label{equ:pdb2}
\end{align}

Next, we define a schema sequence $S$ as an ordered list of all attributes in the schema.
Similarly, let $S_{:j}$ represent all prefix attributes of the $j$th attribute in $S$ and $S_{:1} = \emptyset$ for the purpose of uniform representation. This schema sequence allows us further decompose the set of violations. Let $\Phi_{A_j}$ represent the set of DCs in $\Phi$ that can be fully expressed with the first $j$ attributes in $S$, but cannot be expressed with only the first $j-1$ attributes.

\example{\label{example:dcattribute}
Continue with Example~\ref{example:dc}, 
given a schema sequence $S=[age, edu\_num, edu, cap\_gain, cap\_loss]$, 
we can verify that 
$\Phi_{A_3} = \{\phi_1\}$, 
as the attributes $\{edu\_num, edu\}$ for $\phi_1$ are covered by the first 3 attributes in $S$, but not the first 2 attributes. 
}

Notice that for a DC $\phi \in \Phi_{A_j}$, given a tuple $t_i$, its number of violations $|V(\phi, t_i \mid D_{:i})|$ only depends on the values of the first $j$ attributes in $S$ (i.e., $S_{:j+1}$).
As a result, we can rewrite the weighted sum of violations from  Eqn.~\eqref{equ:pdb2} as follows:
\begin{align}
&\sum_{\phi \in \Phi} w_\phi \times |V(\phi, t_i \mid D_{:i})| = \sum_{j=1}^k \sum_{\phi \in \Phi_{A_j}} w_\phi \times |V(\phi, t_i \mid D_{:i})| \nonumber \\
&=\sum_{j=1}^k \sum_{\phi \in \Phi_{A_j}} w_\phi \times |V(\phi, t_i[S_{:j+1}] \mid D_{:i}[S_{:j+1}])|
\label{equ:pdb3}
\end{align}

Based on the same schema sequence $S$, the tuple probability $\Pr[t_i]$ can be written as $\prod_{j=1}^k \Pr \left( t_i[A_j] \mid t_i[S_{:j}] \right)$ by the chain rule. 

Finally, we have the database probability in Eqn.~\eqref{equ:pdb2} expressed as
\begin{align}
\Pr(D) \propto &\prod_{j=1}^k \prod_{i=1}^n \Big[ \Pr(t_i[A_j] \mid t_i[S_{:j+1}]) \times \nonumber \\
&\exp (-\sum_{\phi \in \Phi_{A_j}} w_\phi \times |V(\phi, t_i[S_{:j+1}] \mid D_{:i}[S_{:j+1}] )|) \Big]
\label{equ:pdb4}
\end{align}

Eqn.~\eqref{equ:pdb4} in fact presents an iterative process to sample a database instance $D$ based on (i) the schema sequence ($j\in [1,k]$), and (ii) the tuple id sequence $(i\in [1,n])$. 
Unlike the tuple id sequence, the schema sequence specifies an ordering of attributes, where each attribute solely depends on the prefix attributes to make correct prediction.
However, it is challenging to find the optimal schema sequence~\cite{DBLP:conf/aistats/Chickering95}, and hence we apply a greedy heuristic algorithm to derive a good one.
In this work, we assume $\Pr[t_i]$ are the same for all tuples. 
Therefore, we just need to learn $k$ (conditional) probabilities $\Pr(t[A_j]|t[S_{:j+1}])$, the weight of DCs $w_\phi$, and the number of DC violations with respect to the prefix tuples. We will use the following example to illustrate the sampling process. 

\begin{figure}[t]
	\center
	\includegraphics[width=\columnwidth]{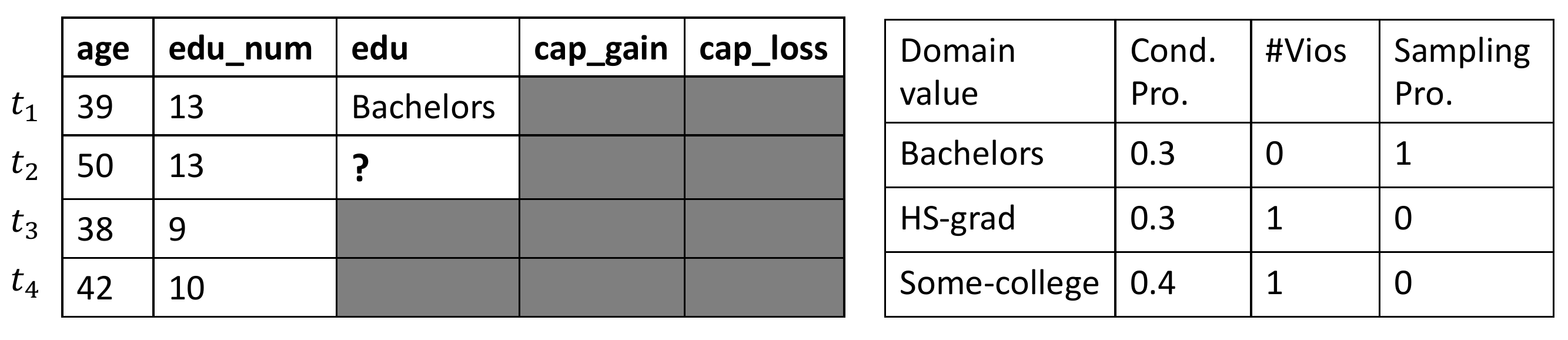}
	\vspace{-.75em}
	\caption{Sampling values in an instance (Example~\ref{exp:overall}).}
	\label{fig:iterative-example}
\end{figure}

\example{\label{exp:overall}
Continue with Examples~\ref{example:dc} and~\ref{example:dcattribute}. Consider all three DCs be hard with infinitely large weight $w_\phi$. Suppose we have already privately learned the conditional distributions from the true data. The construction of $D'$ of 4 tuples works as follows. 

We start with the first attribute $age$. From $t_1$ to $t_4$, we sample a value independently based on the distribution $\Pr(t[age])$. 
Then, we move on to the second attribute, $edu\_num$. 
There is no DC between $age$ and $edu\_num$, each cell from $t_1$ to $t_4$ is filled with a sample based on the conditional distribution $\Pr(t[edu\_num] \mid t[age])$. 

Next, for the third attribute $edu$ (shown in Figure~\ref{fig:iterative-example}), DC $\phi_1$ becomes active as all its relevant attributes ($edu\_num,edu$) have been seen in the sequence. A cell value \texttt{Bachelors} is directly sampled for $t_1[edu]$ from the the conditional distribution $\Pr(t[edu]\mid t[edu\_num=13,age=39])$. 
For $t_2[edu]$, let's say 
the noisy conditional distributions of $edu$ given $age=50$ and $edu\_num=13$ are: (\texttt{Bachelors}, 0.3), (\texttt{HS-grad}, 0.3) and (\texttt{Some-college}, 0.4). 
Consider the infinitely large weight for $\phi_1$,  $edu$ values other than \texttt{Bachelors} will cause violations to $t_1$ and hence their probabilities become very small. 
Therefore, \texttt{Bachelors} is sampled with high probability. 

\revise{After all cells are filled, we get a synthetic  instance $D^\prime$.
Optionally, the Markov Chain Monte Carlo (MCMC) sampling~\cite{mcbook} could be applied to improve the accuracy by randomly choosing a cell $t_i[A_j]$ to re-sample, conditioning on all other cells $D^\prime\setminus \{t_i[A_j]\}$. 
This step repeats for a fixed number of times or till convergence.
}
}

\stitle{System overview.} 
Algorithm~\ref{alg:overall} describes the overall process of our solution \system. 
\system first chooses a schema sequence $S$ based on the schema $R$, domain $\mathcal{D}$, and DCs $\Phi$ (Line~\ref{alg:overall:seq}). 
Then it finds a suitable parameter set $\Psi$ for the subsequent algorithms to ensure the overall privacy loss is bounded by $(\epsilon,\delta)$-DP (Line~\ref{alg:overall:conf}). 
The algorithms $\textsc{TrainModel}(\cdot)$ and $\textsc{LearnWeight}(\cdot)$ privately learn the tuple distribution and weights of the DCs from the private true data $D^*$ (Lines~\ref{alg:overall:train}-\ref{alg:overall:weight}). 
Last, \system applies a constraint-aware sampling algorithm to generate a synthetic database instance. 
We first present the key algorithms (Algorithms~\ref{alg:sequence},~\ref{alg:model} and~\ref{alg:syn}) when the weights of DCs are given in \S~\ref{sec:solution}, 
and then explain how to learn the DC weights (Algorithm~\ref{alg:weight_learning}) in \S~\ref{sec:weight_learning}. 
Last, privacy analysis and parameter search (Algorithm~\ref{alg:paras_search}) are explained in \S~\ref{sec:analysis}.

\revise{
Our system assumes the inputs are static, since we rely on the database instance to learn the generative process (i.e., Algorithms~\ref{alg:model},~\ref{alg:syn} and~\ref{alg:weight_learning}), and on the DCs to learn the weights and attribute sequence. 
However, \system can tolerate small input changes as long as the data distribution and DCs are intact.
For now, if DC changes resulting in a different sequence, we re-run \system; if the changes significantly shift the distribution, we re-run the generative process. 
Future work can apply general DP techniques~\cite{DBLP:conf/nips/CummingsKLT18} for dynamically growing databases for better utility.
}

\begin{algorithm}[t]
\caption{Constraint-aware differentially private data synthesis}\label{alg:overall}
\begin{algorithmic}[1]
\Require Private instance $D^*$, schema $R$, domain $\mathcal{D}$
\Require DCs $\Phi$, privacy budget $(\epsilon,\delta)$
\Procedure {\system}{$D^*, R, \mathcal{D}, \Phi, \epsilon,\delta$}
\State $S \gets \textsc{Sequencing}(R,\mathcal{D},\Phi)$ \Comment{Algorithm~\ref{alg:sequence}} \label{alg:overall:seq}
\State $\Psi \gets$ \textsc{SearchDParas}($\epsilon,\delta, \mathcal{D}, S$)\Comment{Algorithm~\ref{alg:paras_search}}
\label{alg:overall:conf}
\State $M \gets \textsc{TrainModel}(D^*, S,\mathcal{D}, \Psi)$ \Comment{Algorithm~\ref{alg:model}}
\label{alg:overall:train}
\State $W \gets$  
\textsc{LearnWeight}($D^*,\Phi,S,M,\Psi$)\Comment{Algorithm~\ref{alg:weight_learning}}
\label{alg:overall:weight}
\State $D^\prime \gets \textsc{Synthesize}(S,M, \Phi,\mathcal{D}, W)$ \Comment{Algorithm~\ref{alg:syn}}
\label{alg:overall:syn}
\State \textbf{return} $D^\prime$
\EndProcedure
\end{algorithmic}
\end{algorithm}

\section{\system with Known DC Weights
}\label{sec:solution}

For simplicity of presentation, in this section, we consider the weights of the constraints are given (e.g., the weights for hard DCs are set infinitely large). 
We first present our private learning algorithm for the tuple probability and then the database sampling algorithm. 
Last, we show our choice of schema sequence in \system. 

\subsection{
Private Learning of Tuple Probability
}\label{sec:model}

Recall Equ.~\eqref{equ:tupleprob} that,
given a schema sequence $S=[A_1,A_2,\ldots,A_k]$, 
the tuple probability becomes $\Pr[t] = \Pr(t[A_1]) \cdot \prod_{j=2}^k \Pr \left( t[A_j] \mid t[A_1,\ldots,A_{j-1}] \right)$. Instead of learning a single distribution over the full domain of a tuple, we learn the probability distribution of the first attribute in the sequence and $(k-1)$ number of conditional probabilities. For the first attribute, we apply Gaussian mechanism (\S~\ref{sec:preliminaries:dp}) to learn its distribution. For each of remaining $(k-1)$ condition probabilities, we learn it as a discriminative model. In particular, for each conditional probability $\Pr \left( t[A_j] \mid t[A_1,\ldots,A_{j-1}] \right)$, we train a discriminative sub-model that uses context attributes $(A_1,\ldots,A_{j-1})$ to predict the target attribute $A_j$. We denote this sub-model by $M_{X,y}$, where 
$X=S_{:j}$ and $y=S[j]$.
We also apply the tuple embedding to privately learn a unified representation with a fixed dimensionality for each attribute in the tuple (\S~\ref{sec:preliminaries:embedding}). 
The training of each discriminative sub-model on the samples from the true data is optimized and privatized using DPSGD (\S~\ref{sec:preliminaries:dp}).

Algorithm~\ref{alg:model} describes how \system privately learns the probability distribution of the first attribute in the sequence $S$, denoted by $M_{\emptyset,S[1]}$, and the parameters in the $(k-1)$ discriminative sub-models $M_{S_{:j},S[j]}$ for $j\in [2,k]$. 
It takes as input of the true database instance $D^*$ with domain $\mathcal{D}$, the schema sequence $S$ (to be discussed in \S~\ref{sec:sequencing}), 
as well as learning parameters (number of iterations $T$, batch size $b$, learning rate $\eta$, and quantizing $q$ bins for numerical attributes) and noise parameters ($\sigma_g$ and $\sigma_d$ for Gaussian noise, $L_2$ norm clip threshold for gradients $C$). The configuration of these parameters is presented in \S~\ref{sec:analysis} to ensure the overall privacy loss of \system is bounded by the given budget $(\epsilon,\delta)$. 

\begin{algorithm}[t]
\caption{Probabilistic data model training}\label{alg:model}
\begin{algorithmic}[1]
\Require $D^*,\mathcal{D},S$\Comment{True instance, domain, schema sequence}
\Require $n, k, \eta, q$\Comment{cardinality, dimensions, lr, quantization} 
\Require $\sigma_g, \sigma_d$\Comment{Noise scales in $\Psi$} 
\Require $C, T, b$\Comment{$L_2$ norm clip/\#iterations/batch size in $\Psi$} 
\Procedure {TrainModel}{$D^*, S, \mathcal{D} ,\Psi$}
\State $H \gets$ counts of (quantized) values in $D^*$ for 1st attr. $S[1]$ \label{alg:model:hist_start}
\State Add noise drawn from $\mathcal{N}(0, 2\sigma_g^2)$ to each count in $H$
\State $M_{\emptyset, S[1]} \gets$ distribution of $S[1]$ based on $H$, and add it to $M$\label{alg:model:hist_end}
\State Initialize embedding for attribute $S[1]$
\For {$j \in [2,k]$}\label{alg:model:disc_start}
    \State $X=S_{:j}$, load embedding\label{alg:model:embed_context}\Comment{Context attributes}
    \State $y=S[j]$, initialize embedding\label{alg:model:embed_target}\Comment{Target attribute}
    \State Initialize discriminative model $M_{X, y}$\Comment{\cite{aimnet}}\label{alg:model:init_model}
    \State $\mathcal{L}(\theta_y, t) \gets$ loss function on imputing target $y$\label{alg:model:loss}
    \For {$e \in [T]$}\label{alg:model:iter_start}\Comment{For each of iteration}
    	\State $D_e \gets$ random sample on $D^*[X,y]$ with prob $b/n$
    	\State For each $t \in D_e$, compute $g_e(t) \gets \nabla_{\theta_y} \mathcal{L}(\theta_y, t)$\label{alg:model:gradient}
    	\State $\bar{g}_e(t) \gets \max(1, \frac{\norm{g_e(t)}_2}{C})$\Comment{Clip gradient}\label{alg:model:clip}
    	\State $\tilde{g}_e \gets (\sum_{t\in D_e}\bar{g}_e(t) + \mathcal{N}(0, \sigma_d^2C^2\mathbb{I})) / b$\Comment{Add noise}\label{alg:model:noise}
    	\State $\theta_y \gets \theta_y - \eta \times \tilde{g}_e$\label{alg:model:descent}\Comment{Gradient descent}
    \EndFor\label{alg:model:iter_end}
    \State Add $M_{X, y}$ to $M$\label{alg:model:add}
    \State Save embedding and attention weights for $S_{:j+1}$\label{alg:model:save}
\EndFor\label{alg:model:disc_end}

\State \textbf{return} $M$
\EndProcedure
\end{algorithmic}
\end{algorithm}

Following the attribute order in $S$, 
we start with the first attribute $S[1]$ and apply Gaussian mechanism to the true distribution of $S[1]$ (Line~\ref{alg:model:hist_start}-\ref{alg:model:hist_end}). 
If the first attribute has a continuous domain,  we partition its domain into $q$ bins. Starting from the second attribute in $S$, we train the discriminative model. We first load the initial values of the parameters of each sub-model from previous training if they exist (Line~\ref{alg:model:embed_context}). 
Depending on the data type of the target attribute, a cross entropy (for categorical target attribute) or mean squared (for numerical target attribute) loss function on predicting the target attribute value is also set before model training (Line~\ref{alg:model:loss}).

Each discriminative model $M_{S_{:j},S[j]}$ 
is learned via backpropagation for $T$ iterations (Line~\ref{alg:model:iter_start}-\ref{alg:model:iter_end}).
At each iteration, we randomly sample a set of training tuples $D_e$, with sampling probability $b/n$ (i.e., $\mathbb{E}(|D_e|) = b$),
and on each of the training tuple, the gradient w.r.t model parameters is computed (Line~\ref{alg:model:gradient}).
We clip the $L_2$ norm of the gradient by the threshold $C$ (Line~\ref{alg:model:clip}),
and add noise to clipped gradient (Line~\ref{alg:model:noise}) with sensitivity equal to clipping threshold $C$, 
before updating the parameters via gradient descent (Line~\ref{alg:model:descent}).
After one discriminative model is trained, we add it to our probabilistic data model $M$ (Line~\ref{alg:model:add}).
Since we iteratively expand the context attributes as more sub-models are trained, we save the currently trained embeddings of attributes $[X, y]$ (Line~\ref{alg:model:save}), 
and reuse in the initialization of context attributes of the next sub-model (Line~\ref{alg:model:embed_context}).
The final output from Algorithm~\ref{alg:model} is the probabilistic data model $M$, which will be used to sample tuple values in \S~\ref{sec:sampling}.

Algorithm~\ref{alg:model} consists of $1+(k-1)\times T$ rounds of access to the true database instance $D^*$. 
Each access is privatized using the Gaussian mechanism or the DPSGD.
By the composibility of differential privacy (\S~\ref{sec:preliminaries:dp}),
Algorithm~\ref{alg:model} satisfies differential privacy. 
We will analyze the privacy cost in \S~\ref{sec:analysis}.
\revise{
The time complexity is linear to $n+b(k-1)T$, which is the expected number of tuples that are sampled for training.
An efficiency optimization is to train each $M_{X,y}$ in parallel without reusing previously trained embeddings (Line~\ref{alg:model:embed_context}), and we 
\ifpaper 
evaluate this trade-off in our full paper~\cite{full_paper}.
\else
will evaluate this trade-off in \S~\ref{sec:exp:optimization}.
\fi
}

\subsection{Constraint-Aware Database Sampling}\label{sec:sampling}

After we have privately learned the tuple probability, the next step is to sample a database instance $D'$ of size $n$ based on the learned data model $M$ and the given DC weights as summarized in Algorithm~\ref{alg:syn}. 

Given a schema sequence $S$, 
we first independently sample a value for the first attribute in $S$ of all the $n$ tuples based on its noisy probability distribution represented by $M_{\emptyset,S[1]}$ (Line~\ref{alg:syn:sample1}).  
Depending on $S[1]$'s data type, categorical values are sampled directly; 
while for numerical values, we first sample a bin, and randomly take a value from the domain represented by the bin.

From the second attribute in $S$ onward, for each attribute $A_j$ and each tuple $t_i$, we sample a value for $t_i[A_j]$ conditioned on (1) the attributes of $t_i$ that have been assigned a value, i.e., $t_i[S_{:j}]=c$, and (2) the tuples that have been sampled before $t_i$, i.e. $D'_{:i}[S_{:j+1}]$. 
For each $v$ from the domain of $A_j$ (\revise{or a selected set of values of size $d$ if $A_j$ has a continuous or extremely large domain size}),
we first extract the conditional probability 
\begin{equation*}
\Pr(t[A_j]={v}\mid t[S_{:j}]=c)    
\end{equation*}
from the learned discriminative sub-model $M_{S_{:j},S[j]}$, and denote it by $p_{v|c}$ (Line~\ref{alg:syn:condprob}).
If the target attribute $A_j$ has a discrete domain, the conditional probability $p_{v|c}$ takes the  probability that $M_{S_{:j},S[j]}$ predicts the target attribute $A_j=v$ given the context attributes $S_{:j}=c$.
If the target attribute $A_j$ has a continuous domain, the discriminative model is based on regression model and outputs a Gaussian distribution
mean $\mu$ and std $\sigma$ given the context attributes $S_{:j}=c$. We sample $d$ number of candidates from this distribution and assign each candidate $v$ with a probability $p_{v|c}\propto
\{\frac{1}{\sigma\sqrt{2\pi}}\exp(-\frac{1}{2}(\frac{v-\mu}{\sigma})^2)\}$. The other values in the domain are assigned with probability 0. We denote the candidate set by $\mathcal{D}(S[j])$.

Next,
we compute the number of DC violations
$vio_{\phi,v|D'}$ 
if we assign $t_i[A_j]=v$:
\begin{equation*}
|V(\phi, t_i[S_{:j}]=c \wedge 
t_i[A_j]={v}
\mid D'_{:i}[S_{:j+1}])|
\end{equation*}
for each DC violation $\phi\in \Phi_{A_j}$ (Line~\ref{alg:syn:vio}).
Last, we sample a value $v$ based on the combined probability
\begin{equation*}
P[v] \propto 
p_{v|c} \cdot 
\exp (-\sum_{\phi \in \Phi_{A_j}} w_\phi \times vio_{\phi,v|D'}) )
\end{equation*}
and update the $j$th attribute of $t_i$ (Line~\ref{alg:syn:sample}).
The final output is a synthetic database instance $D^\prime$ of size $n$ with the same schema as the true database instance $D^*$.

\begin{algorithm}[t]
\caption{\revise{Constraint-aware database instance sampling}}\label{alg:syn}
\begin{algorithmic}[1]
\Require $S,M,\Phi,\mathcal{D}$\Comment{Schema sequence, data model, DCs, domain}
\Require $W,L, N$\Comment{Weight vector (Alg.~\ref{alg:weight_learning}), sample size, \#round}
\Procedure {Synthesize}{$S, M, \Phi, \mathcal{D}$}
\State $D^\prime[S[1]] \gets$ sample from distribution $M_{\emptyset, S[1]}$\label{alg:syn:sample1}
\For {$j \in [2,k]$} \Comment{Schema sequence $S$}\label{alg:syn:round_begin}
    \For{$i \in [1,n]$}\Comment{Tuple id sequence }
        \State  
        $c \gets t_i[S_{:j}]$ \Comment{Values for context attributes of $t_i$}
        \State $\{p_{v|c} \mid v\in \mathcal{D}(S[j])\} \gets M_{S_{:j}=c,S[j]}$ \label{alg:syn:condprob}
         \For {$v \in \mathcal{D}(S[j])$ and $\phi\in \Phi_{S[j]}$
         }\label{alg:syn:begin_for} 
            \State $vio_{\phi,v|D'} \gets$ num. of vio. of $\phi$ if $t_i[S[j]]=v$ \label{alg:syn:vio}
        \EndFor\label{alg:syn:end_for}
        \State  Update $t_i[S[j]]=v$ where $v$ is sampled with $P[v] \propto p_{v|c} \cdot \exp (-\sum_{\phi \in \Phi_{S[j]}} w_\phi \times  vio_{\phi,v|D'})$
        \label{alg:syn:sample}
    \EndFor
    \State \revise{Resample $m$ random cells $t_r[S[j]]$ or till convergence\label{alg:syn:repeat}}
\EndFor\label{alg:syn:round_end}
\State \textbf{return} $D^\prime$
\EndProcedure
\end{algorithmic}
\end{algorithm}

Without the constraint-aware sampling (Line~\ref{alg:syn:begin_for}-\ref{alg:syn:end_for}), 
the sampling process results in a set of i.i.d. tuple samples. 
This resulted instance can fail to preserve even simple constraints such as FDs (e.g., $\phi_1$) or single-tuple DCs (e.g., $\phi_3$), 
because not all the domain values appear in the true data $D^*$. 
Such values can be sampled due to noisy distribution and hence lead to DC violations. By adjusting the sampling probability based on the violations caused by the new cell value of a tuple (Line~\ref{alg:syn:sample}), we can control the additional number of violations due to the noisy distribution learned.

\revise{
General MCMC sampling requires re-sampling of the entire full $D^\prime$ with all attributes, and hence at least $k-1$ more conditional distributions need to be learned.
However, in the private setting with a fixed privacy budget, learning more distributions will compromise the accuracy of each learned distribution.
Therefore, \system uses a constrained MCMC based on the same set of conditional distributions. 
As we loop over each attribute (Line~\ref{alg:syn:round_begin}-\ref{alg:syn:round_end}), it re-samples random cell values for this attribute, conditioned on all other sampled values (Line~\ref{alg:syn:repeat}).
}

\revise{
The time complexity of checking one DC violations for all $n$ values is
$\mathcal{O}(dn)$ (for an unary DC) or $\mathcal{O}(dn^2)$ (for a binary DC).
This can be optimized by exploiting the property of hard functional dependencies, and 
\ifpaper 
we evaluate this optimization in our full paper~\cite{full_paper}.
\else
we will evaluate one optimization in \S~\ref{sec:exp:optimization}.
\fi
In addition, when $m>0$ for MCMC, the sampling algorithm has an additional cost of $\mathcal{O}(mkd+|\Phi|dnm)$.
The overall complexity of constraint-aware sampling is $\mathcal{O}(nkd + |\Phi|dn^2 + mkd + |\Phi|dnm)$.

\ifpaper
\else
We expect the synthetic database instance has similar number of violations as the truth. We provides a theoretical analysis in Appendix~\ref{app:analysis}.
\fi
}

\subsection{Constraint-Aware Sequencing}\label{sec:sequencing}

\begin{algorithm}[h]
\caption{Constraint-aware attribute sequencing}
\label{alg:sequence}
\begin{algorithmic}[1]
\Require $R,\mathcal{D}, \Phi$\Comment{Input schema, domain, and DCs}
\Procedure {Sequencing}{$R, \mathcal{D}, \Phi$}
\State $\Sigma \gets$  FDs from $\Phi$ sorted by increasing domain size of LHS \label{alg:sequence:sort_fd}
\State Initialize $S \gets []$
\ForAll {$X \rightarrow Y \in \Sigma$}\label{alg:sequence:fd_start}
	\State Sort attributes $X$ by domain size
	\State For all $A\in [X, Y]$, append $A$ to $S$ if $A\not\in S$
\EndFor\label{alg:sequence:fd_end}
\State Append attributes in $(R - S)$ to $S$ in an order of increasing domain size, and \textbf{return} $S$
\label{alg:sequence:append} 
\EndProcedure
\end{algorithmic}
\end{algorithm}

Given a fixed privacy budget, 
the goal is to identify a good schema sequence, where the set of attributes that can well discriminate attribute $A_j$ should appear before $A_j$ in the sequence. 
Unlike prior work~\cite{DBLP:journals/jmlr/ColomboM14, DBLP:conf/icdm/YaramakalaM05} that spend part of the privacy budget in learning a good sequence, we make use of the input DCs $\Phi$ and the domain $\mathcal{D}$. This heuristic approach incurs no privacy cost since the true database instance $D^*$ is not queried.



Specifically, we propose a rule-based, instance-independent method to ensure that for an FD $X \rightarrow Y$ in $\Phi$, we have $X$ ahead of $Y$ in $S$ (unless $Y\rightarrow X$ too).  
Algorithm~\ref{alg:sequence} describes the process of finding a schema sequence $S$.
For the list of FDs $\Sigma=[X_1 \rightarrow Y_1, \ldots, X_m \rightarrow Y_m]$,
we sort the list $\Sigma$ by the minimal domain size of an attribute from $X$ (i.e., $\exists A^1 \in X_1, \forall A^2 \in X_2$, $|\mathcal{D}(A^1)| \leq |\mathcal{D}(A^2)|$)  (Line~\ref{alg:sequence:sort_fd}).
For each FD, we greedily add its left hand side and right hand side attributes into the final schema sequence $S$ (Line~\ref{alg:sequence:fd_start}-\ref{alg:sequence:fd_end}).
For the rest of attributes that do not participate in FDs, we order them by ascending domain size and append to $S$ (Line~\ref{alg:sequence:append}).
\revise{The complexity is $\mathcal{O}(k|\Sigma|+\log k)$, consisted of sorting FDs and attributes.
}

Our sequencing algorithm relies on the given FDs as a subset of DCs.
In cases that $\Phi$ does not include any FDs (i.e., $\Sigma=\emptyset$), Algorithm~\ref{alg:sequence} returns a sequence based on the domain size. 
Following this sequence, each discriminative sub-model (\S~\ref{sec:model}) will have the smallest possible domain size for its context attributes (cross-product of all context attributes' domain sizes), and hence each sub-model can be more accurately learned.
For example, consider $[A_1,A_2,A_3]$ with domain sizes 2, 3, 5, respectively. The overall context attribute domain size is 8 (=2+6), instead of 20 on the reversed sequence.

\stitle{Optimizations for extreme domain sizes.}
For attributes with small domain size, we can group adjacent attributes in the schema sequence into one hyper attribute, and learn one discriminative sub-model instead of multiple sub-models.
As a result, less privacy budget will be consumed.
For example, applying Algorithm~\ref{alg:sequence} on 
the BR2000 dataset~\cite{DBLP:conf/sigmod/ZhangCPSX14} with 38k tuples resulted in a schema sequence starting with 7 binary attributes.
In this case, we can create a hyper attribute of domain size $2^7$ to replace the group of the binary attributes. 
After the synthetic hyper attribute value is generated, we can un-group it to individual attributes and check violations if any. 
On the other end, the distribution of attributes with very large domain size may not be learned well, due to insufficient amount of training data. 
For example, the Tax dataset~\cite{DBLP:journals/pvldb/ChuIP13} with 30k tuples has one $zip$ attribute with domain size of 18k.
The training sample of size $b\times T$ in Algorithm~\ref{alg:model} may not cover all values in the domain, and hence learned distribution can have large variance.
In this case, we can apply Gaussian mechanism to its true distribution, and sample independently without relying on the context attributes.

\section{Learning DC Weights}\label{sec:weight_learning}

\begin{algorithm}[t]
\caption{Learning DC weights}\label{alg:weight_learning}
\begin{algorithmic}[1]
\Require $D^*, \Phi, S$\Comment{True instance, DCs, schema sequence}
\Require $\sigma_w, T_w, L_w$\Comment{Noise scale/\#iteration/sample size in $\Psi$}
\Require $b_w, S_w$\Comment{Batch size in $\Psi$, sensitivity\ifpaper \else (Lemma~\ref{lemma:sensitivity_weights})}\fi
\Procedure {LearnWeight}{$ D^*, \Phi, S, M, \Psi$}
\State Initialize weight vector $W$ of length $|\Phi|$ if unknown 
\State Take a random sample $\hat{D}$ from $D^*$ with a probability $L_w/n$ 
\State Drop tuples from the sample if $|\hat{D}|> L_w$ \label{alg:weight_learning_beginsample}
\State Compute violation matrix $V$ of size ($|\hat{D}| \times |\Phi|$) from $\hat{D}$\label{alg:weight_learning_beginvio}
\State Add noise drawn from $\mathcal{N}(0, S_w^2\sigma_w^2)$ to each value in $V$
\State Set negative values in $V$ to zero\label{alg:weight_learning_endvio}
\For{$A_j \in S$ and $e\in [T_w]$}
\label{alg:weight_learning_beginlearn}
	    \State $ids \gets$ sample $b$ ids from $[1,L_w]$ with prob $b_w/L_w$
        \For {each  $i\in ids$}
            \State $O \gets \exp
            (-\sum_{\phi_l\in \Phi_{A_j}} W[l] \cdot V[i][l])$ \label{alg:weight_learning:obj}
            \State Update $W$ via back propagation by max $O$\label{alg:weight_learning:bp}
	\EndFor
\EndFor\label{alg:weight_learning_endlearn}
\State \textbf{return} $W$
\EndProcedure
\end{algorithmic}
\end{algorithm}

\system so far assumes the weights of DCs $W$ are known.
For example, the weights for hard DCs (no violations in the true data) are set to be infinitely large.
However, for soft DCs, the weights are usually unknown and need to be estimated.
We follow the intuition that if a DC is observed with many violations in the training data, then its weight will be set small. 
Otherwise, if there is no violation, then its weight will be set large.
Based on this intuition, we design Algorithm~\ref{alg:weight_learning} to first privately learn the number of violations to each DC and then estimate the weights as a post-processing step.  

We transform the given data instance $D$ into a violation matrix $V$ of size $|D|\times |\Phi|$, where each value $V[i][l]$ represents the number of violations to the $l$th DC in $\Phi$ caused by tuple $t_i$ with respect to all other tuples in $D$, i.e.,
$V(\phi_l, t_i \mid  D-\{t_i\}).$
Based on the transformed data, the objective is to maximize the exponential part represented in Eqn.~\eqref{equ:pdb}.
However, the violation matrix based on the full true instance is highly sensitive to the change of one tuple. For binary DCs that involve two tuples, changing one tuple can incur up to $\mathcal{O}(n)$ additional number of violations.

To bound the sensitivity of the violation matrix, we sample a small set of tuples $\hat{D}$ of size $L_w$ as the training example (Line~\ref{alg:weight_learning_beginsample}).
Each tuple from the true instance $D^*$ is independently sampled with probability $L_w/n$ (i.e., $\mathbb{E}(|\hat{D}|)=L_w$).
If the resulted sample has a size greater than $L_w$, we randomly drop tuples to crop the size to $L_w$. 
This allows us to bound the sensitivity of the violation matrix, \revise{and also reduces the time complexity from $\mathcal{O}(|\Phi|n^2)$ to $\mathcal{O}(|\Phi|L_w^2)$.}
\ifpaper
The sensitivity analysis on $S_w$ can be found in our full paper~\cite{full_paper}.
\else
\begin{lemma}\label{lemma:sensitivity_weights}
The $L_2$ sensitivity of the violation matrix for $\Phi$ that contains only unary and binary DCs is 
$S_w=|\phi_{u}| + |\phi_{b}| \times \sqrt{L_w^2 - L_w}$,
where $|\phi_{u}|$ and $|\phi_{b}|$ represent the number of unary DCs and the number of binary DCs in $\Phi$, respectively.
\end{lemma}
\begin{proof}
Consider a pair of neighboring instances by changing one tuple. 
If a DC is an unary DC, then the differing tuple can change the violation count by 1. 
If the DC is a binary DC, the differing tuple may violate all other $L_w - 1$ tuples in the instance. 
Thus, the $L_2$ norm of the maximum violation count change $S_w$ is:
\begin{equation*}
\begin{split}
    S_w & = (|\phi_{u}| \times \sqrt{1}) + (|\phi_{b}| \times \sqrt{1^2 + 1^2 + \cdots + 1^2 + (L_w-1)^2})\\
    & = |\phi_{u}| \times 1 + |\phi_{b}| \times \sqrt{L_w-1 + (L_w-1)^2}\\
    & = |\phi_{u}| + |\phi_{b}| \times \sqrt{L_w^2 - L_w}\qedhere
\end{split}
\end{equation*}
\end{proof}
\fi

\eat{
Consider only unary and binary DCs, we can prove (Appendix~\ref{app:proofs:sw}) that the sensitivity of the violation matrix is as follows.
\begin{lemma}\label{lemma:sensitivity_weights}
The sensitivity of the violation matrix $S_w$ is $|\phi_{u}| + |\phi_{b}| \times \sqrt{L_w^2 - L_w}$,
where $|\phi_{u}|$ and $|\phi_{b}|$ represents the number of unary DCs and the number of binary DCs respectively.
\end{lemma}
}

Hence, we apply Gaussian mechanism to perturb the violation matrix $V$ over the samples and post-process all the negative noisy counts to zeros (Lines~\ref{alg:weight_learning_beginvio}-\ref{alg:weight_learning_endvio}). Then we loop over each attribute $A_j\in S$ for $T_w$ iterations (Line~\ref{alg:weight_learning_beginlearn}). For each $A_j$, we sample $b$ rows from the noisy $V$ to update weights $W$ for the set of active DCs related to $A_j$ (Lines~\ref{alg:weight_learning_beginlearn}-\ref{alg:weight_learning_endlearn}). 
We will analyze the privacy cost in \S~\ref{sec:analysis}.
\revise{The time complexity of this post-processing step is $\mathcal{O}(|\Phi|bT_w)$ in terms of the number of tuples that are used for learning.}

\section{Privacy Analysis}\label{sec:analysis}

\ifpaper

\begin{algorithm}[t]
\caption{Searching DP parameters}\label{alg:paras_search}
\begin{algorithmic}[1]
\Require $\epsilon, \delta, \mathcal{D}, S$\Comment{Privacy budget, domain, schema sequence}
\Procedure {SearchDParas}{$\epsilon, \delta, \mathcal{D}, S$}
    \State Initialize parameter configuration $\Psi$ with a default setting $(\sigma_g.min, \sigma_d.min,\sigma_w.min, b.max,T.max,|S|,L_w.max,\ldots)$ \label{alg:paras_search:range}
	\State $\Psi.b_w \gets 1$ if DC weights are unknown
	\While {$\epsilon_{\Psi}(\delta)> \epsilon$} 
	\ifpaper \Comment{\revise{Cost of \system~\cite{full_paper}}} \else \Comment{Eqn.~\eqref{eqn:epsilon_psi}}\fi
	\label{alg:paras_search:loss}
		\State If $\Psi.T > T_{min}$, then decrease $\Psi.T$ \label{algo:paras_search:tunestart}
		\State If $\Psi.\sigma_d < {\sigma_d}_{max}$, then increase $\Psi.\sigma_d$
		\State If $\Psi.\sigma_g < {\sigma_g}_{max}$, then increase $\Psi.\sigma_g$
		\State If $\Psi.b > b_{min}$, then decrease $\Psi.b$
		\State ...\label{algo:paras_search:tuneend}
	\EndWhile
	\State \textbf{return} $\Psi $
\EndProcedure
\end{algorithmic}
\end{algorithm}

\else

\begin{algorithm}[t]
\caption{Searching DP parameters}\label{alg:paras_search}
\begin{algorithmic}[1]
\Require $\epsilon, \delta, \mathcal{D}, S$\Comment{Privacy budget, domain, schema sequence}
\Procedure {SearchDParas}{$\epsilon, \delta, \mathcal{D}, S$}
	\State $C \gets 1, \sigma_d \gets 1.1, \eta \gets 10^{-4}$\Comment{norm clip, noise scale, lr}\label{alg:paras_search:fix}
	\State $\sigma_g \in [0.1/|\mathcal{D}(S[1])|, 4\sqrt{\log(1.25/\delta)}/\epsilon]$, $\sigma_d \in [1, 1.5]$ \label{alg:paras_search:begin_v}
	\State $b \in [16, 32]$, $T \in [n/\text{min}(b), 5n/\text{min}(b)]$\label{alg:paras_search:end_v}
	\State Initialize $\sigma_g, \sigma_d$ to the minimal, and $T, b$ to the maximal\label{alg:paras_search:range}
	\If {DC weights unknown}
		\State $\epsilon_w, L_w\gets 100$, $\sigma_w \gets \sqrt{2 \log{(1.25/\delta_w)}}/\epsilon_w$\label{alg:paras_search:sigma_w}
		\State $b_w \gets 1$, $T_w \gets L_w/b_w$
	\EndIf
	
	\While {$\epsilon_{\Psi}(\delta)> \epsilon$} 
	\Comment{Eqn.~\eqref{eqn:epsilon_psi}}
	\label{alg:paras_search:loss}
		\State If $T > T_{min}$, then decrease T\label{algo:paras_search:tunestart}
		\State If $\sigma_d < {\sigma_d}_{max}$, then increase $\sigma_d$
		\State If $\sigma_g < {\sigma_g}_{max}$, then increase $\sigma_g$
		\State If $b > b_{min}$, then decrease $b$\label{algo:paras_search:tuneend}
	\EndWhile
	\State \textbf{return} $\Psi $, a set consisting of all above parameters
\EndProcedure
\end{algorithmic}
\end{algorithm}

\fi

\system involves at most three processes that require access to the true database instance:
\squishlist
\item[$M_1$:] Learning the distribution of the first attribute in the schema sequence (Algorithm~\ref{alg:model} Line~\ref{alg:model:hist_start}-\ref{alg:model:hist_end});
\item[$M_2$:] Training $k-1$ discriminative models 
(Algorithm~\ref{alg:model} Line~\ref{alg:model:disc_start}-\ref{alg:model:disc_end}); \item[$M_3$:] Learning the DC weights if unknown (Algorithm~\ref{alg:weight_learning}).
\squishend

Each process has been privatized using the Gaussian mechanism or DPSGD. 
\revise{The other steps (Algorithm~\ref{alg:syn} and Algorithm~\ref{alg:sequence}) not accessing the true database do not incur privacy loss.}
Hence, we can show \system achieves DP by simple sequential composition~\cite{DBLP:conf/icalp/Dwork06} and post-processing property~\cite{DBLP:conf/eurocrypt/DworkKMMN06} of DP. 
\ifpaper
In our full paper~\cite{full_paper}, we give a tighter privacy bound using \emph{R\'enyi DP} (RDP)~\cite{mironov2017renyi} and prove the privacy of \system.
\else
However, this does not give us the tightest privacy bound. Instead,
we apply \emph{R\'enyi DP} (RDP)~\cite{mironov2017renyi}, a generalized privacy notion and its advanced composition techniques for the privacy analysis of \system. 

\begin{definition}[R\'enyi-DP~\cite{mironov2017renyi}]
A randomized algorithm $\mathcal{M}$ with domain $\mathcal{D}$ is ($\alpha, \epsilon$)-RDP at order $\alpha > 1$,
for any pair of neighboring database instances $D,D' \in \mathcal{D}$ that differ in one tuple. 
Let $P_{D}$ and $P_{D^\prime}$ be the output probability density of  $\mathcal{M}(D)$ and $\mathcal{M}(D^\prime)$, respectively. 
It holds that:
$\frac{1}{\alpha - 1} \log\mathbb{E}_{x\sim \mathcal{M}(D')} \left( \frac{P_{D}(x)}{P_{D^\prime}(x)} \right)^\alpha \leq \epsilon$.
\end{definition}

We state the RDP cost of \system as follows.
Both the post-processing and composability properties apply to RDP.
Specifically, if a sequence of adaptive mechanisms $\mathcal{M}_1$, $\mathcal{M}_2$, $\cdots$, $\mathcal{M}_k$ satisfy ($\alpha,\epsilon_1$)-, ($\alpha,\epsilon_2$)-, $\cdots$, ($\alpha,\epsilon_k$)-RDP, 
then the composite privacy loss is ($\alpha, \sum_{i=1}^k\epsilon_i$)-RDP. 

As we applied the Gaussian mechanism on sampled data, we summarize the RDP privacy loss of a generalized mechanism, the \emph{sampled Gaussian mechanism} (SGM)~\cite{DBLP:journals/corr/abs-1908-10530}.
\begin{lemma}\label{lemma:sgm}
Given a database $D$ and query $f: \mathcal{D} \rightarrow \mathbb{R}^d$,
returning $f(\{x\in D \mid \text{x is sampled with probability r}\})+\mathcal{N}(0,S_f^2\sigma^2 \mathbb{I}^d)$ results in the following RDP cost 
for an integer moment $\alpha$ \footnote{Analysis on general fractional moments is in related work~\cite{DBLP:journals/corr/abs-1908-10530}.}
\begin{equation*}
    R_{\sigma, r}(\alpha) = 
    \begin{cases}
     \frac{\alpha}{2\sigma^2} &  r = 1\\ 
     \sum_{k=0}^{\alpha} \binom{\alpha}{k}(1-r)^{\alpha-k} r^{k}\exp(\frac{\alpha^2 - \alpha}{2\sigma^2}) &  0<r<1
    \end{cases}
\end{equation*}
\end{lemma}

We analyze the RDP cost of each step in \system and result in the following total cost.

\begin{theorem}\label{theorem:privacy}
The total RDP cost of \system with parameter configuration set $\Psi =\{\sigma_g,\sigma_d,\sigma_w,b,T,k,L_w,i_{w}\ldots\}$ (Algorithm~\ref{alg:overall}) is 
    \begin{equation*}
    \begin{split}
       R_{\Psi}(\alpha) = &  \frac{\alpha}{2\sigma_g^2} 
       +          T(k-1) \times \sum_{k=0}^{\alpha} \binom{\alpha}{k}(1-\frac{b}{n})^{\alpha-k} (\frac{b}{n})^{k}\exp(\frac{\alpha^2 - \alpha}{2\sigma_d^2}) \\
        &
        +i_{w}\sum_{k=0}^{\alpha} \binom{\alpha}{k}(1-\frac{L_w}{n})^{\alpha-k} (\frac{L_w}{n})^{k}\exp(\frac{\alpha^2 - \alpha}{2\sigma_w^2}),
    \end{split}
    \end{equation*}
   where $i_{w}$ is a binary indicator for $M_3$ (DC weight learning). 
\end{theorem}

\begin{proof}
\system has the following adaptive SGMs:
\squishlist
\item For $M_1$, the sampling rate is 1, $R_{M_1}(\alpha) = \alpha/2\sigma_g^2$.
\item For $M_2$, the sampling rate is set to $b/n$, and SGM is applied $T\times (k-1)$ times. 
Thus, $R_{M_2}(\alpha) =  T(k-1) \times \sum_{k=0}^{\alpha} \binom{\alpha}{k}(1-\frac{b}{n})^{\alpha-k} (\frac{b}{n})^{k}\exp(\frac{\alpha^2 - \alpha}{2\sigma_d^2})$.
\item For $M_3$, the sampling rate is $L_w/n$.
Thus, $R_{M_3}(\alpha) =  \sum_{k=0}^{\alpha} \binom{\alpha}{k}(1-\frac{L_w}{n})^{\alpha-k} (\frac{L_w}{n})^{k}\exp(\frac{\alpha^2 - \alpha}{2\sigma_w^2})$.
\squishend
By the composition property~\cite{mironov2017renyi} of RDP, the total RDP cost is:
\begin{equation*}
    \begin{split}
       R_{\system}(\alpha) = &  \frac{\alpha}{2\sigma_g^2} + \sum_{k=0}^{\alpha} \binom{\alpha}{k}(1-\frac{L_w}{n})^{\alpha-k} (\frac{L_w}{n})^{k}\exp(\frac{\alpha^2 - \alpha}{2\sigma_w^2}) + \\
        & T(k-1) \times \sum_{k=0}^{\alpha} \binom{\alpha}{k}(1-\frac{b}{n})^{\alpha-k} (\frac{b}{n})^{k}\exp(\frac{\alpha^2 - \alpha}{2\sigma_d^2})
    \end{split}
    \end{equation*}
\end{proof}

By the tail bound property of RDP~\cite{mironov2017renyi}, we can convert the RDP cost of \system to $(\epsilon,\delta)$-DP, where $\epsilon$ is computed by 
\begin{equation} \label{eqn:epsilon_psi}
\epsilon_{\Psi}(\delta) = \min_{\alpha} R_{\Psi}(\alpha) + \frac{\log(1/\delta)}{\alpha-1},    
\end{equation}
for a given $\delta$. The order $\alpha$ is usually searched within  a range~\cite{waites2019pyvacy}.
\fi

In practice, the overall privacy budget ($\epsilon, \delta$) is specified as an input to \system, 
and one needs to judiciously set the privacy parameters in $\Psi$.
Setting these parameters is non-trivial as they are volatile to input datasets. 
To automatically assign parameters, \system provides a parameter search algorithm, summarized in Algorithm~\ref{alg:paras_search}.
It takes the privacy budget ($\epsilon,\delta$) and outputs a set of parameters $\Psi$ that ensures that the overall privacy cost does not exceed $(\epsilon,\delta)$.
It starts with a default setting based on prior experimental heuristics~\cite{DBLP:journals/jmlr/BergstraB12, waites2019pyvacy} and the domain information $\mathcal{D}$. 
The noise parameters including ($\sigma_g,\sigma_d,\sigma_w,b,T,L_w$) are  boldly set to give the best possible accuracy (Line~\ref{alg:paras_search:range}). 
If this privacy cost of this configuration is higher than $\epsilon$ (Line~\ref{alg:paras_search:loss}), then we use a priority order
to decide which parameter to tune (Lines~\ref{algo:paras_search:tunestart}-\ref{algo:paras_search:tuneend}). This process is repeated till the privacy loss is capped at our total budget. 
\revise{
The time complexity is linear to the size of parameter space.
}

\ifpaper
The parameter settings can be found in our full paper~\cite{full_paper}.
\else
\fi

\section{Evaluation}
\label{sec:evaluation}

\newcounter{exp}

In this section, we evaluate the synthetic data generated by \system with three utility metrics: (i) consistency with DC constraints in the true data; (ii) usefulness in training classification models; and (iii) accuracy in answering $\alpha$-way marginal queries. 
We show that:
\squishlist
\item \system preserves data consistency, while state-of-the-art methods fail to preserve most DCs. \system is practically efficient.
\item While \system is not designed for particular tasks, it can achieve comparable and even better quality in the learning and query task, compared to methods that are designed for these tasks.
\item The constraint-aware sampling and sequencing are effective to keep data consistency.
\item \revise{\system scales linearly with the number of DCs.} 
\squishend

\subsection{Evaluation Setup}\label{sec:exp:setup}

\ifpaper

\begin{table}[t]
\centering
\caption{\revise{Description of the datasets.}}
\label{tab:data}
\vspace{-.75em}
\begin{tabular}{|l|c|c|c|c|c|}
\hline
Dataset & $n$      & $k$  & Domain size & Hard DCs  & DC IDs\footnote{\revise{The sets of DCs are listed in the full paper~\cite{full_paper}.}} \\ \hline
Adult   & 32,561 & 15 & $\approx 2^{52}$  & Yes  & $\{\phi_{1-2}^a\}$      \\ \hline
BR200   & 38,000 & 14 & $\approx 2^{16}$  & No & $\{\phi_{1-3}^b\}$    \\ \hline
Tax     & 30,000 & 12 & $\approx 2^{71}$  & Yes & $\{\phi_{1-6}^t\}$     \\ \hline
TPC-H   & 20,000 & 9  & $\approx 2^{42}$  & Yes  & $\{\phi_{1-4}^h \}$    \\ \hline
\end{tabular}
\end{table}

\else

\begin{table*}[t]
\centering
\caption{Description of the datasets that are used in the experiments.}
\vspace{-.75em}
\label{tab:data}
\resizebox{\textwidth}{!}{%
\begin{tabular}{|l|l|l|l|l|l|}
\hline
\multicolumn{1}{|c|}{Dataset} & \multicolumn{1}{c|}{$n$} & \multicolumn{1}{c|}{$k$} & \multicolumn{1}{c|}{Domain size} & \multicolumn{1}{c|}{Hard DCs} & \multicolumn{1}{c|}{DCs                                                                                                         (omitting the universal quantifier) }\\ \hline
Adult                         & 32,561                           & 15                           & \multicolumn{1}{c|}{$\approx 2^{52}$} & Yes & \begin{tabular}[c]{@{}l@{}}$\phi_1^a: \neg (t_i[edu]=t_j[edu] \land t_i[edu\_num] \neq t_j[edu\_num])$\\ $\phi_2^a: \neg (t_i[cap\_gain]>t_j[cap\_gain] \land t_i[cap\_loss] < t_j[cap\_loss])$\end{tabular}                                                     \\ \hline
BR2000                        & 38,000                           & 14                           & \multicolumn{1}{c|}{$\approx 2^{16}$} & No & \begin{tabular}[c]{@{}l@{}} $\phi_1^b: \neg (t_i[a13] = t_j[a13] \land t_i[a11] < t_j[a11] \land t_i[a3] > t_j[a3])$\\ $\phi_2^b: \neg (t_i[a12] \neq t_j[a12] \land t_i[a13] \leq t_j[a13] \land t_i[a5] \geq t_j[a5]) $\\ $\phi_3^b: \neg (t_i[a5] \leq t_j[a5] \land t_i[a3] > t_j[a3] \land t_i[a12] \neq t_j[a12] \land t_i[a11]>t_j[a11]) $\end{tabular}                           \\ \hline
Tax                           & 30,000                           & 12                           & \multicolumn{1}{c|}{$\approx 2^{71}$} & Yes & \begin{tabular}[c]{@{}l@{}}$\phi_1^t:\neg (t_i[zip]=t_j[zip] \land t_i[city] \neq t_j[city])$\\ $\phi_2^t: \neg (t_i[areacode]=t_j[areacode] \land t_i[state] \neq t_j[state])$\\ $\phi_3^t: \neg (t_i[zip]=t_j[zip] \land t_i[state]\neq t_j[state])$\\ $\phi_4^t: \neg (t_i[state]=t_j[state] \land t_i[has\_child]=t_j[has\_child] \land t_i[child\_exemp] \neq t_j[child\_exemp])$\\ $\phi_5^t:\neg (t_i[state]=t_j[state] \land t_i[marital]=t_j[marital] \land t_i[single\_exemp] \neq t_j[single\_exemp])$\\ $\phi_6^t:  \neg (t_i[state]=t_j[state] \land t_i[salary] > t_j[salary] \land t_i[rate] < t_j[rate]) $\end{tabular} \\ \hline
TPC-H                         & 20,000                                 & 9                              & \multicolumn{1}{c|}{$\approx 2^{42}$} & Yes & \begin{tabular}[c]{@{}l@{}}$\phi_1^h: \neg(t_i[c\_custkey]=t_j[c\_custkey] \land t_i[c\_nationkey) \neq t_j[c\_nationkey]$\\ $\phi_2^h: \neg(t_i[c\_custkey]=t_j[c\_custkey] \land t_i[c\_mktsegment) \neq t_j[c\_mktsegment]$\\ $\phi_3^h: \neg(t_i[c\_custkey]=t_j[c\_custkey] \land t_i[n\_name) \neq t_j[n\_name]$\\ $\phi_4^h: \neg(t_i[n\_name) = t_j[n\_name] \land t_i[n\_regionkey] \neq t_j[n\_regionkey])$\end{tabular}                                                     \\ \hline
\end{tabular}
}
\end{table*}

\fi

%
%
\stitle{Datasets.} We choose 4 different datasets with mixed data types and DCs, listed in Table~\ref{tab:data}. 
First,
the Adult dataset~\cite{Dua:2019} consists of 15 census attributes and 2 hard DCs. Second,
the BR2000 dataset~\cite{DBLP:conf/sigmod/ZhangCPSX14} has a smaller domain size than the Adult dataset, but it has 3 soft DCs with unknown weights. 
The third dataset, Tax~\cite{DBLP:journals/pvldb/ChuIP13}, has a very large domain size, e.g., $zip$ ($\approx 2^{15}$) and $city$ ($\approx 2^{14}$) and 6 hard DCs.
Last, TPC-H~\cite{tpch}, a synthethic dataset that joins 
three tables (Orders, Customer and Nation) and removes  unique attributes such as $orderkey$ and $comment$. The final table consists of 20,000 orders with 9 numerical and categorical attributes. The set of hard DCs are obtained by the foreign key and primary key constraints.


\stitle{Baselines.}
Four state-of-the-arts to allow the synthesis of relational data with DP guarantees are considered: 
1) PrivBayes~\cite{DBLP:conf/sigmod/ZhangCPSX14}, a statistical method based on Bayesian network;
2) PATE-GAN~\cite{DBLP:conf/iclr/JordonYS19a}, a GAN-based method that trains a data generator using the PATE's student-teacher model ~\cite{DBLP:conf/iclr/PapernotSMRTE18};
3) DP-VAE~\cite{DBLP:journals/corr/abs-1812-02274}, which samples from the latent space of a privately trained auto-encoder~\cite{DBLP:journals/corr/KingmaW13}; and 
4) The winning solution of the NIST challenge~\cite{nist} (labeled as NIST), which applies probabilistic inference~\cite{DBLP:conf/icml/McKennaSM19} over marginals.

PATE-GAN and DP-VAE require the input dataset to be encoded into numeric vectors, 
and we apply the best encoding scheme empirically~\cite{DBLP:journals/pvldb/FanLLCSD20}.
Additionally, PATE-GAN requires one labeled attribute to train a set of conditional generators, where each generator produces synthetic data conditioning on one value in the domain of the labeled attribute.
We choose the attribute with smallest domain size from each dataset as the labeled attribute, 
and generate the same number of tuples as in the true data,
although it reveals the true histogram of the labeled attribute and favors answering marginal queries.  
Finally, NIST requires a set of marginals as input for inference. We use marginals over every single attribute, and over 10 randomly chosen attribute pairs.  

\stitle{Evaluation Metrics.}
We evaluate a synthetic database instance $D^\prime$ of the same size as the true data $D^*$ using three metrics. 

\noindent \underline{Metric I: DC Violations.} Since all known DCs are binary, we measure the percentage of tuple pairs that violate DCs in an instance $D$ of size $n$, i.e.,  $100\cdot|V(\phi, D)| / \binom{n}{2}$.

\noindent \underline{Metric II: Model training.} We consider 9 classification models 
(LogisticRegression, 
AdaBoost, 
GradientBoost,
XGBoost,
RandomForest,
BernoulliNB,
DecisionTree,
Bagging, and MLP).
On every single attribute of a dataset, we train all models to classify one binary label (e.g., income is more than 50k or not, age is senior or not, occupation is government job or not) using all other attributes as features.
The quality of the learning task on one attribute is represented by the average of all models.
Accuracy and F1 are reported for learning quality. 
Each model is trained using 70\% of the synthetic database instance, and evaluate the accuracy and F1 using the same 30\% of the true database instance.  We also show the results of training and testing on the true dataset labeled as Truth.

\noindent \underline{Metric III: $\alpha$-way marginals.} 
For each attribute combination $\mathbb{A}$, we compute the $\alpha$-way marginal, $h:\mathcal{D} \rightarrow  \mathbb{R}^{|\mathcal{D}(\mathbb{A})|}$ on the synthetic data $D^\prime$ and true data $D^*$, respectively, and then report the total variation distance~\cite{DBLP:books/daglib/0035708} as $\max_{a\in \mathcal{D}(\mathbb{A})} |h(D^\prime)[a]-h(D^*)[a]|$.



\stitle{Implementation details.}
\system was implemented in Python 3.6 \revise{and tested with $m=0$ by default}. 
For the discriminative sub-models, we integrated the code~\cite{holocleancode} from  AimNet in the HoloClean system.
For the baselines, 
we reused the code~\cite{privbayescode, pategancode, nistcode} from their authors with all default parameters.
All the 9 models in the learning task were implemented using standard libraries~\cite{sklearn, DBLP:conf/kdd/ChenG16} and trained with default parameters, 
except that we set $\texttt{random\_state}=0$ whenever possible, for the purpose of reproducibility.
We report the mean and standard deviation of 3 runs for each test.
All experiments were conducted on a machine with 12 cores and 64GB RAM.
\revise{The code, data and evaluation metrics are open sourced on GitHub: \url{https://github.com/cgebest/Kamino}.}

\subsection{End-to-End Evaluation}
We compare \system with all four baselines at a fixed privacy budget $(\epsilon=1, \delta=10^{-6})$.

\stepcounter{exp}
\subsubsection{Experiment \theexp: DC Violations}\label{sec:exp:dcs}

We show that synthetic data generated by \system has a similar number of DC violations as the true database instance.
Table~\ref{tab:dc_vio} lists the percentage of tuple pairs that violate each of the given DC.
On the Adult, Tax and TPC-H datasets, \system incurs zero violations, which is consistent to the observations in the true database instances.
On the BR2000 dataset, the overall numbers of DC violations on the synthetic  instance output by \system are the closest to those on the truth among all approaches.
The baselines fail to preserve most of the DCs.
For instance, the hard DC $\phi_1^a$ on the Adult dataset has about 11.3\%, 32\%, and 20.3\% violations in the synthetic data generated by PrivBayes, DP-AVE and PATE-GAN, respectively.
Although NIST does not have violations like \system, it it because NIST filled the entire $edu\_num$ column with the same value.  
For another instance, all the hard DCs induced by the foreign key and primary key constraints in the TPC-H dataset, are preserved only in \system.

\begin{table}[t]
\centering
\caption{Percentage of tuple pairs that violate DCs. 
\system has the closet DC violations as the truth, while none of the baselines are able to preserves most of the DCs.
}
\vspace{-.75em}
\label{tab:dc_vio}
\resizebox{\columnwidth}{!}{%
\begin{tabular}{|c|c|c|c|c|c|c|}
\hline
DC         & Truth & PrivBayes    & DP-VAE       & PATE-GAN             & NIST          & \system               \\ \hline
$\phi_1^a$ & 0.0   & 11.3$\pm$0.3 & 32.0$\pm$0.2 & 20.3$\pm$0.0         & 0.0$\pm$0.0   & \textbf{0.0$\pm$0.0} \\ \hline
$\phi_2^a$ & 0.0   & 1.4$\pm$0.6  & 13.2$\pm$0.1 & 24.8$\pm$0.1         & 0.0$\pm$0.0   & \textbf{0.0$\pm$0.0} \\ \hline
$\phi_1^b$ & 0.4   & 1.6$\pm$0.0  & 0.0$\pm$0.0  & \textbf{0.4$\pm$0.0} & 0.0$\pm$0.0   & 0.6$\pm$0.0          \\ \hline
$\phi_2^b$ & 0.9   & 2.6$\pm$0.2  & 15.6$\pm$0.2 & 0.2$\pm$0.0          & 28.1$\pm$6.8  & \textbf{0.6$\pm$0.0} \\ \hline
$\phi_3^b$ & 0.5   & 1.4$\pm$0.1  & 0.0$\pm$0.0  & 0.1$\pm$0.0          & 0.0$\pm$0.0   & \textbf{0.3$\pm$0.2} \\ \hline
$\phi_1^t$ & 0.0   & 0.0$\pm$0.0  & 0.0$\pm$0.0  & 0.0$\pm$0.0          & 7.4$\pm$1.3   & \textbf{0.0$\pm$0.0} \\ \hline
$\phi_2^t$ & 0.0   & 0.8$\pm$0.0  & 0.0$\pm$0.0  & 0.8$\pm$0.0          & 0.4$\pm$0.0   & \textbf{0.0$\pm$0.0} \\ \hline
$\phi_3^t$ & 0.0   & 0.0$\pm$0.0  & 0.0$\pm$0.0  & 0.0$\pm$0.0          & 8.0$\pm$1.7   & \textbf{0.0$\pm$0.0} \\ \hline
$\phi_4^t$ & 0.0   & 0.4$\pm$0.0  & 98.9$\pm$0.0 & 2.1$\pm$0.0          & 0.0$\pm$0.0   & \textbf{0.0$\pm$0.0} \\ \hline
$\phi_5^t$ & 0.0   & 0.5$\pm$0.0  & 99.0$\pm$0.0 & 4.0$\pm$0.0          & 0.0$\pm$0.0   & \textbf{0.0$\pm$0.0} \\ \hline
$\phi_6^t$ & 0.0   & 0.4$\pm$0.0  & 24.5$\pm$0.1 & 0.9$\pm$0.0          & 0.0$\pm$0.0   & \textbf{0.0$\pm$0.0} \\ \hline
$\phi_1^h$ & 0.0   & 0.2$\pm$0.0  & 16.7$\pm$0.2 & 5.1$\pm$0.1          & 64.0$\pm$45.2 & \textbf{0.0$\pm$0.0} \\ \hline
$\phi_2^h$ & 0.0   & 0.2$\pm$0.0  & 15.7$\pm$0.1 & 4.4$\pm$0.1          & 53.4$\pm$37.7 & \textbf{0.0$\pm$0.0} \\ \hline
$\phi_3^h$ & 0.0   & 0.2$\pm$0.0  & 15.3$\pm$0.2 & 5.1$\pm$0.1          & 64.0$\pm$45.2 & \textbf{0.0$\pm$0.0} \\ \hline
$\phi_4^h$ & 0.0   & 0.6$\pm$0.0  & 30.1$\pm$0.1 & 1.2$\pm$0.0          & 3.2$\pm$0.0   & \textbf{0.0$\pm$0.0} \\ \hline
\end{tabular}
}
\end{table}

\begin{figure*}[h]
    \centering
    
    \begin{subfigure}[h]{0.24\textwidth}%
        \includegraphics[width=\textwidth]{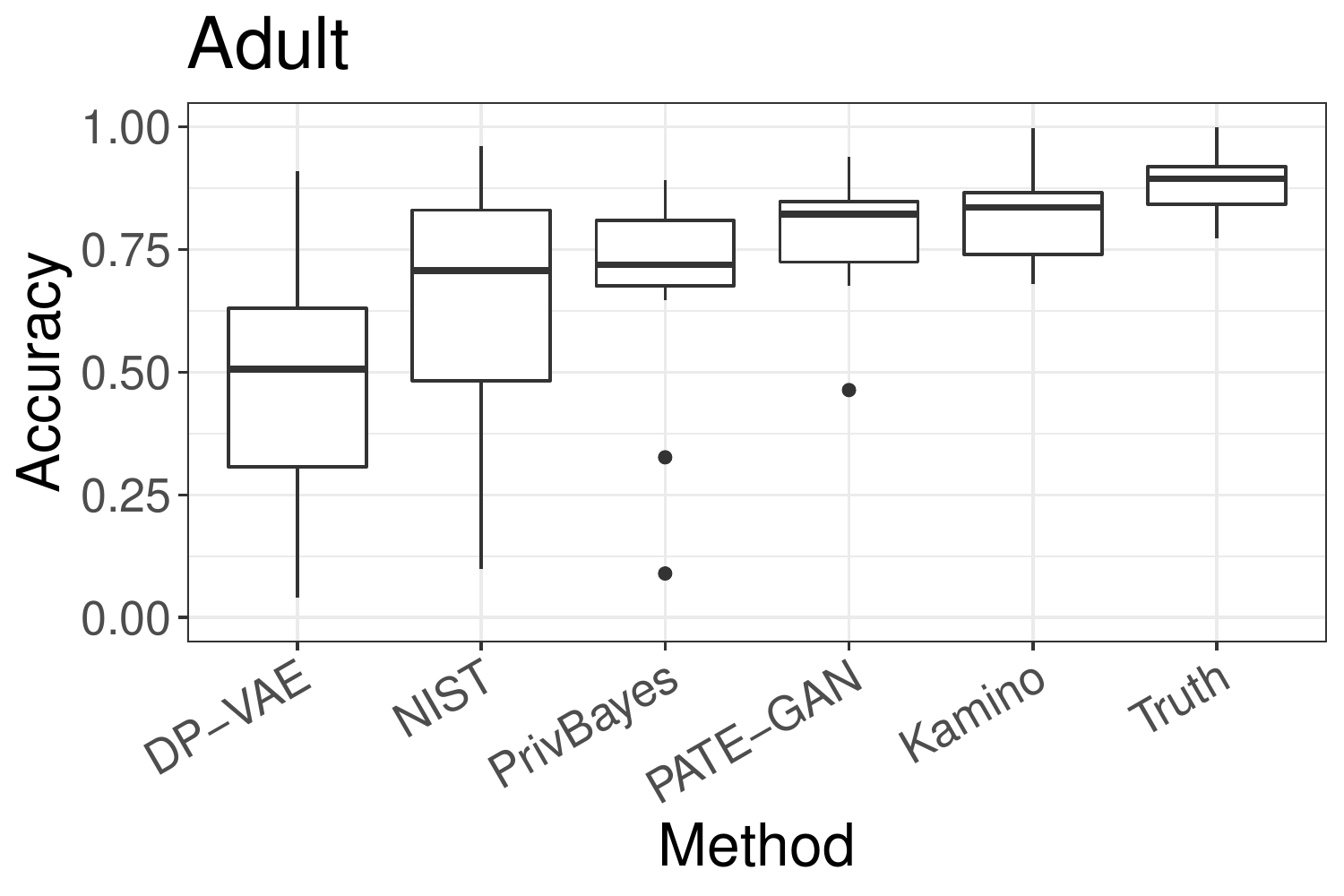}%
    \end{subfigure}%
    \begin{subfigure}[h]{0.24\textwidth}%
        \includegraphics[width=\textwidth]{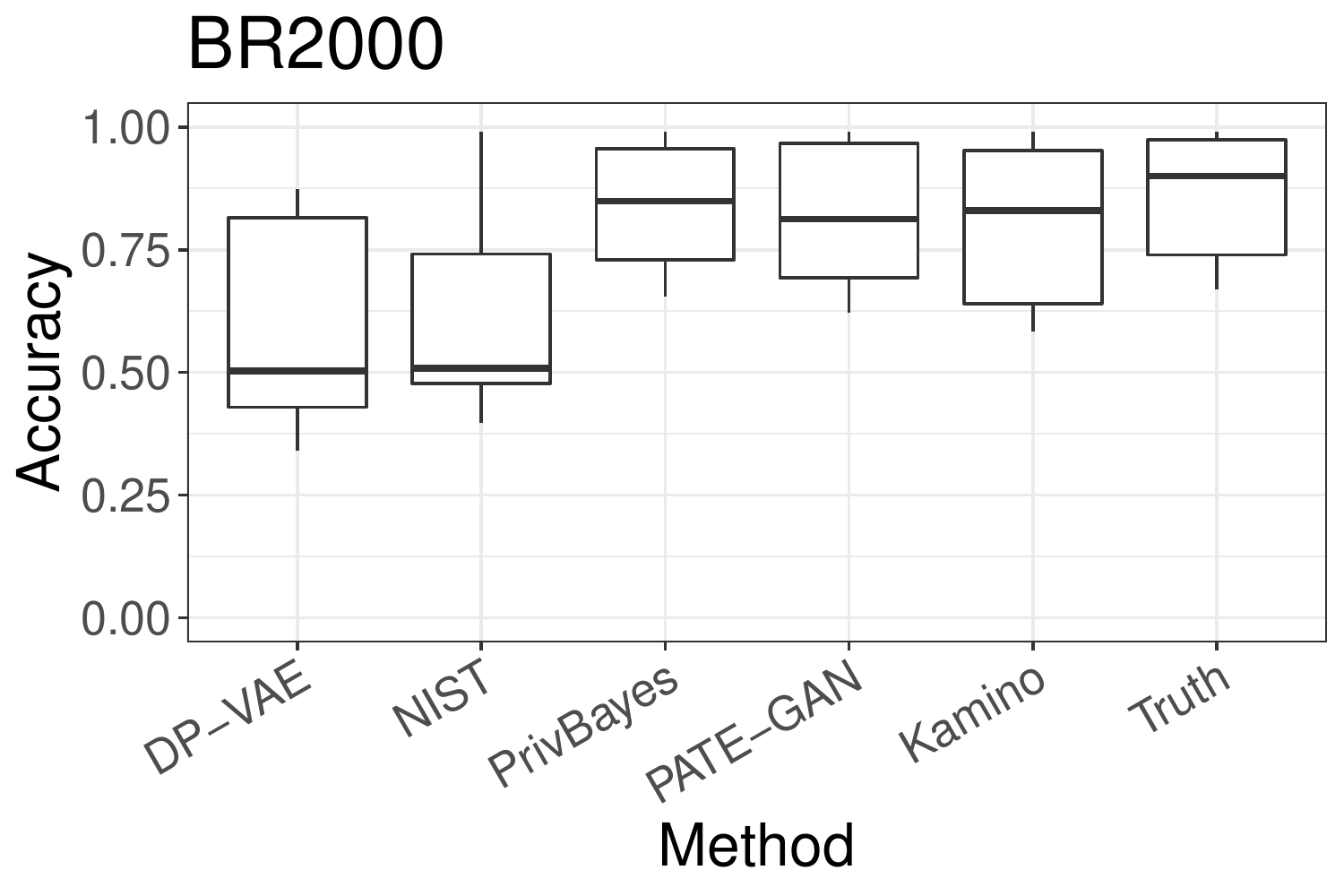}%
    \end{subfigure}
    \begin{subfigure}[h]{0.24\textwidth}%
        \includegraphics[width=\textwidth]{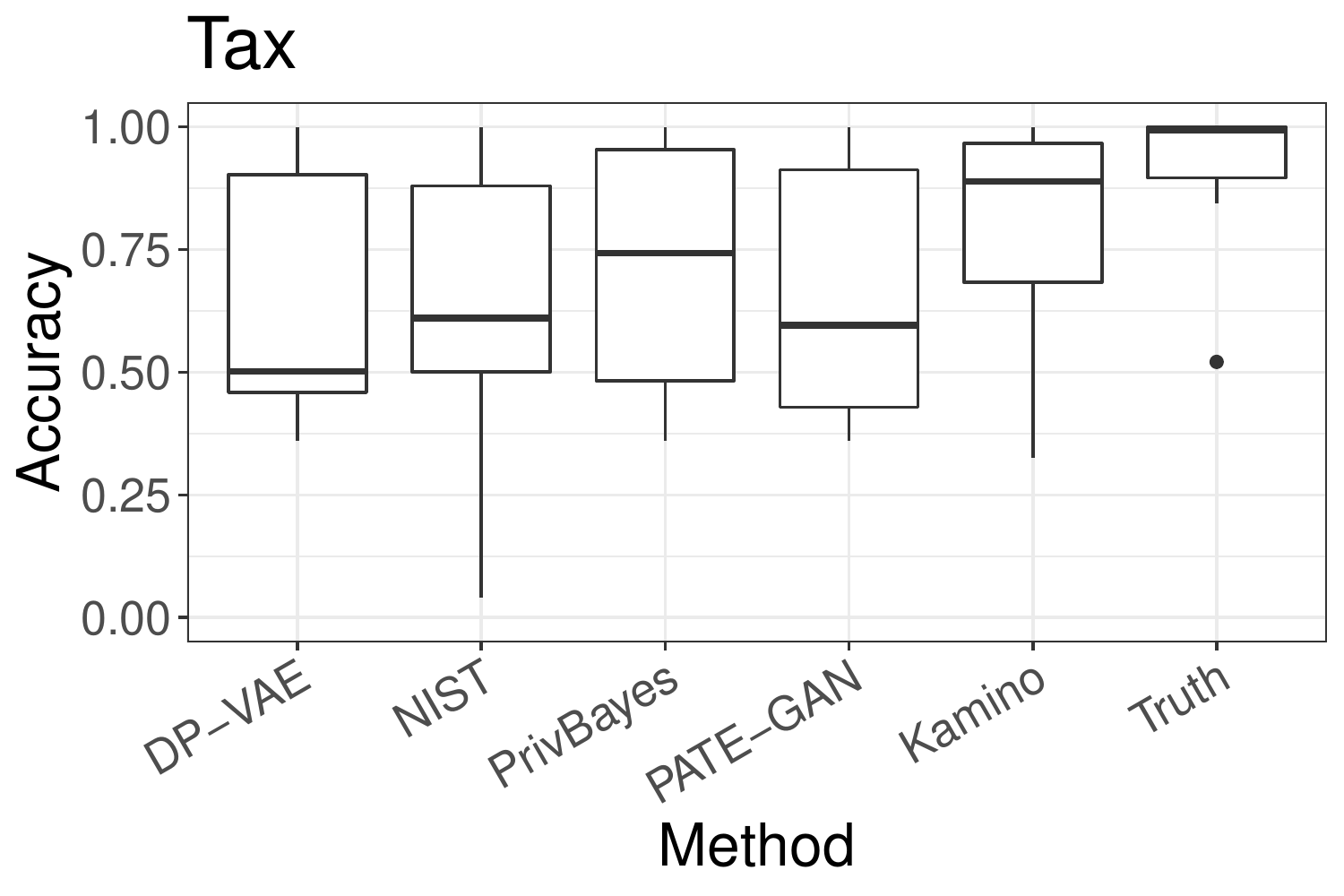}%
    \end{subfigure}%
    \begin{subfigure}[h]{0.24\textwidth}%
        \includegraphics[width=\textwidth]{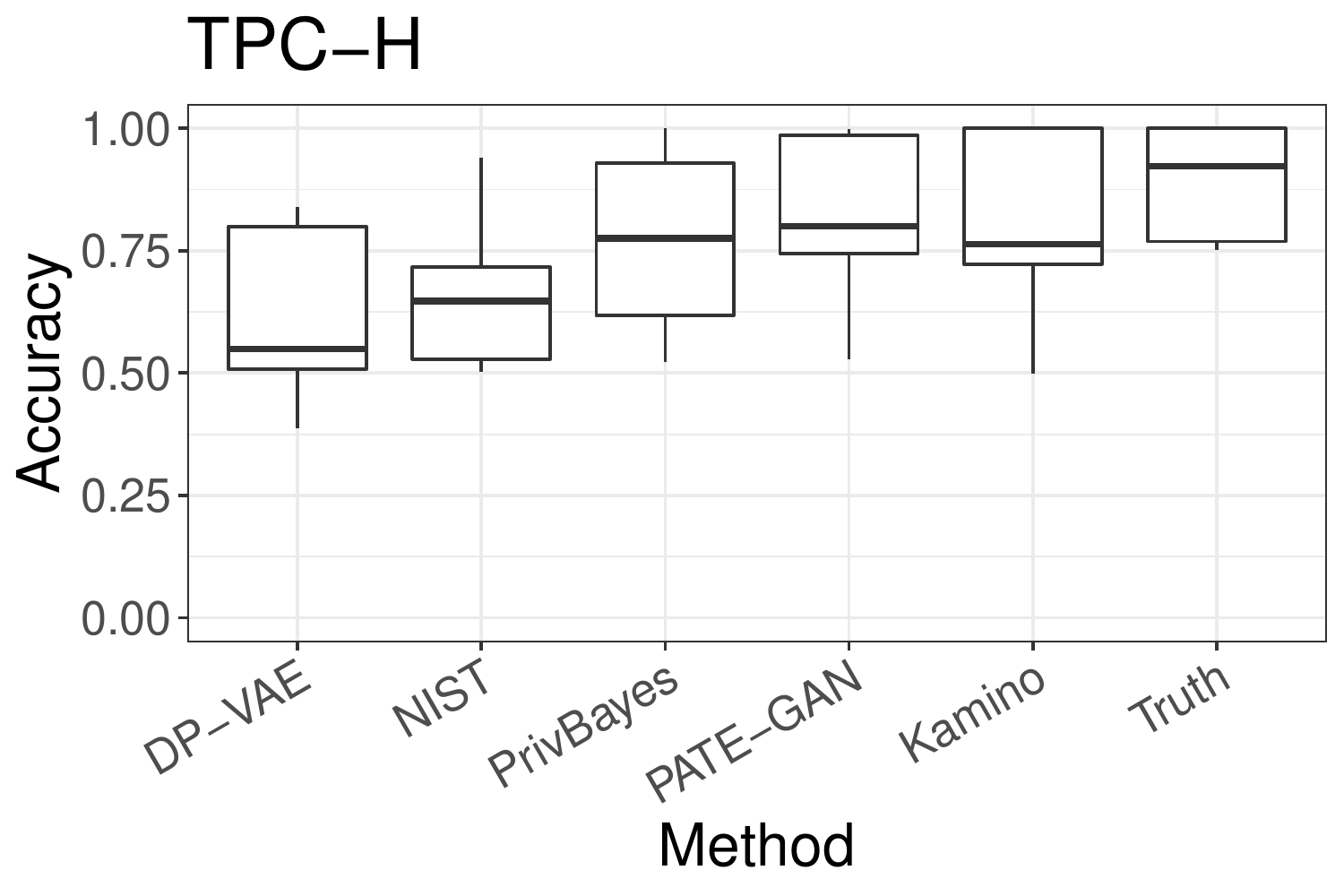}%
    \end{subfigure}%
    
    \begin{subfigure}[h]{0.24\textwidth}%
        \includegraphics[width=\textwidth]{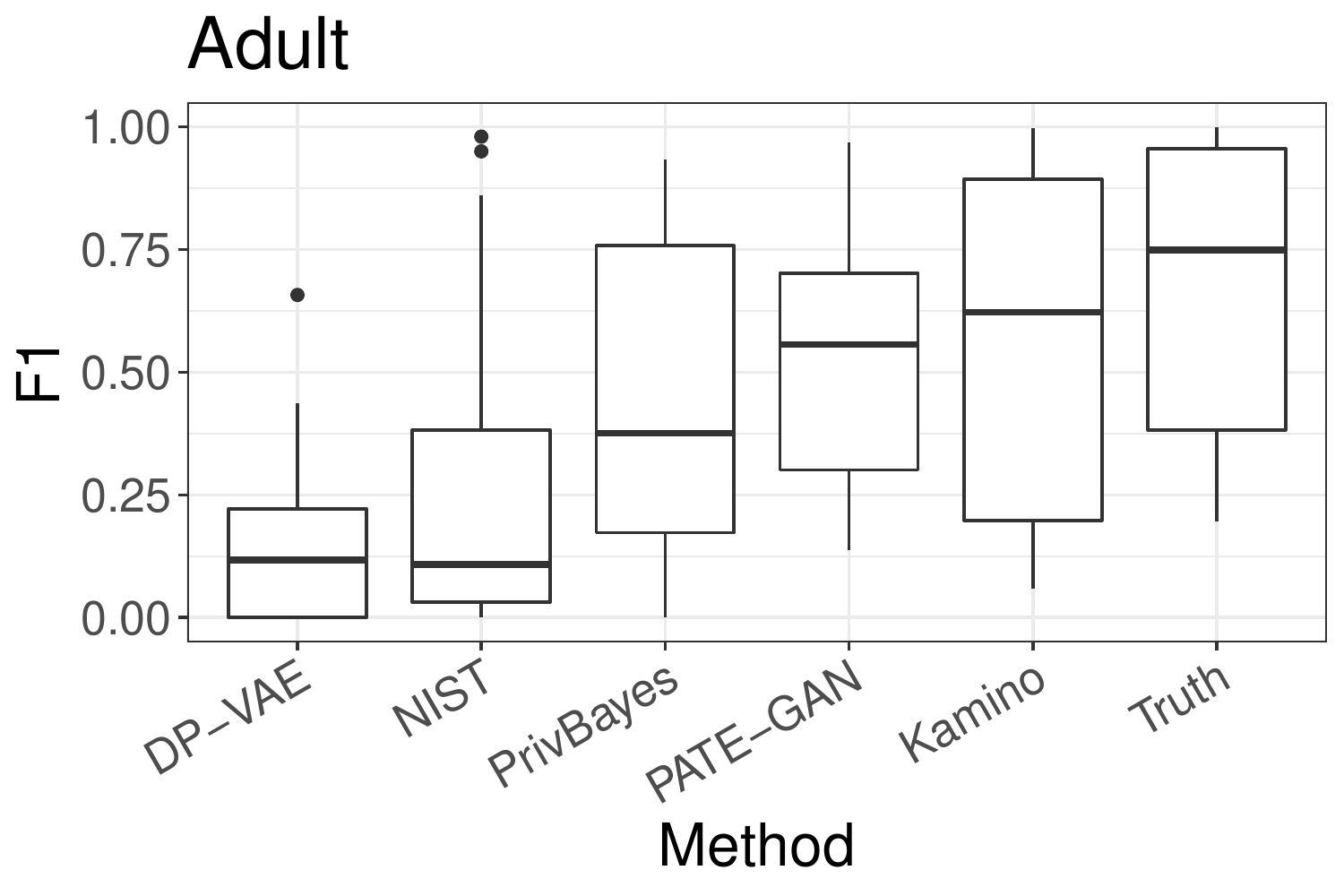}%
    \end{subfigure}%
    \begin{subfigure}[h]{0.24\textwidth}%
        \includegraphics[width=\textwidth]{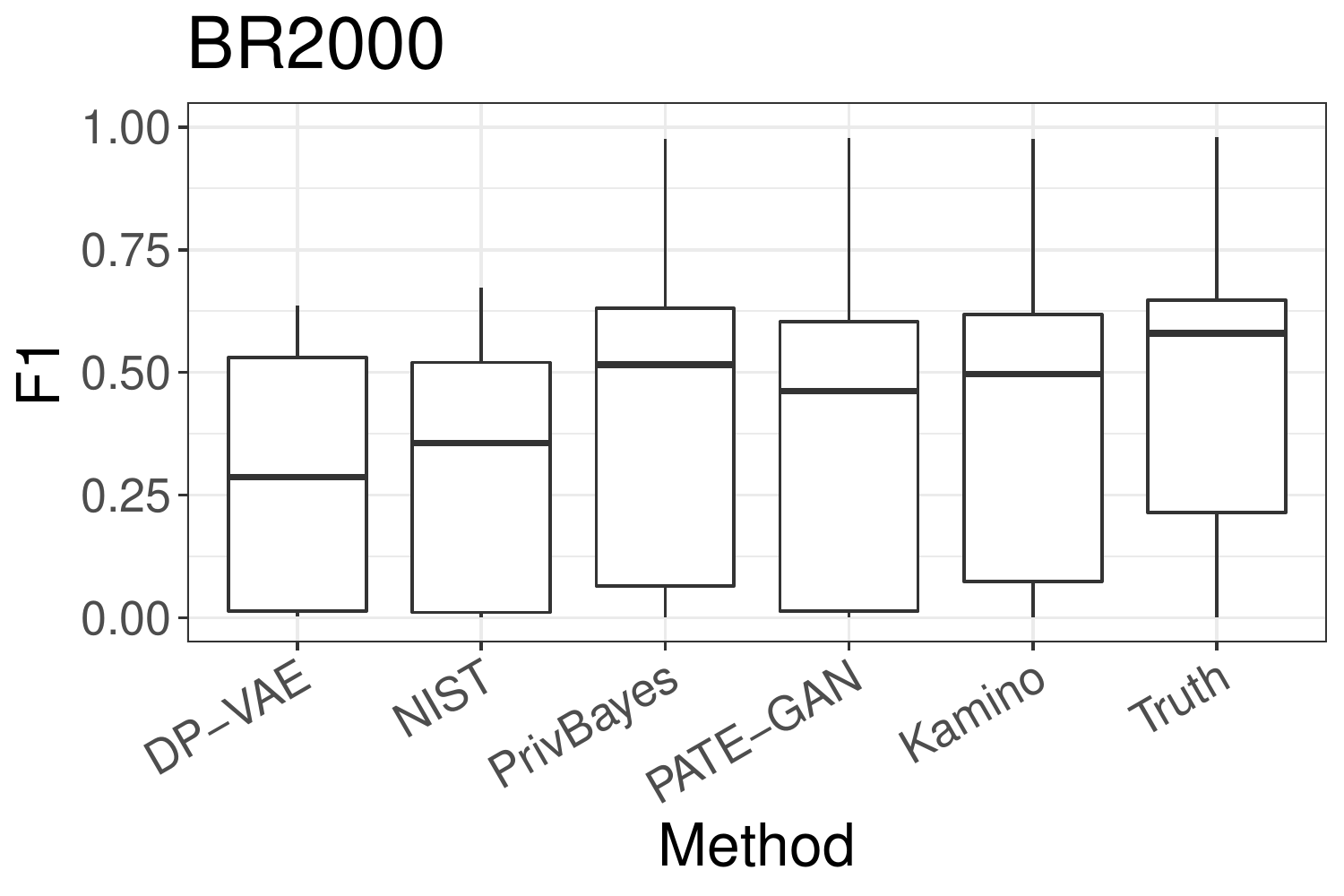}%
    \end{subfigure}
    \begin{subfigure}[h]{0.24\textwidth}%
        \includegraphics[width=\textwidth]{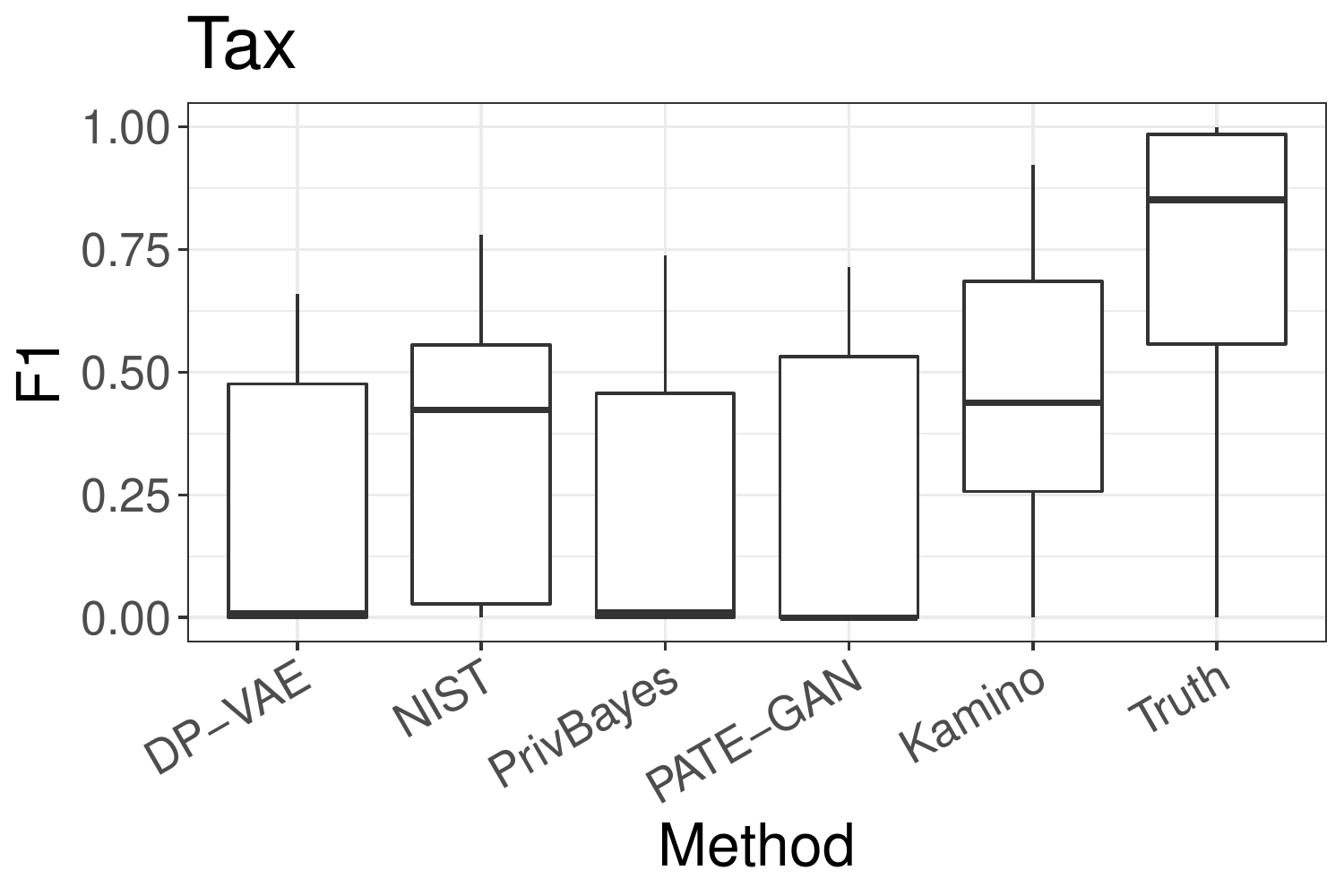}%
    \end{subfigure}%
    \begin{subfigure}[h]{0.24\textwidth}%
        \includegraphics[width=\textwidth]{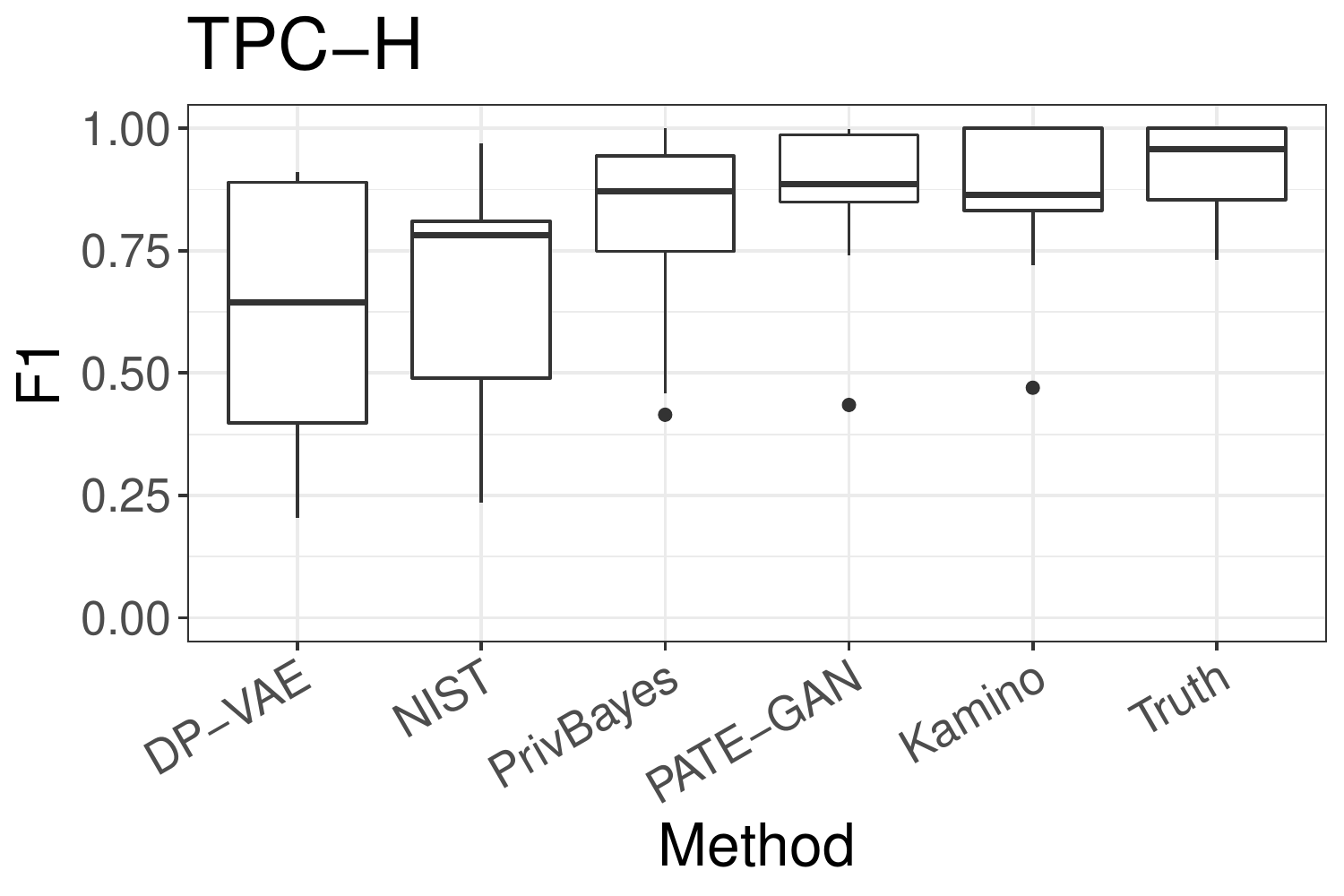}%
    \end{subfigure}%

    \vspace{-.75em}
    \caption{
    Accuracy and F1 of evaluating classification models, which are tested on the true dataset and trained on synthetic data by different methods. Each point represents an averaged classification quality (accuracy or F1) over 9 models for one target attribute using all other attributes as features. Each box represents a set of classifications, one for each attribute in the schema.
    \system achieves the overall best accuracy and F1 scores on most datasets. 
    }
    \label{fig:training}
\end{figure*}

\stepcounter{exp}
\subsubsection{Experiment \theexp: Model Training}

Figure~\ref{fig:training} shows the accuracy and F1 on classifying all attributes.
Each data point in Figure~\ref{fig:training} represents an average of 9 classification models for classifying one target attribute,
and we use the box plot to show classification quality on all attributes for each of the dataset.
As Figure~\ref{fig:training} shows, \system achieves the best overall accuracy and F1 on most datasets:
the mean of all attributes in \system is the closest to the truth, and other quartiles are the best for majority of the tests comparing to the baseline systems.
For instance, on Adult, training and testing on the true database instance gives average accuracy of 0.88.
The models on the synthetic data by \system is 0.82, 
which outperforms PATE-GAN (0.77), PrivBayes (0.68), NIST (0.66), and DP-VAE (0.54). 

\stepcounter{exp}
\subsubsection{Experiment \theexp: $\alpha$-way Marginals}

\begin{figure*}[t]
    \centering
    
    \begin{subfigure}[h]{0.24\textwidth}%
        \includegraphics[width=\textwidth]{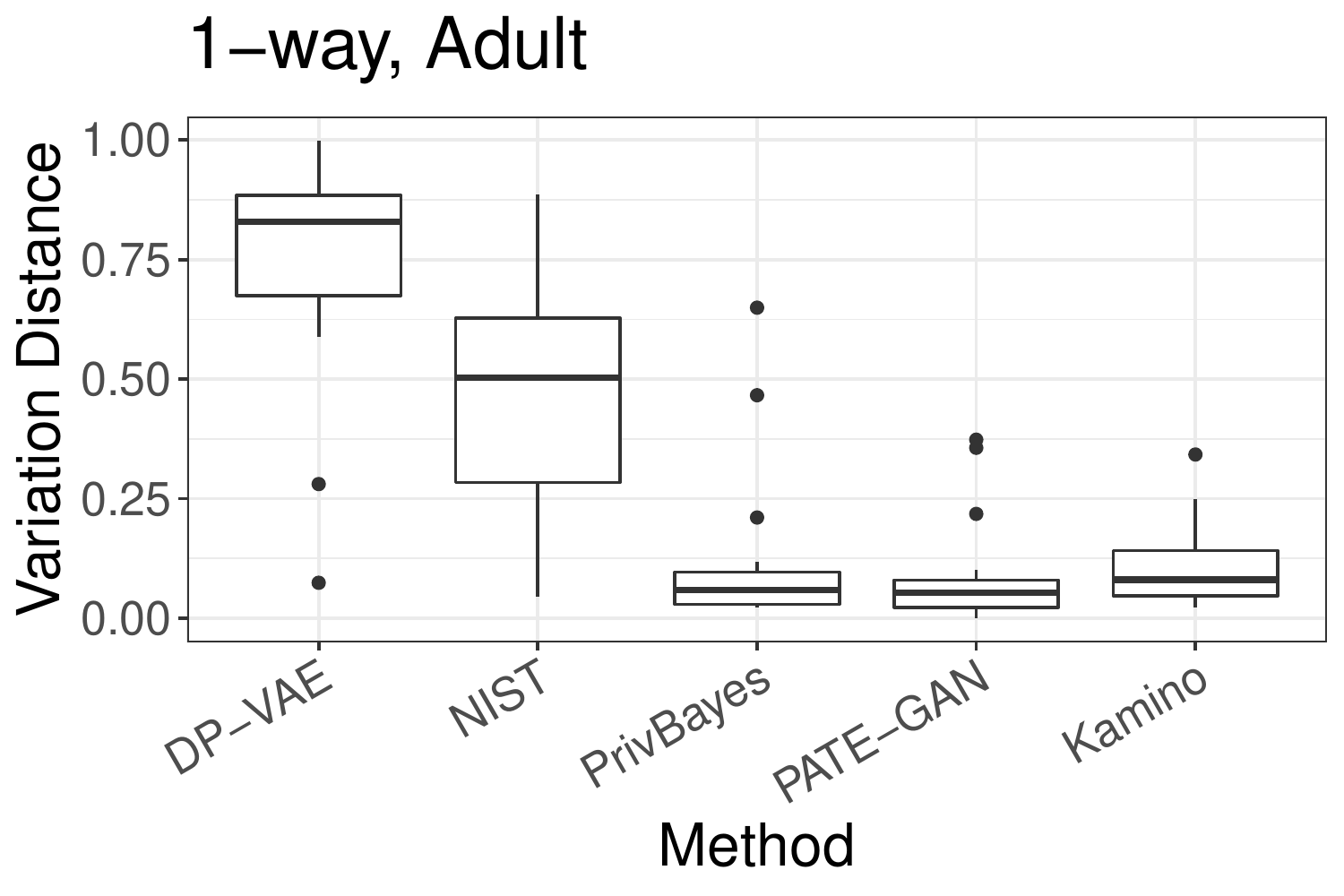}%
    \end{subfigure}%
    \begin{subfigure}[h]{0.24\textwidth}%
        \includegraphics[width=\textwidth]{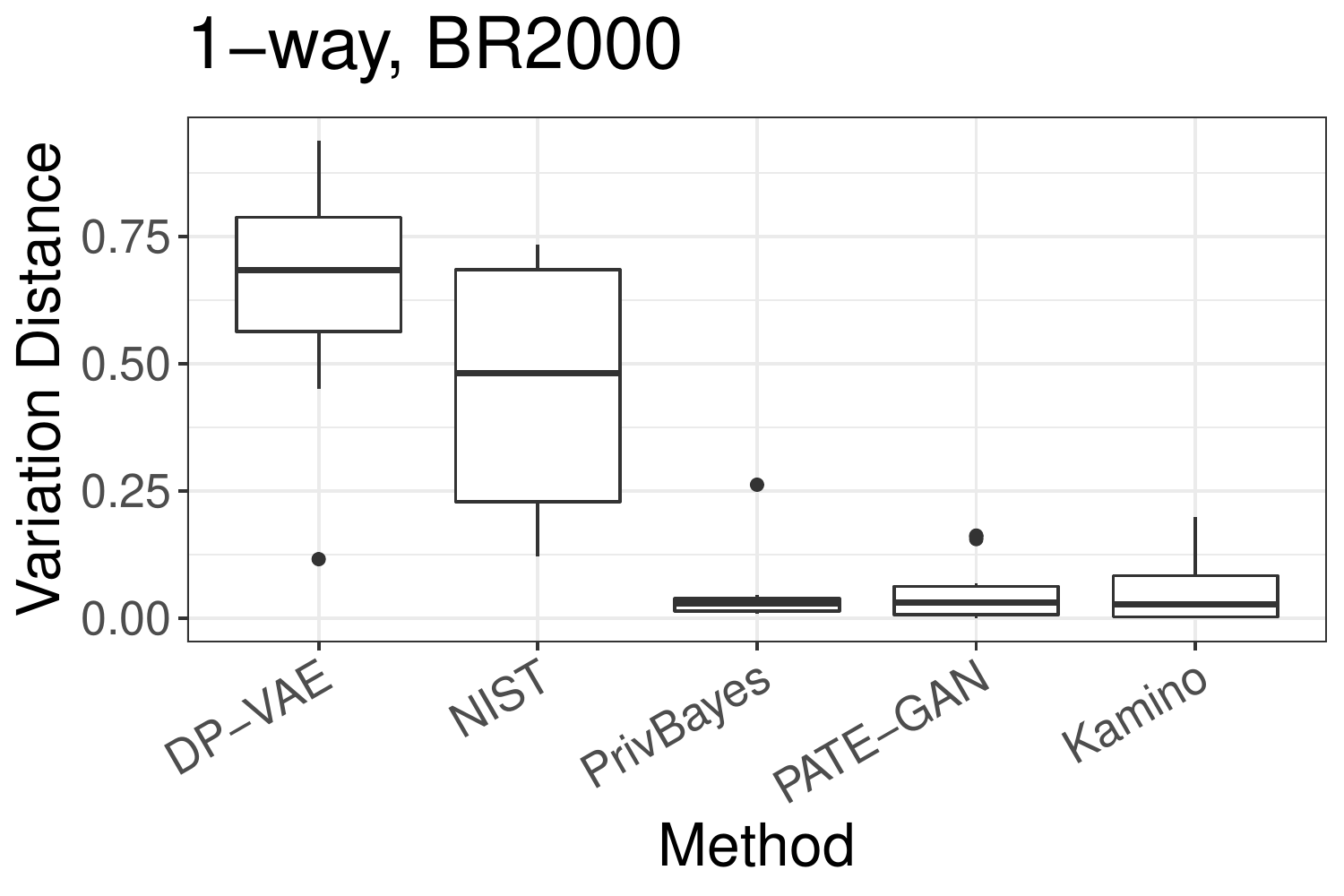}%
    \end{subfigure}
    \begin{subfigure}[h]{0.24\textwidth}%
        \includegraphics[width=\textwidth]{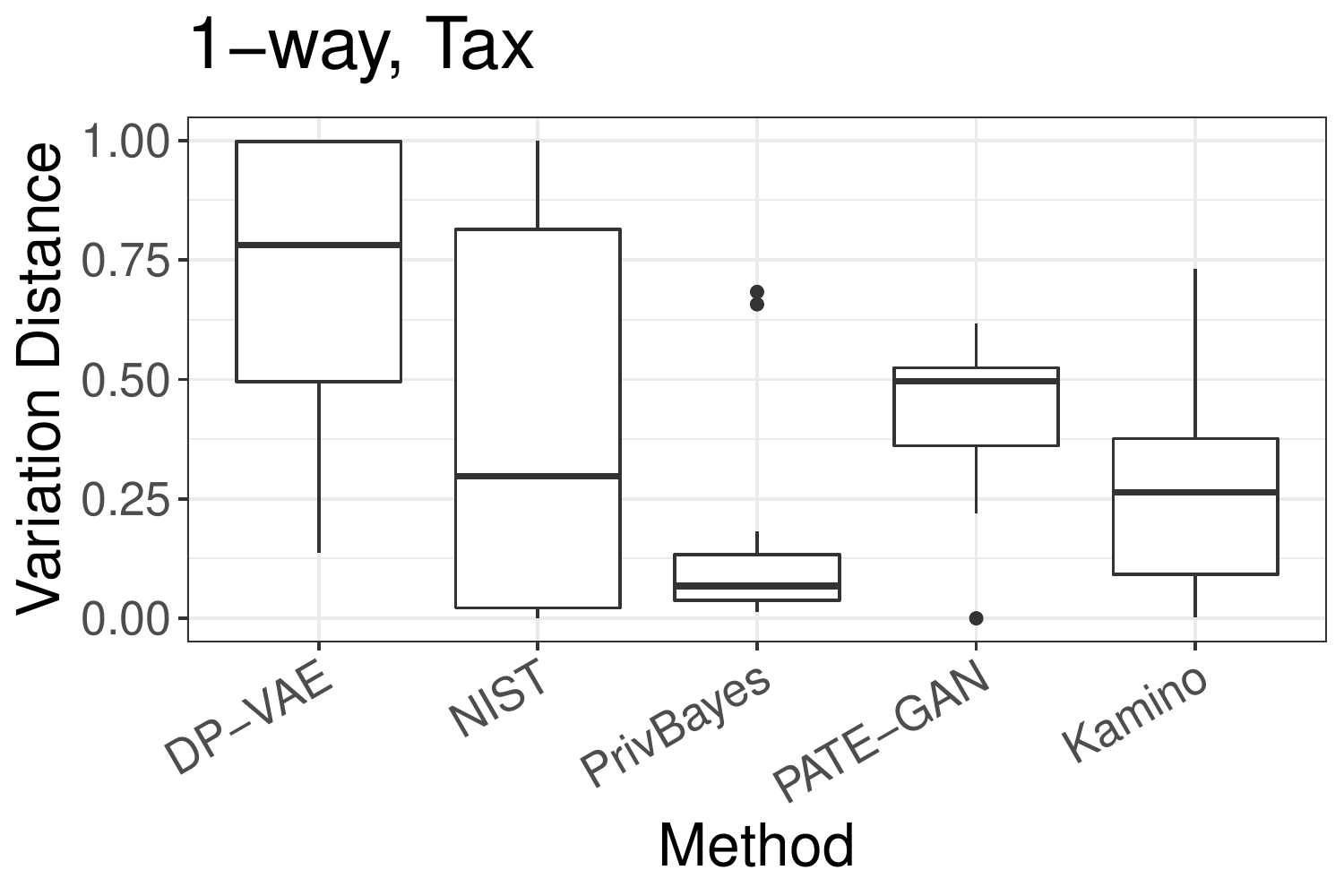}%
    \end{subfigure}%
    \begin{subfigure}[h]{0.24\textwidth}%
        \includegraphics[width=\textwidth]{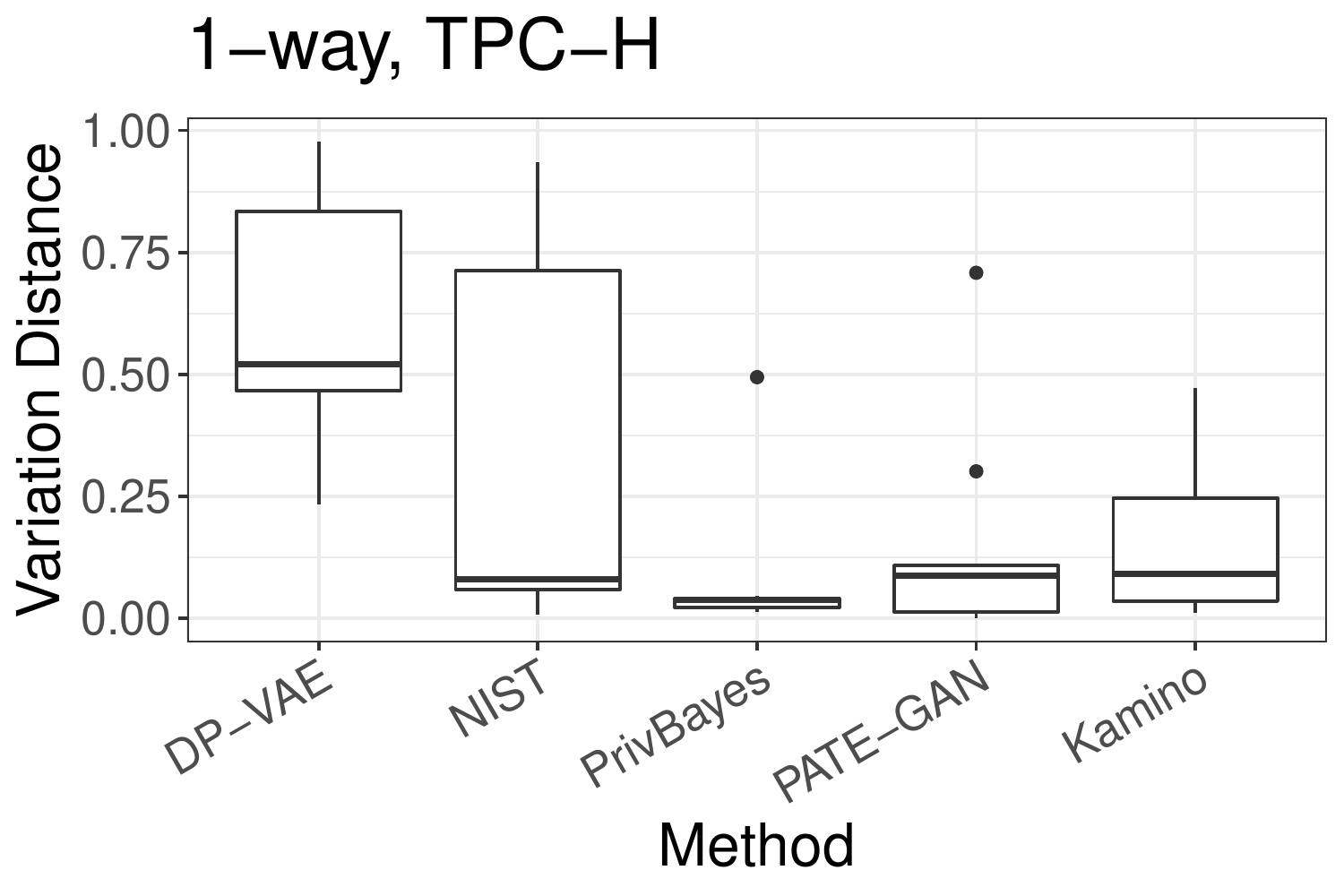}%
    \end{subfigure}%
    
    \begin{subfigure}[h]{0.24\textwidth}%
        \includegraphics[width=\textwidth]{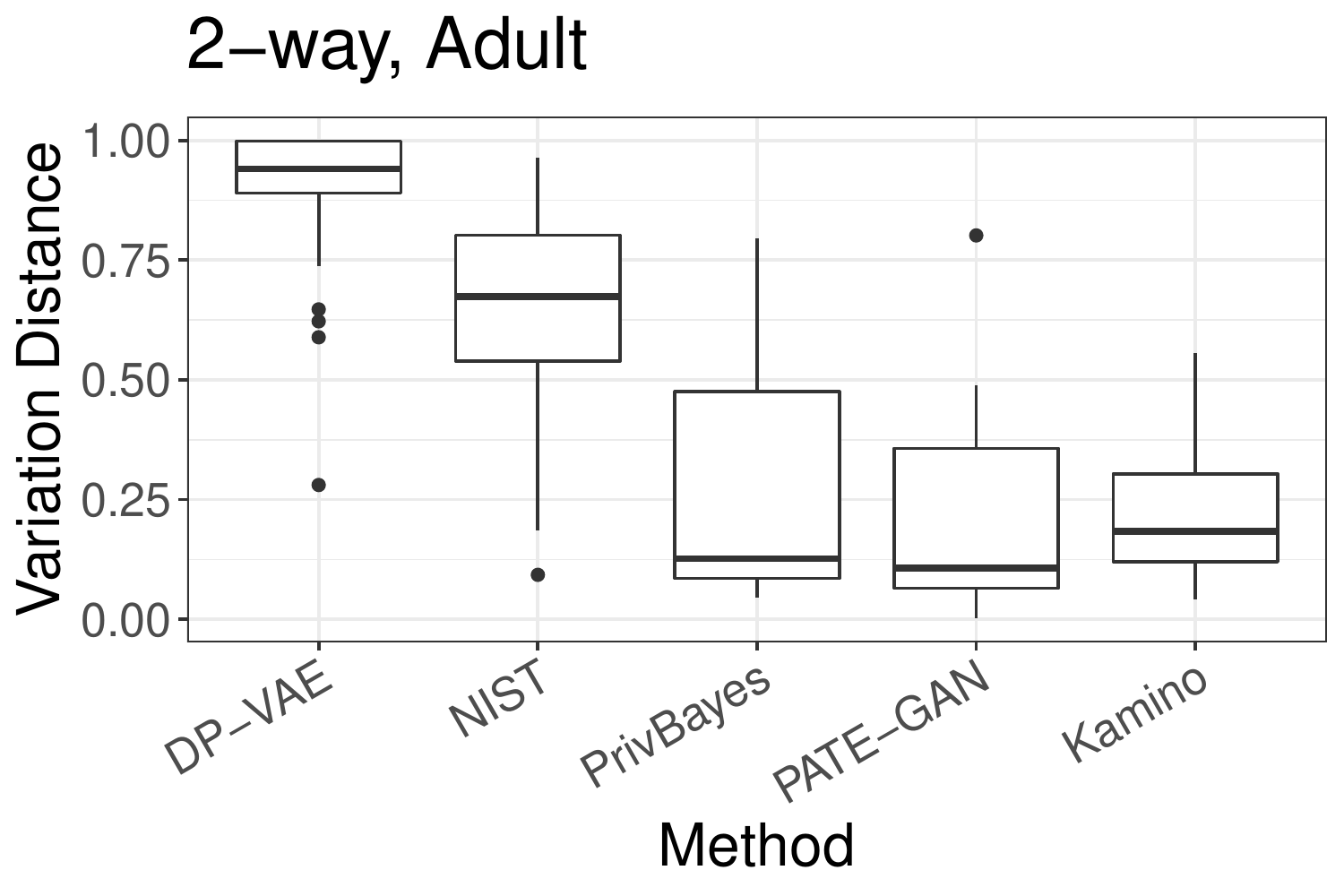}%
    \end{subfigure}%
    \begin{subfigure}[h]{0.24\textwidth}%
        \includegraphics[width=\textwidth]{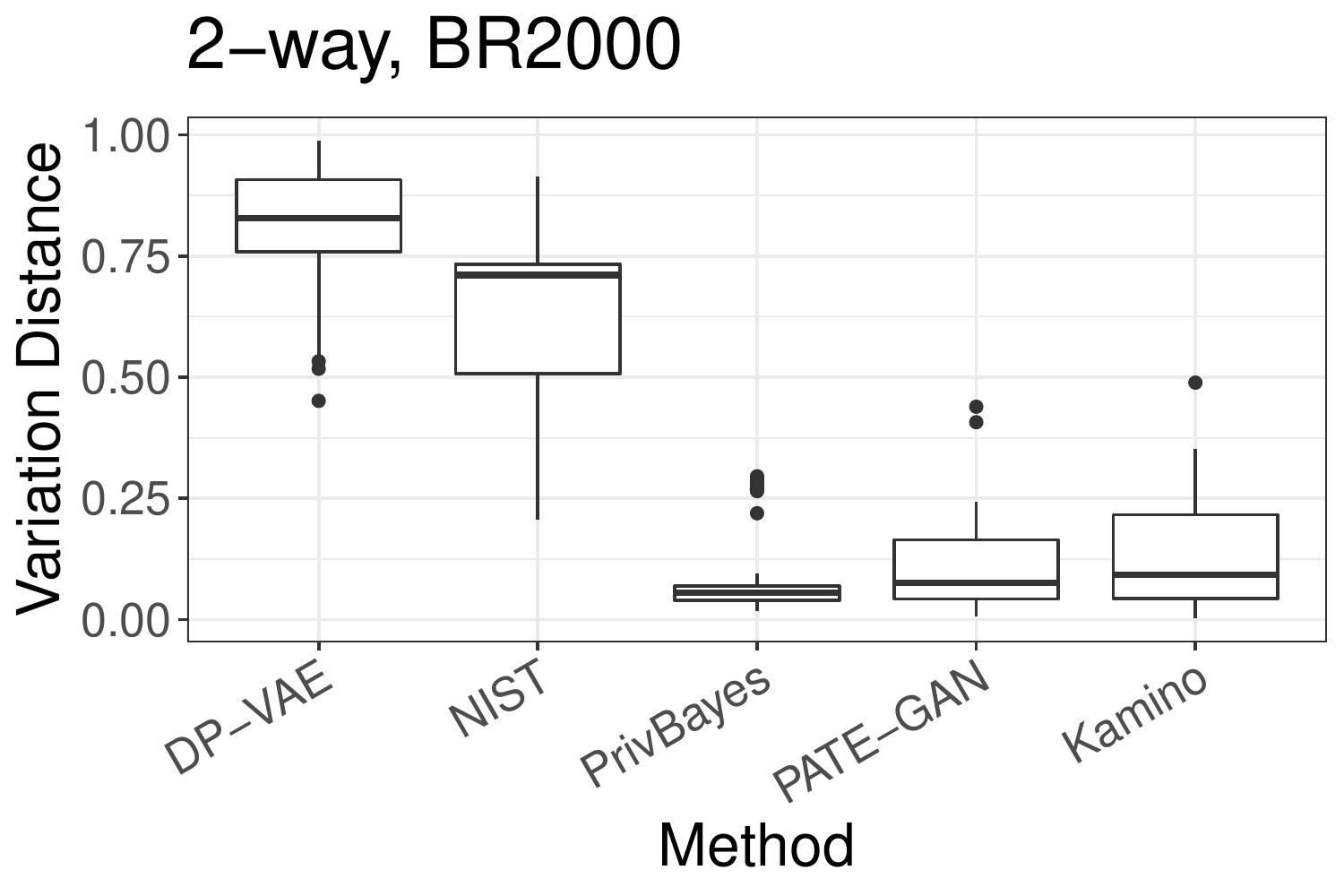}%
    \end{subfigure}
    \begin{subfigure}[h]{0.24\textwidth}%
        \includegraphics[width=\textwidth]{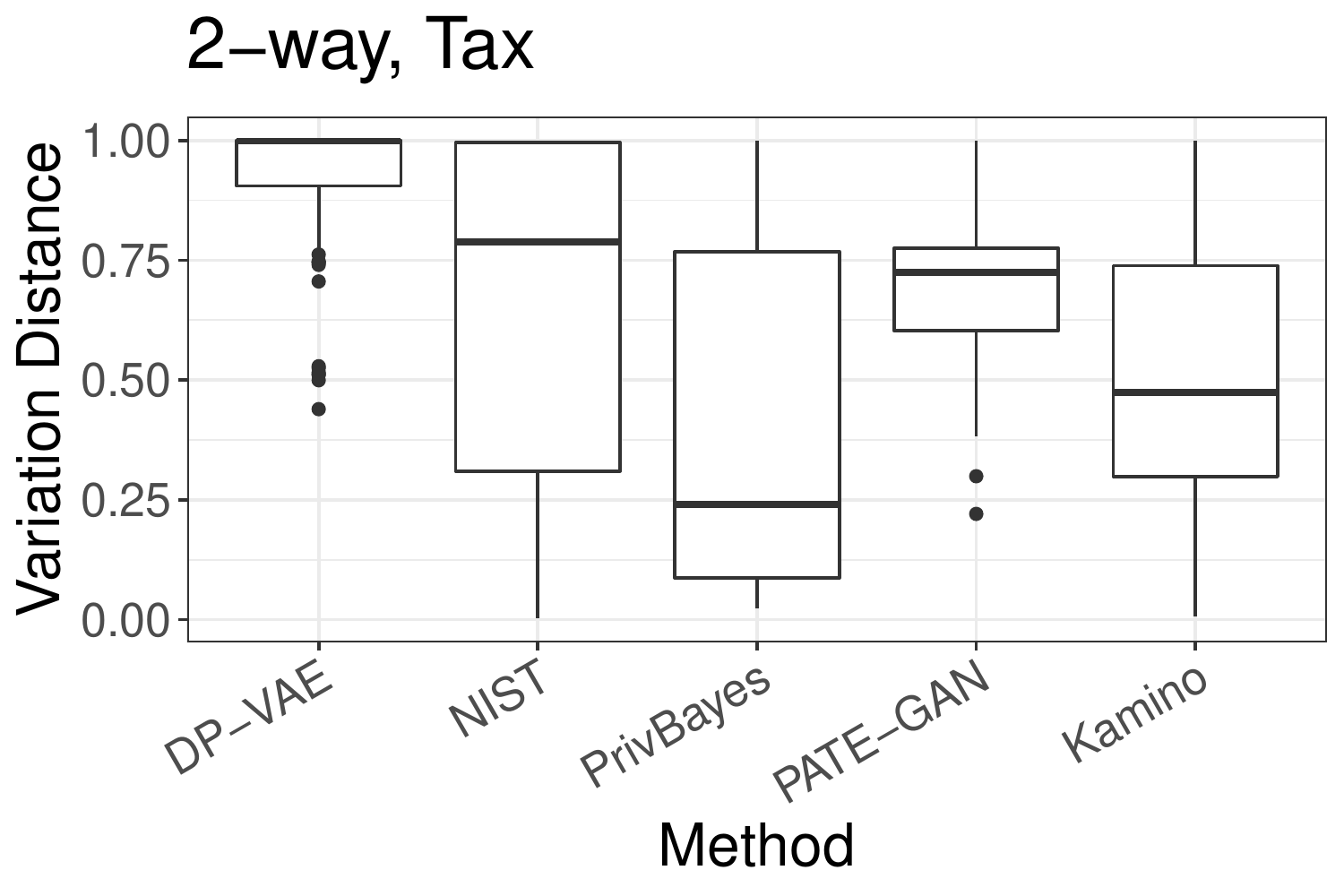}%
    \end{subfigure}%
    \begin{subfigure}[h]{0.24\textwidth}%
        \includegraphics[width=\textwidth]{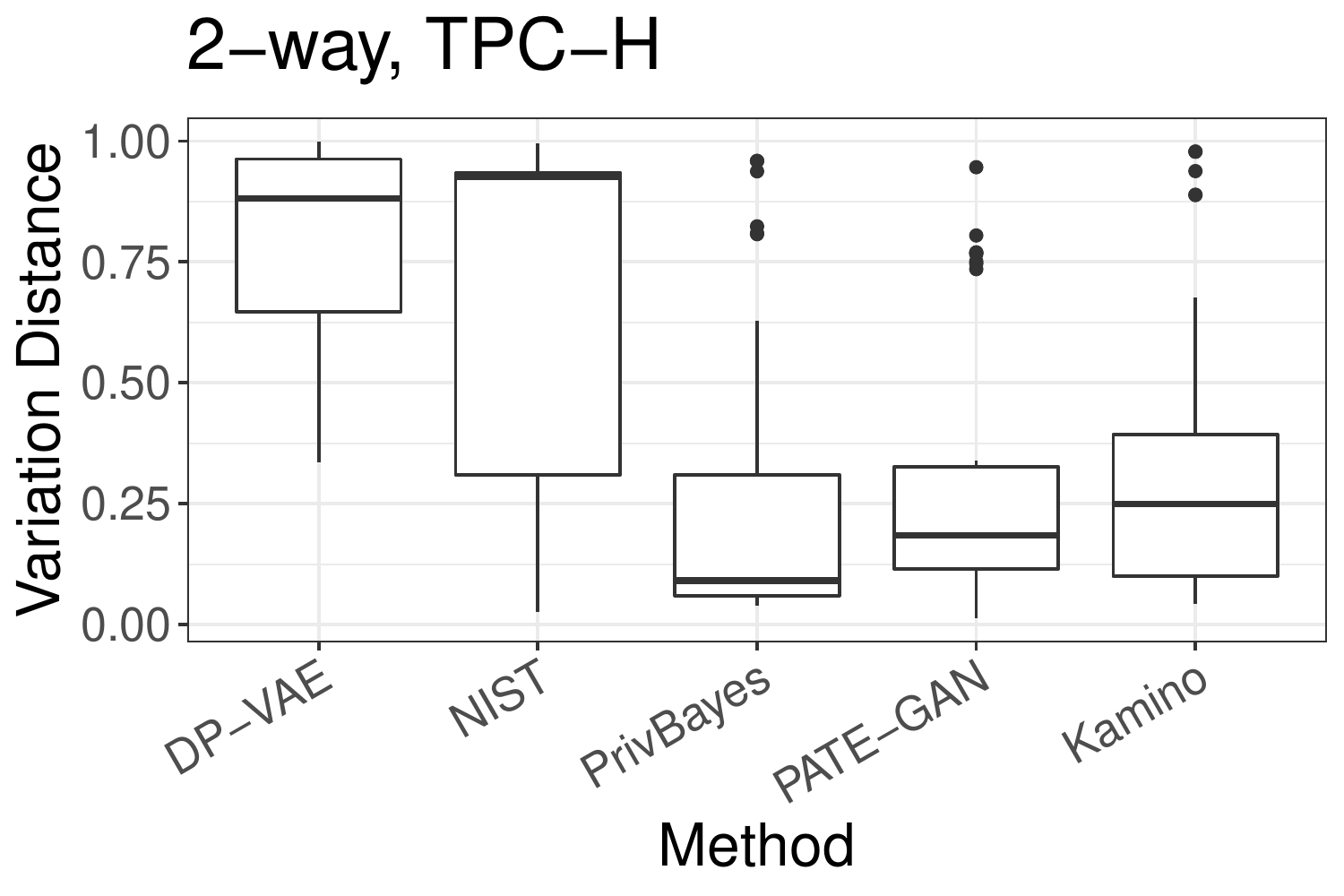}%
    \end{subfigure}%
    
    
    \vspace{-.75em}
    \caption{Total variation distance on $\alpha$-way marginals, where $\alpha=[1,2]$. Each point represents a total variance distance for one attribute set, and each box represents total variance distance for all attribute sets. 
    It shows that \system can achieve overall the best (Adult) or close to the best (BR2000, Tax and TPC-H) variation distance.
    }
    \label{fig:marginals}
\end{figure*}

Figure~\ref{fig:marginals} shows the total variation distance for all attributes or attribute combinations on each of the dataset.
Each data point represents a total variation distance of the distributions between the true database instance and the synthetic database instance, for a certain attribute (1-way) or an attribute set (2-way).
As it shows, \system has the smallest or close to the smallest variation distances.
Taking the first 1-way marginal on the Adult dataset as an example, 
\system has a mean of 0.11, which is second to the smallest mean of PATE-GAN (0.09),
and a maximal distance of 0.34, which is the smallest comparing to PATE-GAN (0.37), PrivBayes (0.65), NIST (0.89), and DP-VAE (1.0).

\stepcounter{exp}
\subsubsection{Experiment \theexp: Execution time}\label{sec:exp:time}
Since \system explicitly checks DC violations during sampling,
it is expected to take longer running time than baseline methods that generate i.i.d samples.
In our evaluation, NIST and PrivBayes were the most efficient on all datasets, and took at most 217$\pm$13 and 1,367$\pm$561 seconds, respectively.
Because of training deep models on encoded data, running time of DP-VAE and PATE-GAN on all datasets fell into the range of 20 minutes to 13 hours.
For \system, the running time on all datasets were in 5-16 hours, which is still practically efficient.

\revise{
Figure~\ref{fig:time_profile} profiles \system's execution time of each process (sequencing, model training, computing violation matrix and learn DC weights for soft DCs, and sampling).
As Figure~\ref{fig:time_profile} shows, performance of \system is dominated by training and sampling, which together take more than 99\% of the total time.
\ifpaper
MCMC re-sampling further increases the sampling time, but it leads to better task qualities. We show the detailed evaluation on MCMC re-sampling in our full paper~\cite{full_paper}.

In addition, the full paper also include evaluations of optimization techniques, which can speed up model training on Adult by 3.5$\times$, and allow \system to complete in 10 hours for a TPC-H dataset that scaled up to 1 million rows. 
\else
\fi
}

\eat{
Table~\ref{tab:exp:perf} shows the running time of generating synthetic datasets by different methods.
\system explicitly checks DC violations at the sampling and weight learning steps,
and hence it requires more execution tome, comparing to the baseline methods doing i.i.d. sampling.
Despite the longer execution time, the longest synthesis by \system completed in about 16 hours on the Tax dataset, which is still practically efficient.

\begin{table}[t]
\centering
\caption{Running time (in second) of generating synthetic datasets by different methods.}
\label{tab:exp:perf}
\vspace{-.75em}
\begin{tabular}{|c|c|c|c|c|}
\hline
Data   & PrivBayes & DP-VAE      & PATE-GAN    & \system \\ \hline
Adult  & 162$\pm$16  & 5,501$\pm$94 & 2,856$\pm$04  &  31,981$\pm$314  \\ \hline
BR2000 &1,367$\pm$561 & 9,106$\pm$07 & 1,782$\pm$05  &  17,594$\pm$1317  \\ \hline
Tax    & 189$\pm$0   & 4,829$\pm$32 & 1,429$\pm$31  &  58,418$\pm$950  \\ \hline
TPC-H  &   7$\pm$0   & 7,834$\pm$43 &  48,587$\pm$6919 &   16,434$\pm$335  \\ \hline
\end{tabular}%
\end{table}
} 

\begin{figure*}[t]
    \centering    
    \begin{subfigure}[h]{0.24\textwidth}%
        \includegraphics[width=\textwidth]{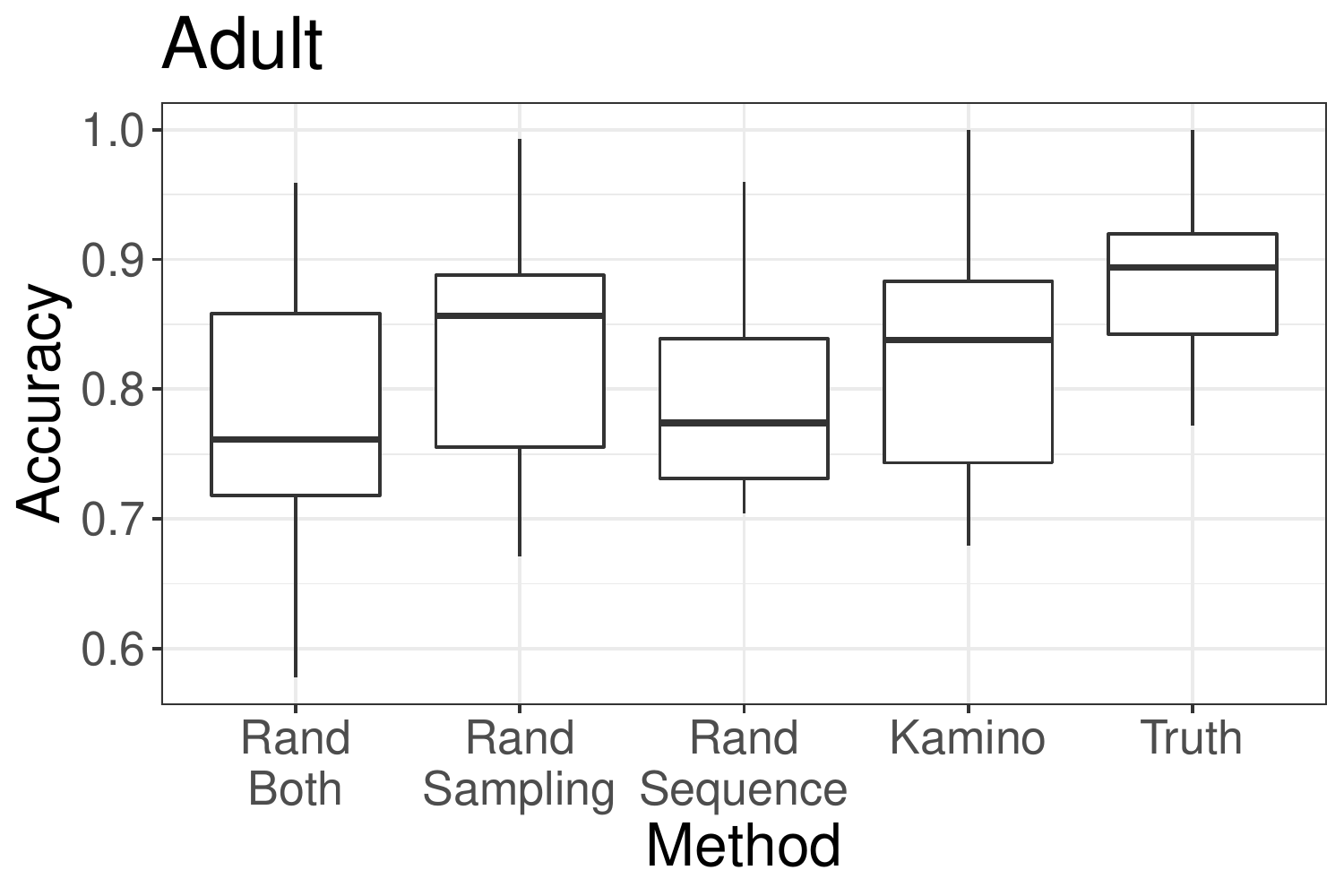}%
        \caption{Accuracy}
    \end{subfigure}%
    \begin{subfigure}[h]{0.24\textwidth}%
        \includegraphics[width=\textwidth]{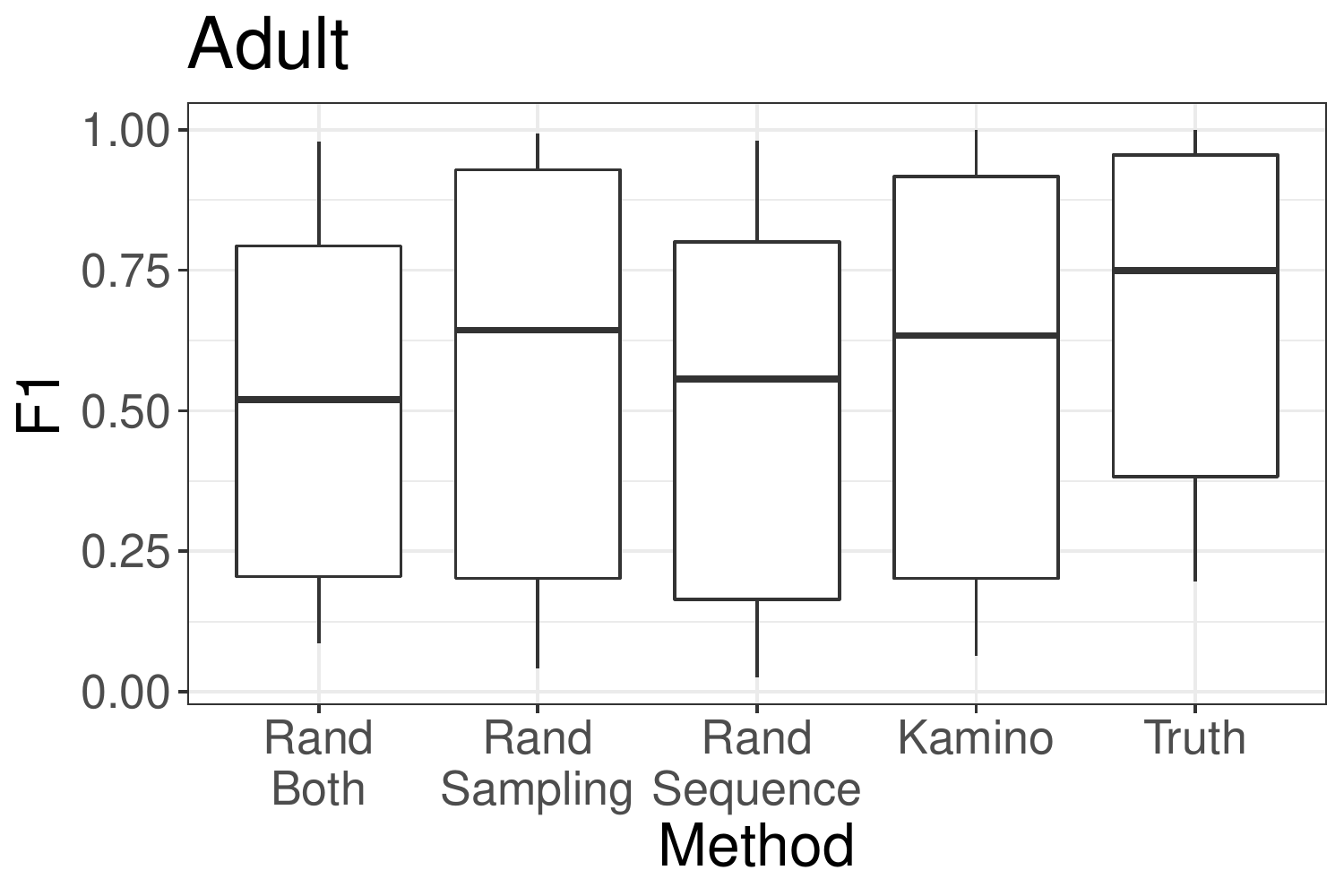}%
        \caption{F1}
    \end{subfigure}%
    \begin{subfigure}[h]{0.24\textwidth}%
        \includegraphics[width=\textwidth]{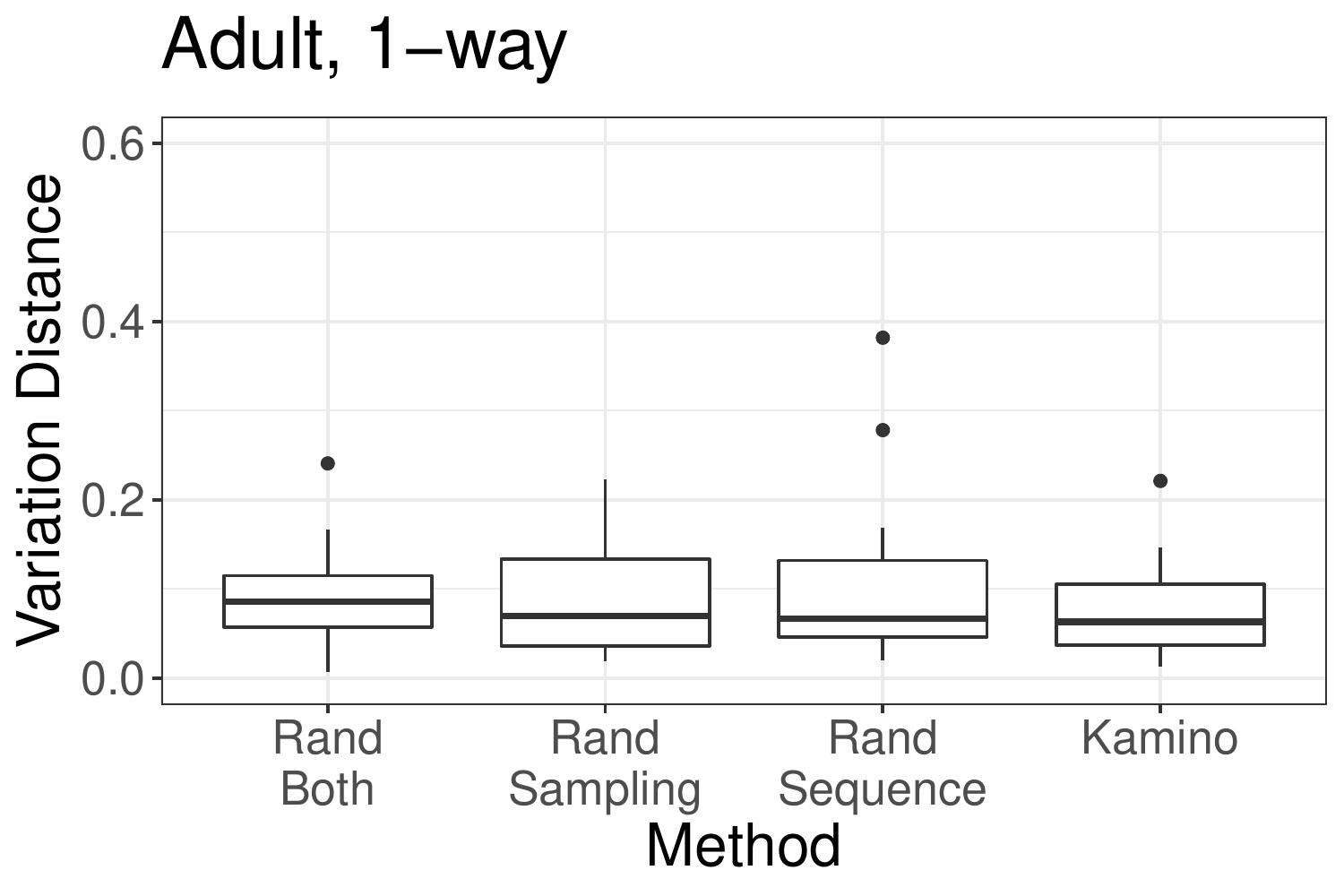}%
        \caption{1-way marginal}
    \end{subfigure}%
    \begin{subfigure}[h]{0.24\textwidth}%
        \includegraphics[width=\textwidth]{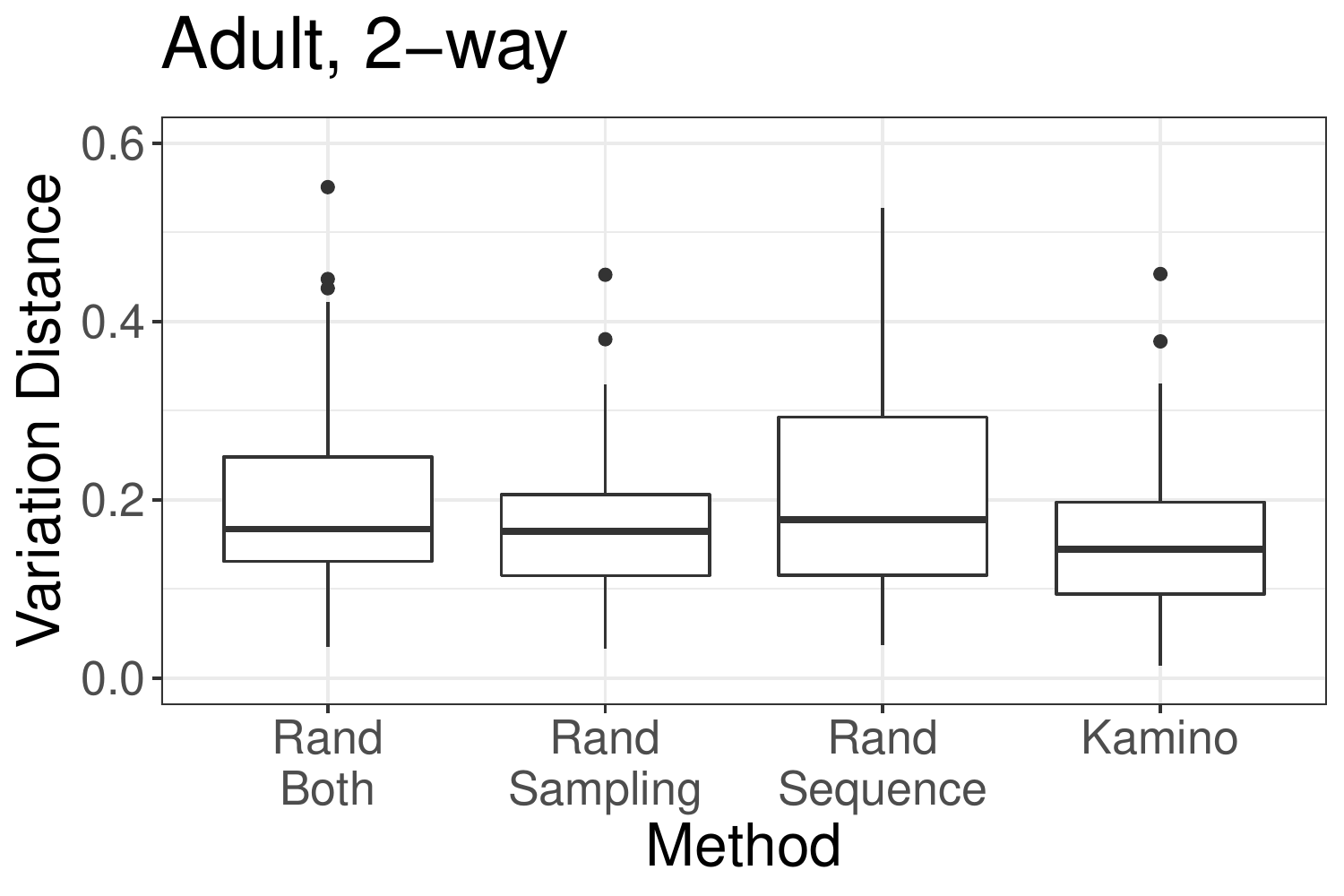}%
        \caption{2-way marginal}
    \end{subfigure}%
    
    \vspace{-.75em}
    \caption{Accuracy and F1 of model training on \system, and sub-optimal \system without constraint-aware sampling, sequencing, and neither, using the Adult dataset as the example. It shows the the \system with constraint-aware components can achieve the best quality in both the learning task and in the query task.}
	\label{fig:training_components}
\end{figure*}

\begin{figure*}[t]
    \centering    
    \begin{subfigure}[h]{0.24\textwidth}%
        \includegraphics[width=\textwidth]{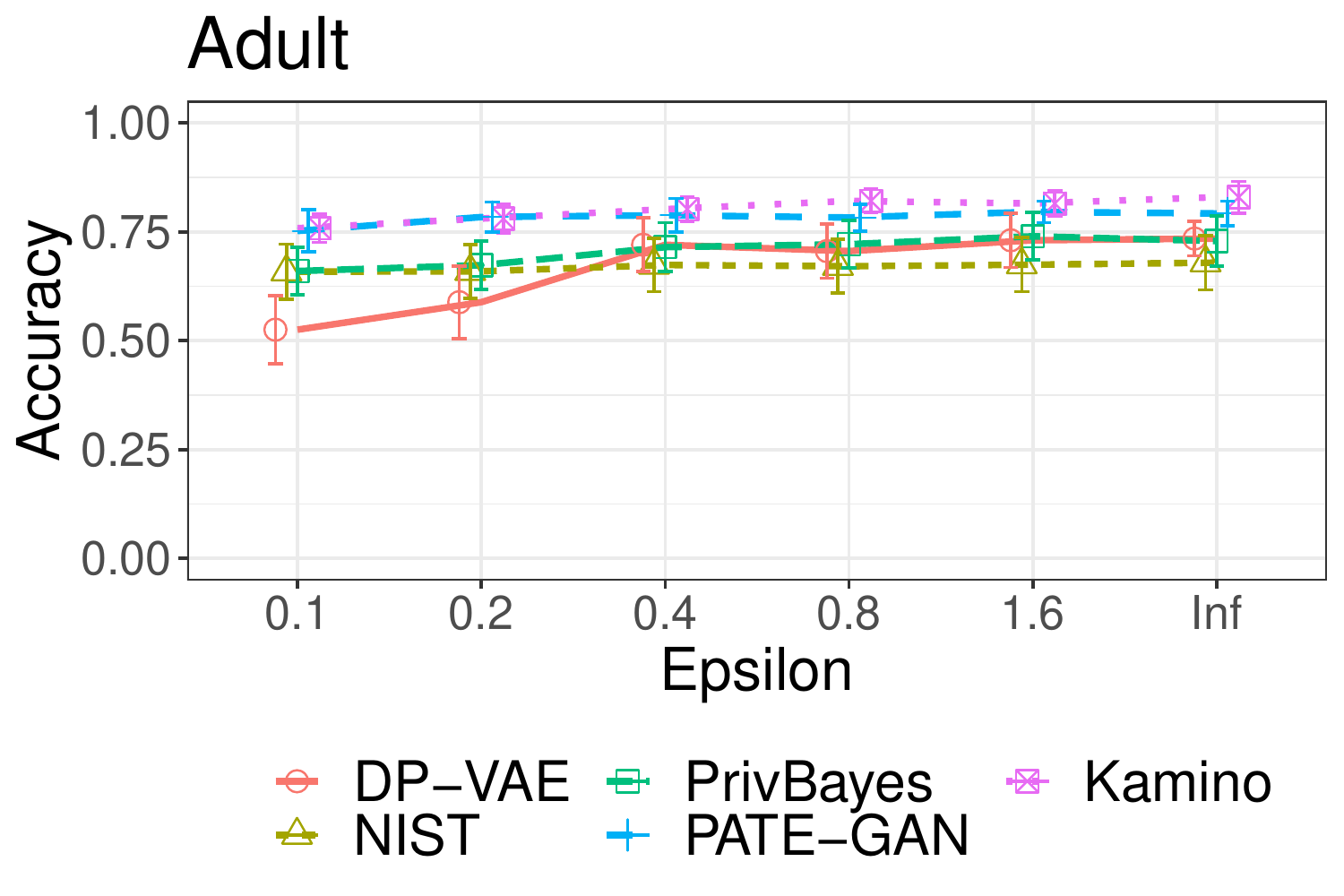}%
        \caption{Accuracy}\label{fig:vary_budget_all:accuracy}
    \end{subfigure}%
    \begin{subfigure}[h]{0.24\textwidth}%
        \includegraphics[width=\textwidth]{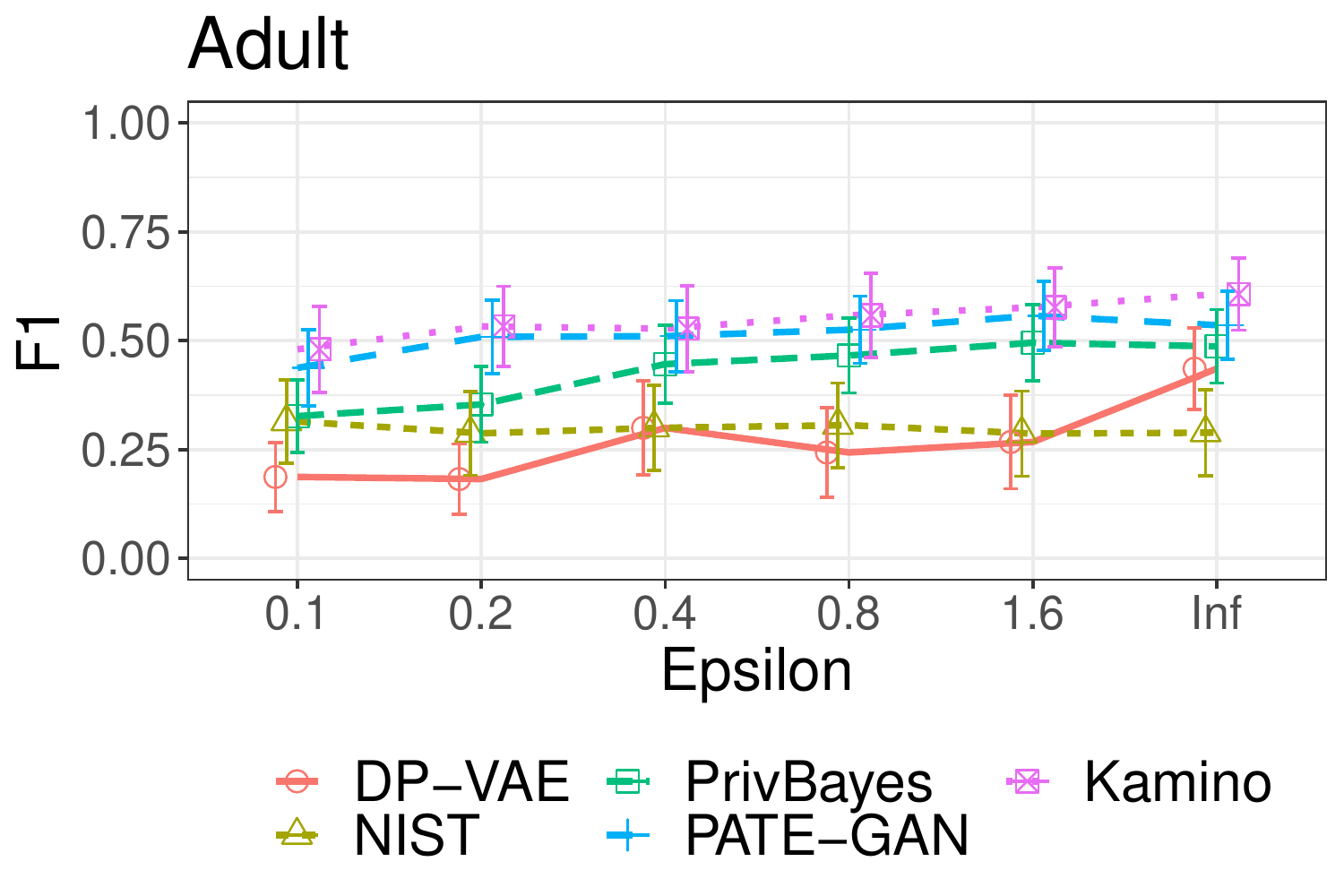}%
        \caption{F1}\label{fig:vary_budget_all:f1}
    \end{subfigure}%
    \begin{subfigure}[h]{0.24\textwidth}%
        \includegraphics[width=\textwidth]{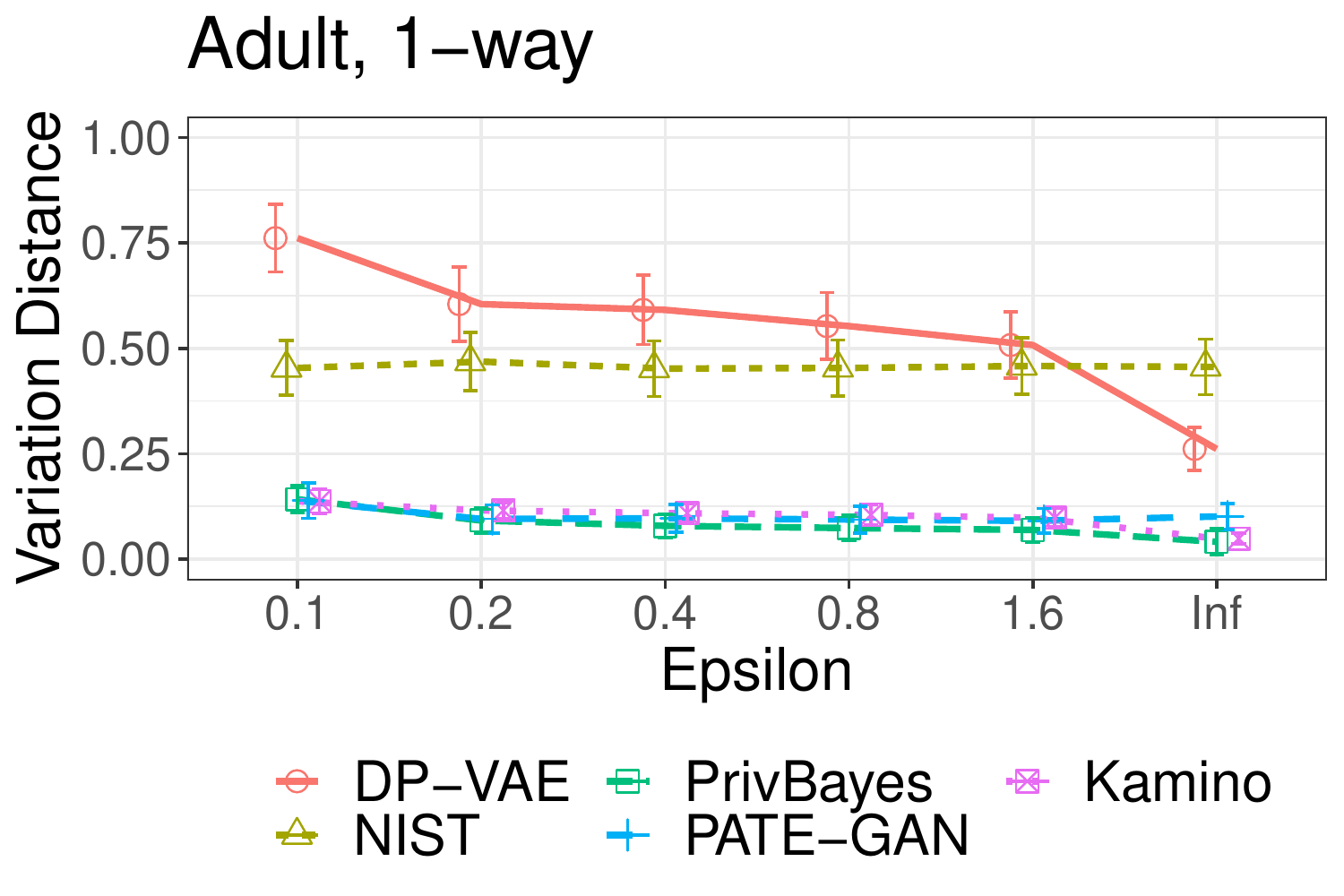}%
        \caption{1-way marginal}\label{fig:vary_budget_all:1-way}
    \end{subfigure}%
    \begin{subfigure}[h]{0.24\textwidth}%
        \includegraphics[width=\textwidth]{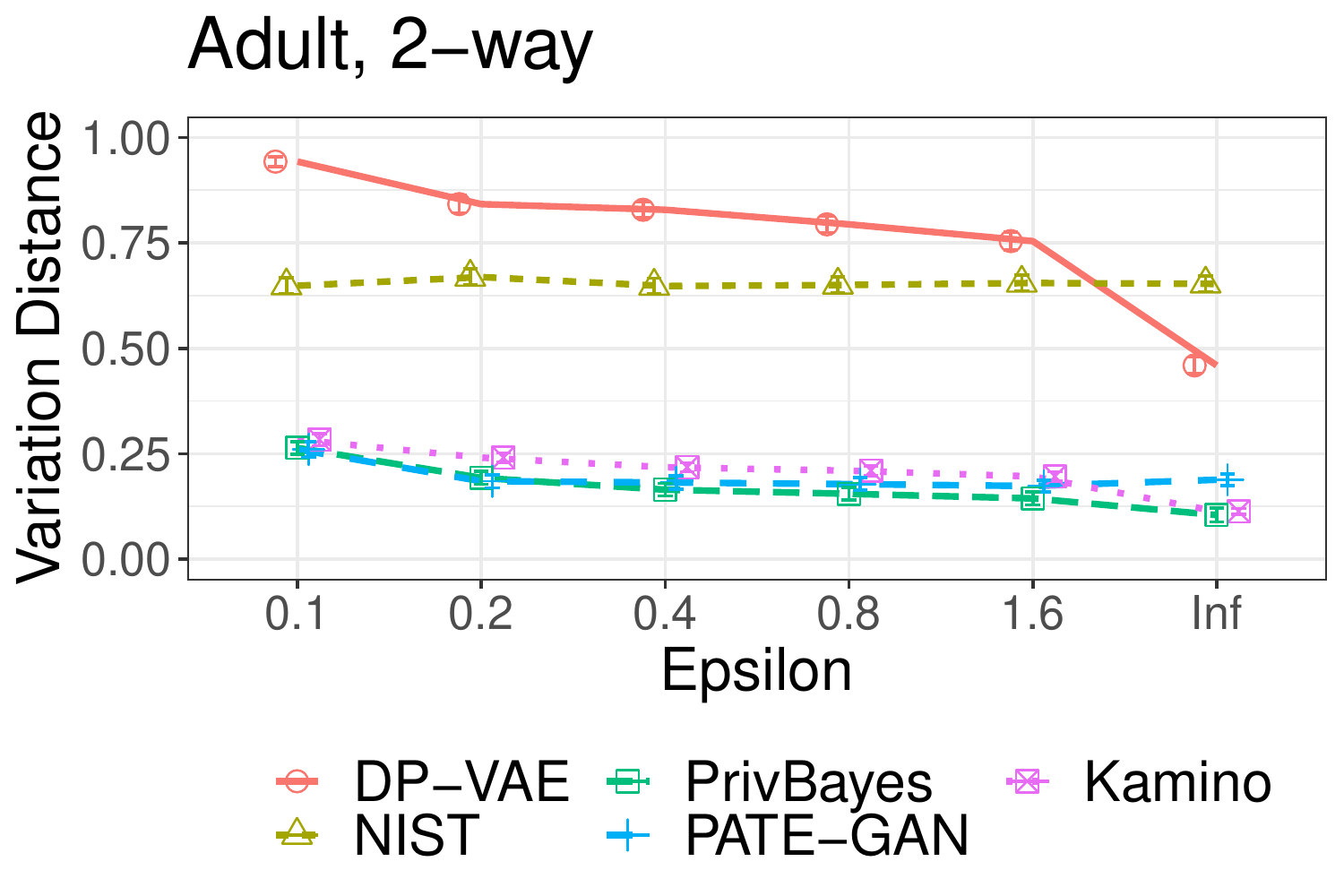}%
        \caption{2-way marginal}\label{fig:vary_budget_all:2-way}
    \end{subfigure}%
    
    \vspace{-.75em}
    \caption{\revise{Task quality of the \system and baselines by varying privacy budget ($\epsilon, 10^{-6}$). }}
	\label{fig:vary_budget_all}
\end{figure*}

\subsection{Component Evaluation}\label{sec:evaluation:component}

\stepcounter{exp}
\subsubsection{Experiment \theexp: Effectiveness of constraint-aware components}
Recall that our approach takes DCs into account when it samples synthetic values (\S~\ref{sec:sampling}) and generate the schema sequence (\S~\ref{sec:sequencing}).
In this experiment, we compare \system with three sub-optimal \system that do not have the constraint-aware components:
\squishlist
\item Replace constraint-aware sampling (Algorithm~\ref{alg:syn}) in \system with sampling tuples independently, labeled as ``RandSampling";
\item  Replace constraint-aware sequencing (Algorithm~\ref{alg:sequence}) by a random sequence, labeled as ``RandSequence";
\item  Replace both components above, labeled as  ``RandBoth".
\squishend

\begin{table}[t]
\centering
\caption{Percentage of DC violations using \system, and sub-optimal \system w/o constraint-aware components.}
\label{tab:dc_vio_adult}
\vspace{-.75em}
\resizebox{\columnwidth}{!}{%
\begin{tabular}{|c|c|c|c|c|c|}
\hline
DC         & Truth & \system    & RandSequence & RandSampling & RandBoth     \\ \hline
$\phi_1^a$ & 0     & 0.0$\pm$0.0 & 0.0$\pm$0.0 & 0.4$\pm$0.0  & 9.1$\pm$8.5 \\ \hline
$\phi_2^a$ & 0     & 0.0$\pm$0.0 &  0.0$\pm$0.0 & 36.8$\pm$0.3 & 26.1$\pm$11.0 \\ \hline
\end{tabular}
}
\end{table}

Table~\ref{tab:dc_vio_adult} compares DC violations of the synthetic data generated by \system and by sub-optimal \system without constraint-aware components.
First, we see that without constraint-aware sampling component (Algorithm~\ref{alg:syn}), 
the synthetic data generated by RandSampling and RandBoth have more violations than the other two methods.
Second, the constraint-aware sequencing component (Algorithm~\ref{alg:sequence}) is also important. 
Take $\phi_1^a: edu \rightarrow edu\_num$ as an example, RandBoth (without the constraint-aware sequencing) results in a higher number of DC violations than RandSampling. 
This is because that $edu$ is not necessarily placed before $edu\_num$ in a random schema sequence, and the noisy model cannot preserve the correlation between these two attributes.
Similar, without constraint-aware components, quality downgrades in both learning and query task shown in Figure~\ref{fig:training_components}. 

We omit the presentation of non-private runs for similar observations.
We believe that the constraint-aware components can also be incorporated into the baseline systems, but we skip the comparison because it requires significant re-design of the baseline systems.

\begin{figure*}[t]
\minipage{0.25\textwidth}
  \includegraphics[width=\textwidth]{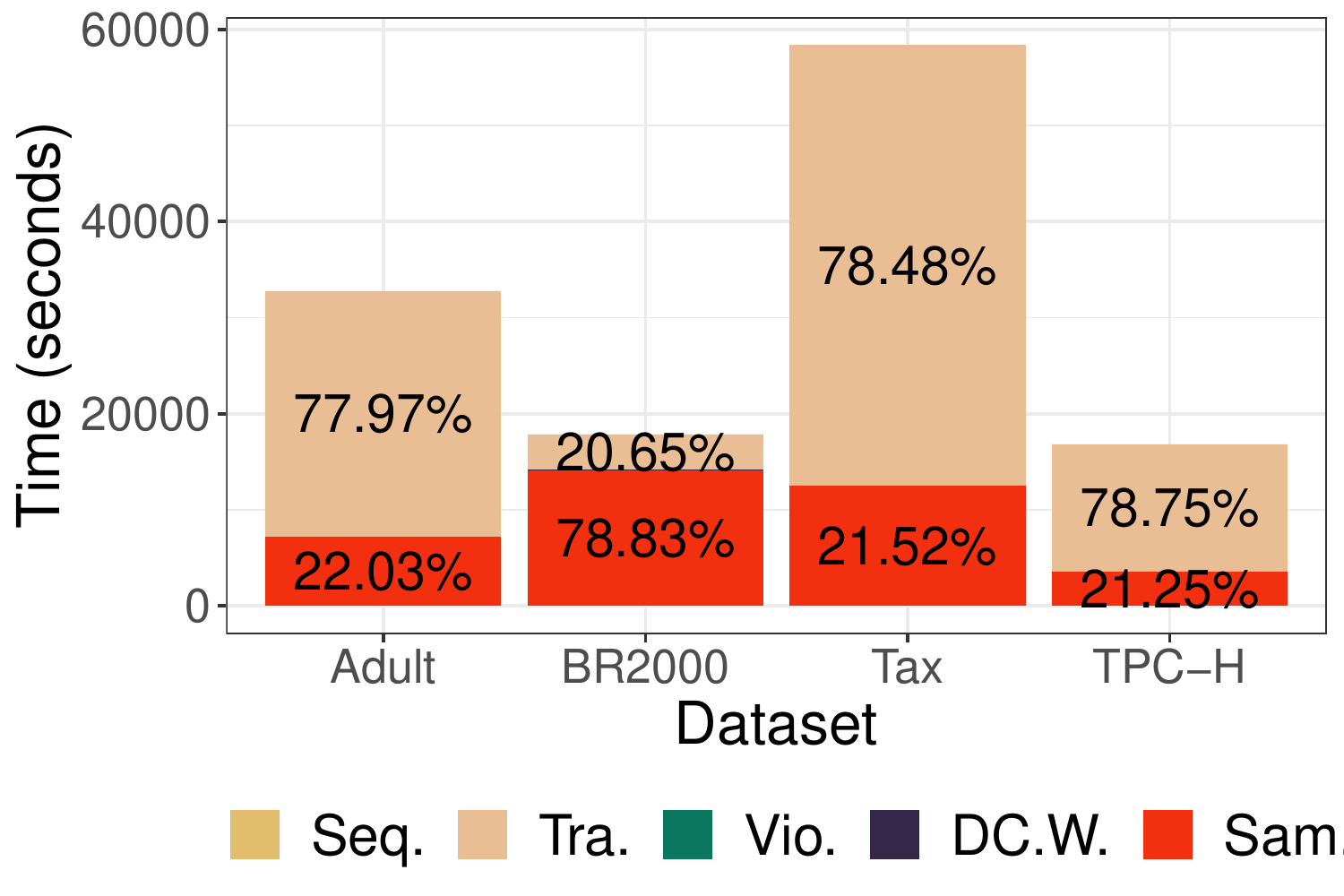}
  \caption{\revise{Time profiling of end-to-end runs on all datasets.}}\label{fig:time_profile}
\endminipage\hfill
\minipage{0.75\textwidth}%
    \begin{subfigure}[h]{0.33\textwidth}%
        \includegraphics[width=\textwidth]{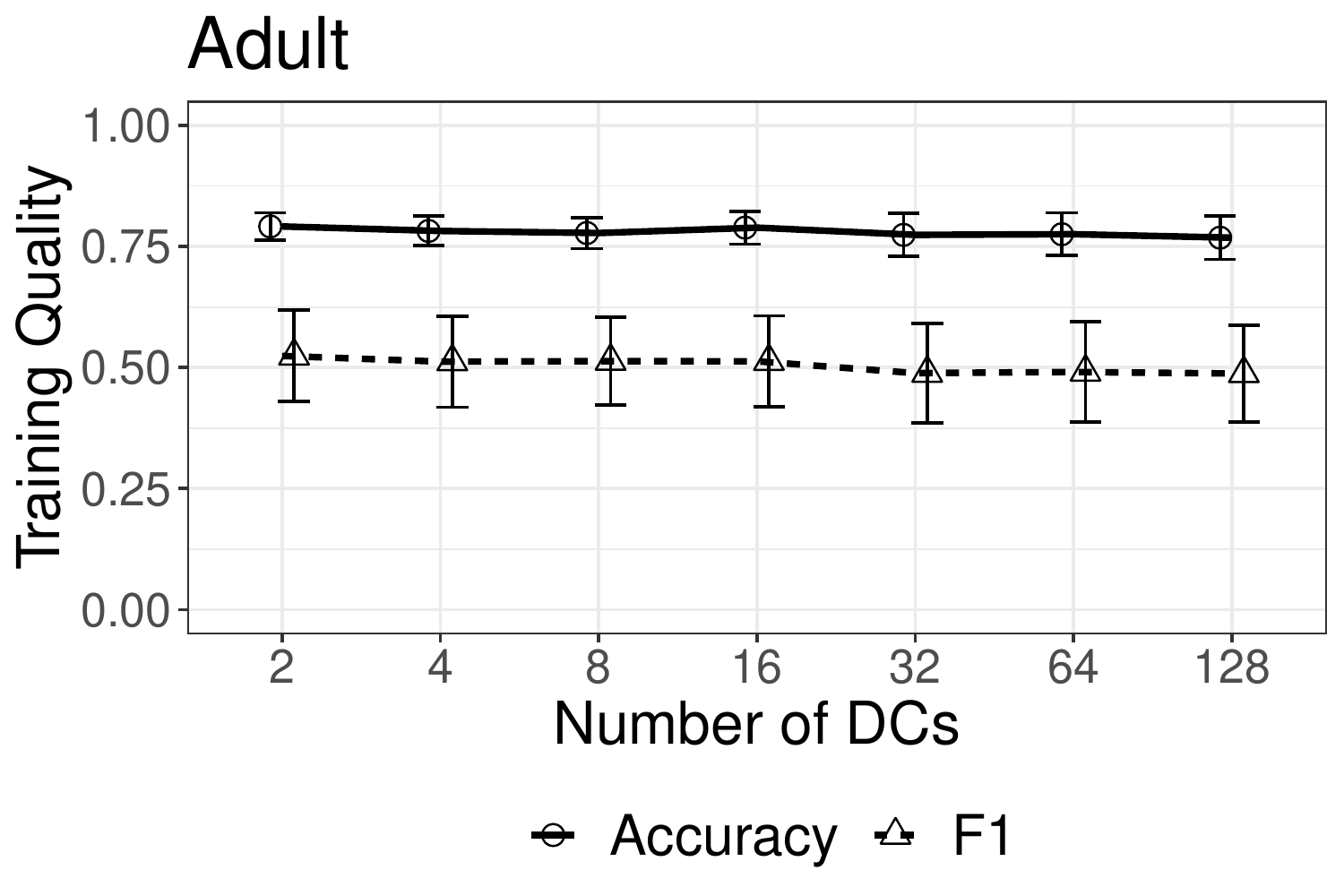}%
        \caption{Model Training}\label{fig:scalability:training}
    \end{subfigure}%
    \begin{subfigure}[h]{0.33\textwidth}%
        \includegraphics[width=\textwidth]{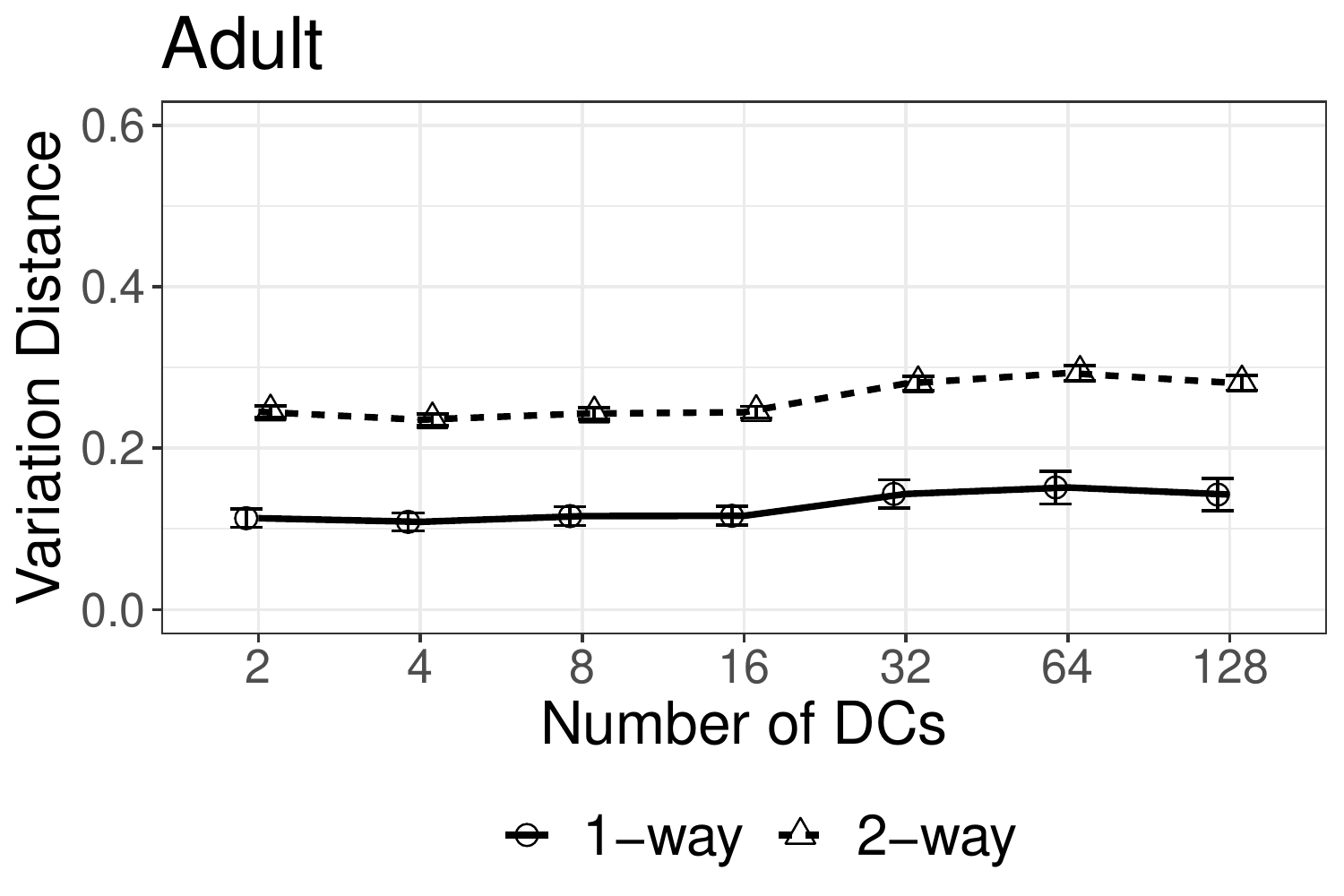}%
        \caption{Marginal Distance}\label{fig:scalability:marginal}
    \end{subfigure}%
    \begin{subfigure}[h]{0.33\textwidth}%
        \includegraphics[width=\textwidth]{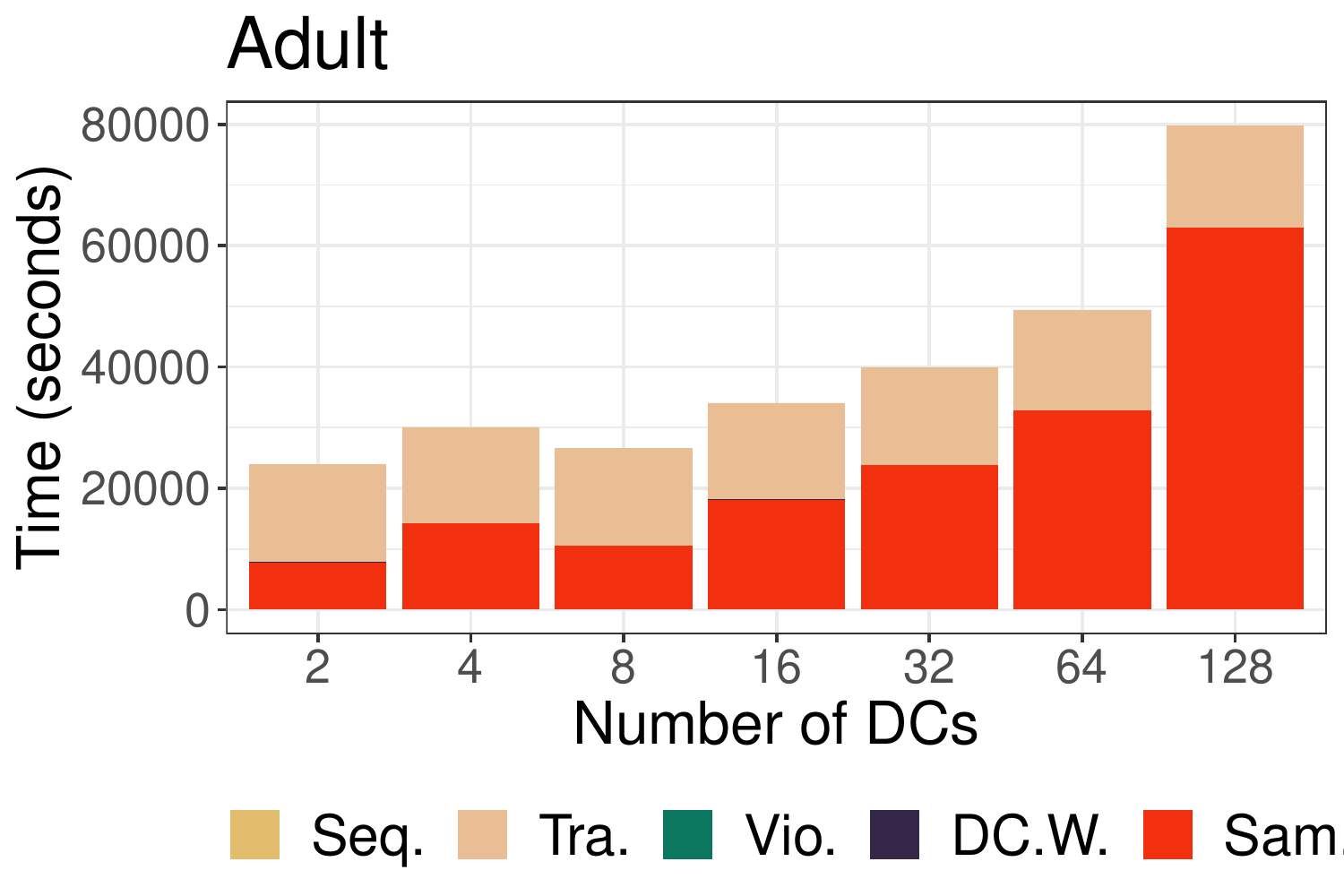}%
        \caption{Time Complexity}\label{fig:scalability:timimg}
    \end{subfigure}%
    \vspace{-.75em}
    \caption{\revise{Task quality and execution time by varying the number of DCs.}}\label{fig:scalability}
\endminipage
\end{figure*}

\revise{
\stepcounter{exp}
\subsubsection{Experiment \theexp: \system vs Accept-Reject Sampling}\label{sec:exp:ar-sampling}

\eat{
\begin{figure}[t]
    \centering    
    \begin{subfigure}[h]{0.5\columnwidth}%
        \includegraphics[width=\textwidth]{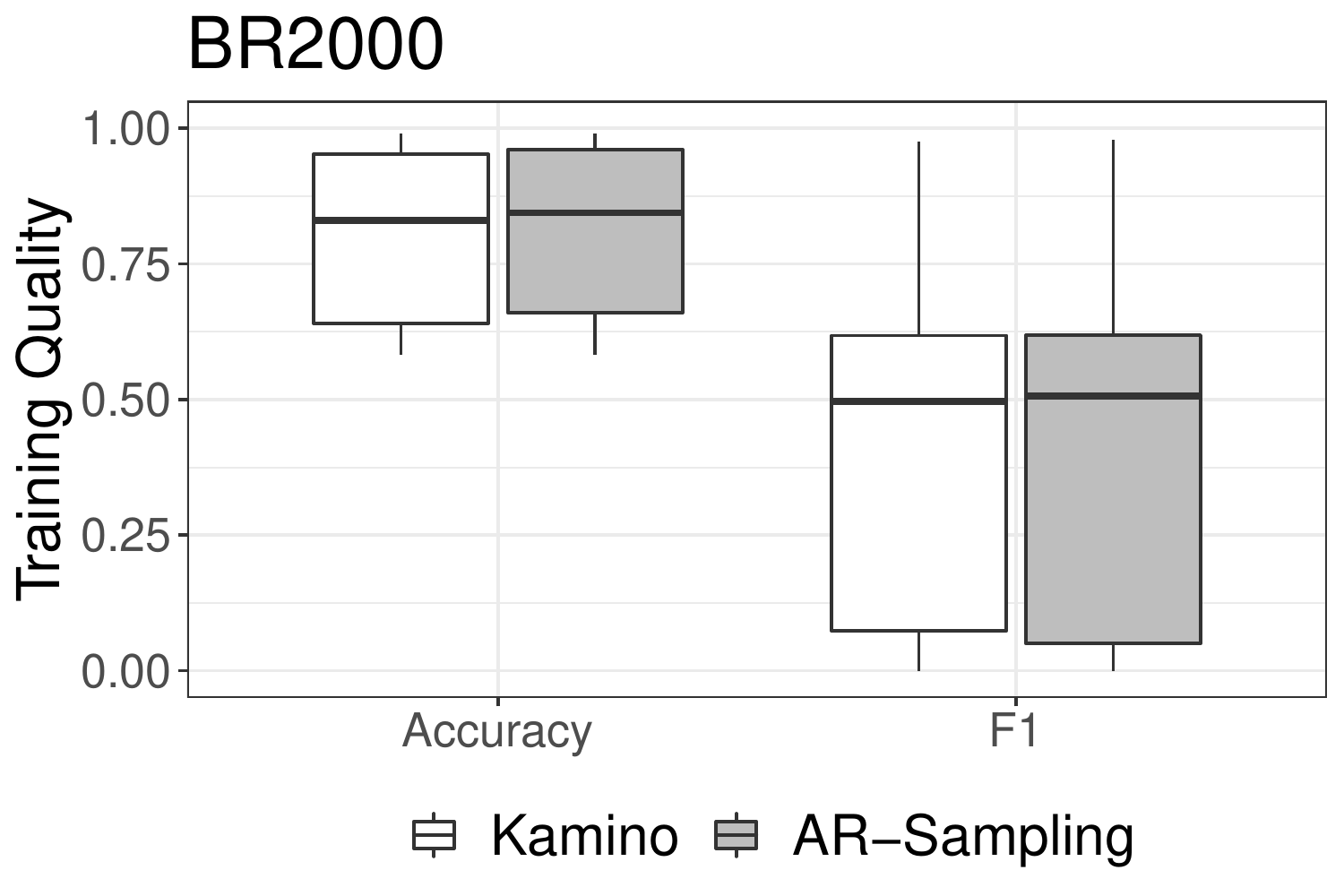}%
        \caption{Model Training}\label{fig:ar:training}
    \end{subfigure}%
    \begin{subfigure}[h]{0.5\columnwidth}%
        \includegraphics[width=\textwidth]{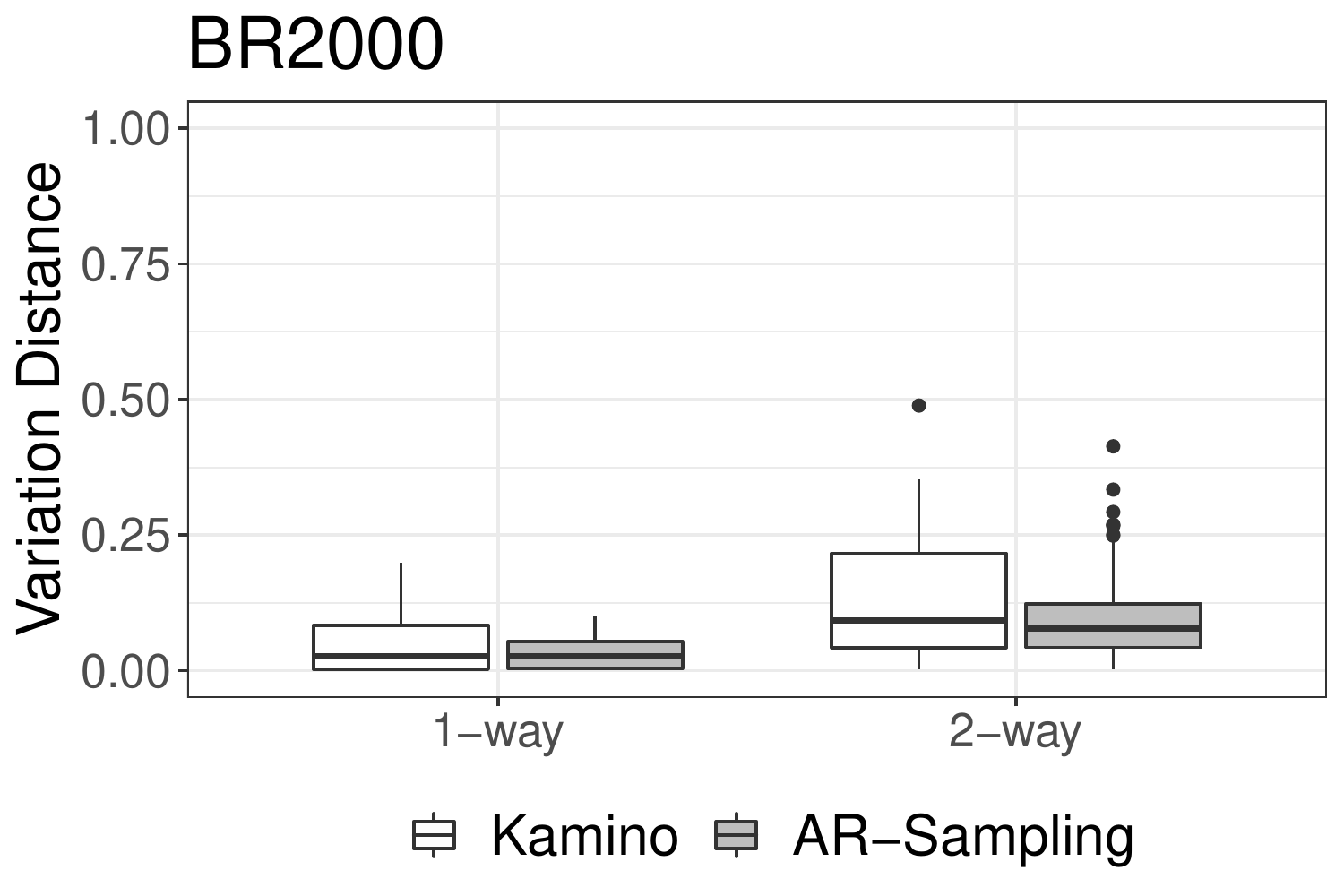}%
        \caption{Marginal Distance}\label{fig:ar:marginal}
    \end{subfigure}%
    \vspace{-.75em}
    \caption{\revise{\system v.s. accept-reject sampling.}}
	\label{fig:ar}
\end{figure}

\begin{figure}[t]
    \centering    
    \begin{subfigure}[h]{0.5\columnwidth}%
        \includegraphics[width=\textwidth]{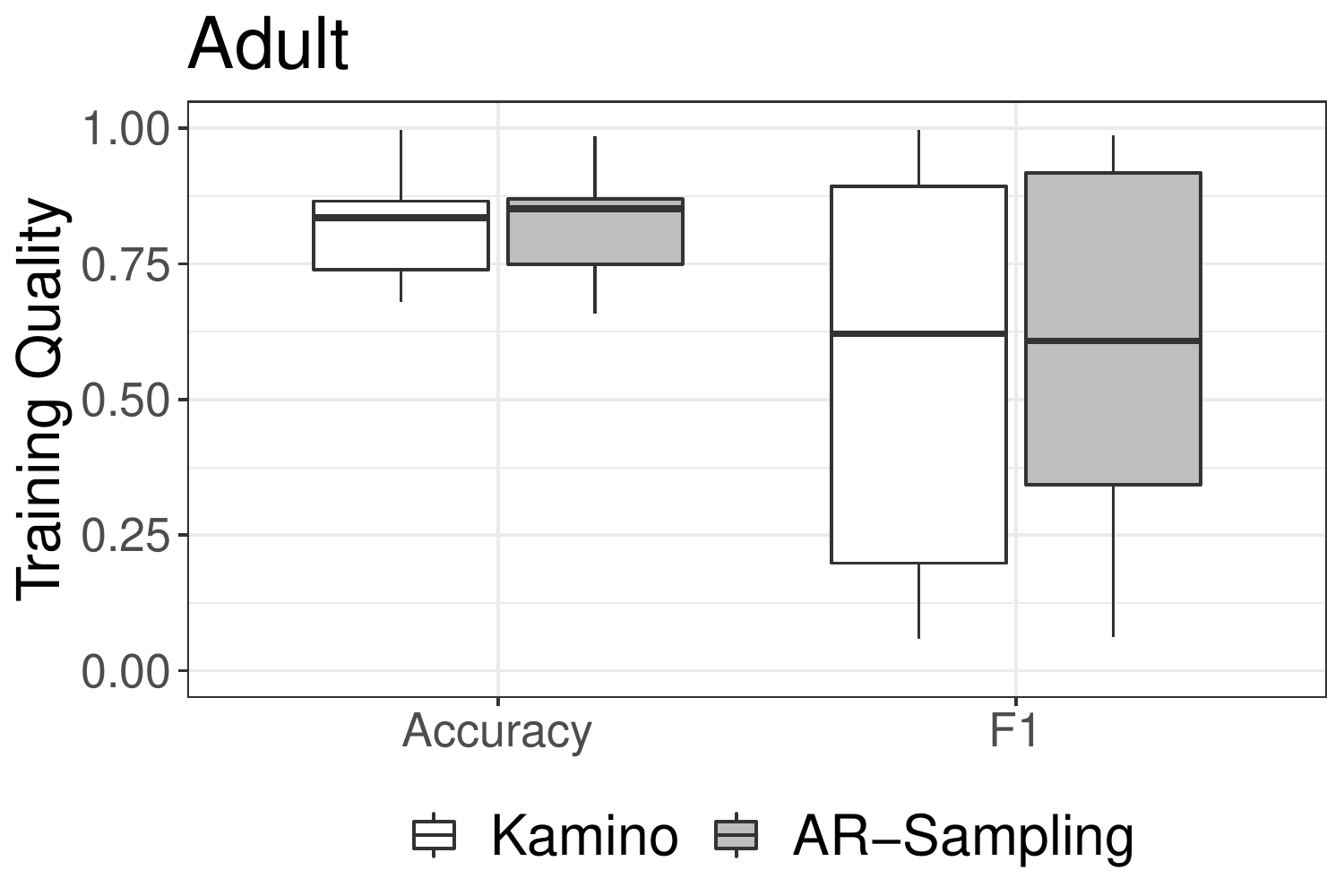}%
        \caption{Model Training}\label{fig:ar:training}
    \end{subfigure}%
    \begin{subfigure}[h]{0.5\columnwidth}%
        \includegraphics[width=\textwidth]{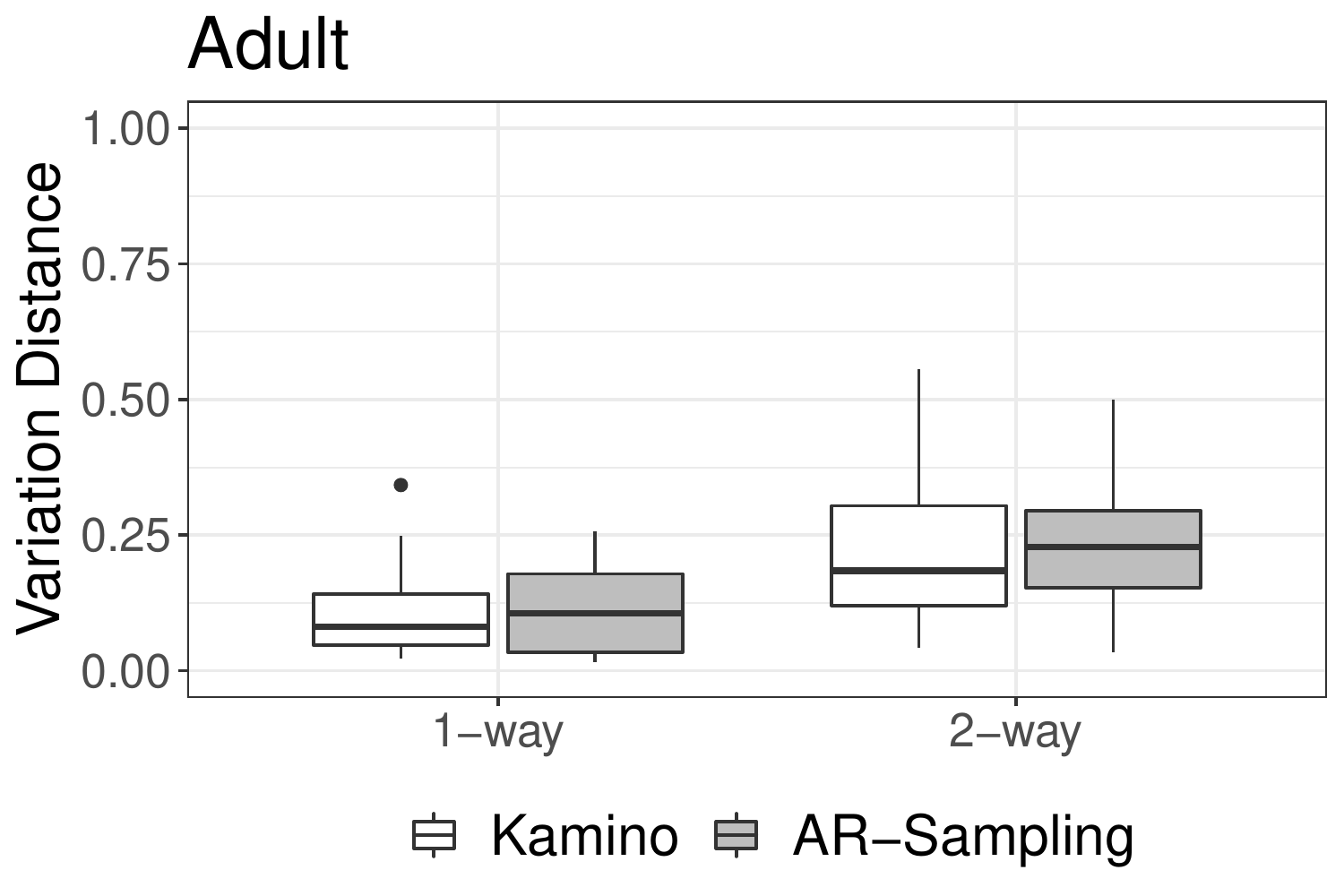}%
        \caption{Marginal Distance}\label{fig:ar:marginal}
    \end{subfigure}%
    \vspace{-.75em}
    \caption{\revise{\system v.s. accept-reject sampling.}}
	\label{fig:ar}
\end{figure}
}

\system's constraint-aware sampling (Algorithm~\ref{alg:syn}) explicitly constructs the target distribution and directly samples from it for filling a cell (Line~\ref{alg:syn:sample}).
Another sampling method is the accept-reject (AR) sampling~\cite{mcbook}, which samples one value at a time, and accepts this value probabilistically based on its violations.
For soft DCs, AR-sampling can be an alternative, but it does not work well for hard DCs.

We first evaluate \system using AR-sampling on the Adult dataset with hard DCs.
AR-sampling does not work well when hard DCs are present.
If a sampled value incurs any violations, then its accept ratio (i.e., $\exp(-\sum_{\phi \in \Phi_{A_j}} w_\phi \times vio_{\phi, v \mid D^\prime})$, where $v$ is the sampled value of attribute $A_j$) diminishes to 0, since $w_\phi=\infty$.
As a result, AR-sampling needs re-sampling multiple times until a value can be accepted, depending on the other cells that have been filled with sampled values.  
For efficiency purpose, we allow at most 300 samples per cell: if no values can be accepted, we take the last sampled value and as a result, violations can occur.  
\system with AR-sampling does produce violations for the two DCs $\phi_a^1$ (0.4$\pm$0.0) and $\phi_a^2$ (37.2$\pm$0.0).
The execution time of \system with AR-sampling takes 7.5 hours, which is 1.9$\times$ longer. 

On the BR2000 dataset with soft DCs, \system with AR-sampling completes in 1.26 hours (0.24 hour for the AR-sampling step) on average, which is faster than the constraint-aware sampling (3.9 hours).
AR-sampling converges faster due to its relatively high accept ratio.
For DC violations and task qualities, we observe that \system with AR-sampling performs similarly with \system.
}

\revise{
\stepcounter{exp}
\subsubsection{Experiment \theexp: Varying Privacy Budget}\label{sec:exp:budget}
We show the impact of the privacy budget in the task qualities using the Adult dataset as the example.
Figure~\ref{fig:vary_budget_all} compares the data usefulness by varying the privacy budget parameter ($\epsilon,\delta$) at different $\epsilon=[0.1,0.2,0.4,0.8,1.6]$ with a constant $\delta=10^{-6}$.
$\epsilon=\infty$ refers to non-private \system and baselines. 
First of all, increasing the privacy budget leads to overall better quality in both the learning and the query tasks.
Consistent with the observations in Figures~\ref{fig:training}-\ref{fig:marginals},
\system always achieves the best in training quality (Figures~\ref{fig:vary_budget_all:accuracy}-\ref{fig:vary_budget_all:f1}) and close to best marginal distances (Figures~\ref{fig:vary_budget_all:1-way}-\ref{fig:vary_budget_all:2-way}) at different privacy budgets. 
The averaged model accuracy over all attributes on \system is 0.8 at privacy budget ($\epsilon=0.2, \delta=10^{-6}$), 
which outperforms DP-VAE (0.54), NIST(0.66), PrivBayes (0.68) and PATE-GAN (0.77) at 5$\times$ larger $\epsilon=1$.

\stepcounter{exp}
\subsubsection{Experiment \theexp: Scalability of DCs}\label{sec:exp:scalability}

In this experiment, we vary the number of DCs from the input to \system.
Due to the lack of large numbers of ground DCs,
we generate the input DCs by discovering approximate DCs~\cite{DBLP:journals/pvldb/PenaAN19} to simulate the knowledge from the domain expert.

Figure~\ref{fig:scalability} shows the task quality and time profiling as increasing the number of soft DCs from 2 to 128, under the fixed privacy budget ($\epsilon=1, \delta=10^{-6}$) on the Adult dataset.
Since the DC weights are noisy and approximately learned using a subset of data (Algorithm~\ref{alg:weight_learning}), increasing the number of DCs implies more noisy adjustment for the sampling probabilities (Algorithm~\ref{alg:syn}).
As a result, task quality is expected to decrease given a finite privacy budget.
Figure~\ref{fig:scalability:training} and Figure~\ref{fig:scalability:marginal} show that as the number of DCs increases t0 128, task quality only degrades by 0.04.

As the number of DCs increases, more time is required to compute the violation matrix, learn DC weights, and to sample.
In particular,  for \system's constraint-aware sampling process (Algorithm~\ref{alg:syn}), introducing more DCs will linearly increases the time to check DC violations for each of the DCs.
Since the total execution time is dominated by the sampling process, the total execution time of \system scales linearly with the number of DCs.
Figure~\ref{fig:scalability:timimg} shows that when the number of DC increases from 2 to 128, the total execution time increases only by 3$\times$.

\ifpaper
\else
\stepcounter{exp}
\subsubsection{Experiment \theexp: Varying the Number of Re-sampling in MCMC}\label{sec:exp:mcmc}

\begin{figure*}[t]
    \centering    
    \begin{subfigure}[h]{0.3\textwidth}%
        \includegraphics[width=\textwidth]{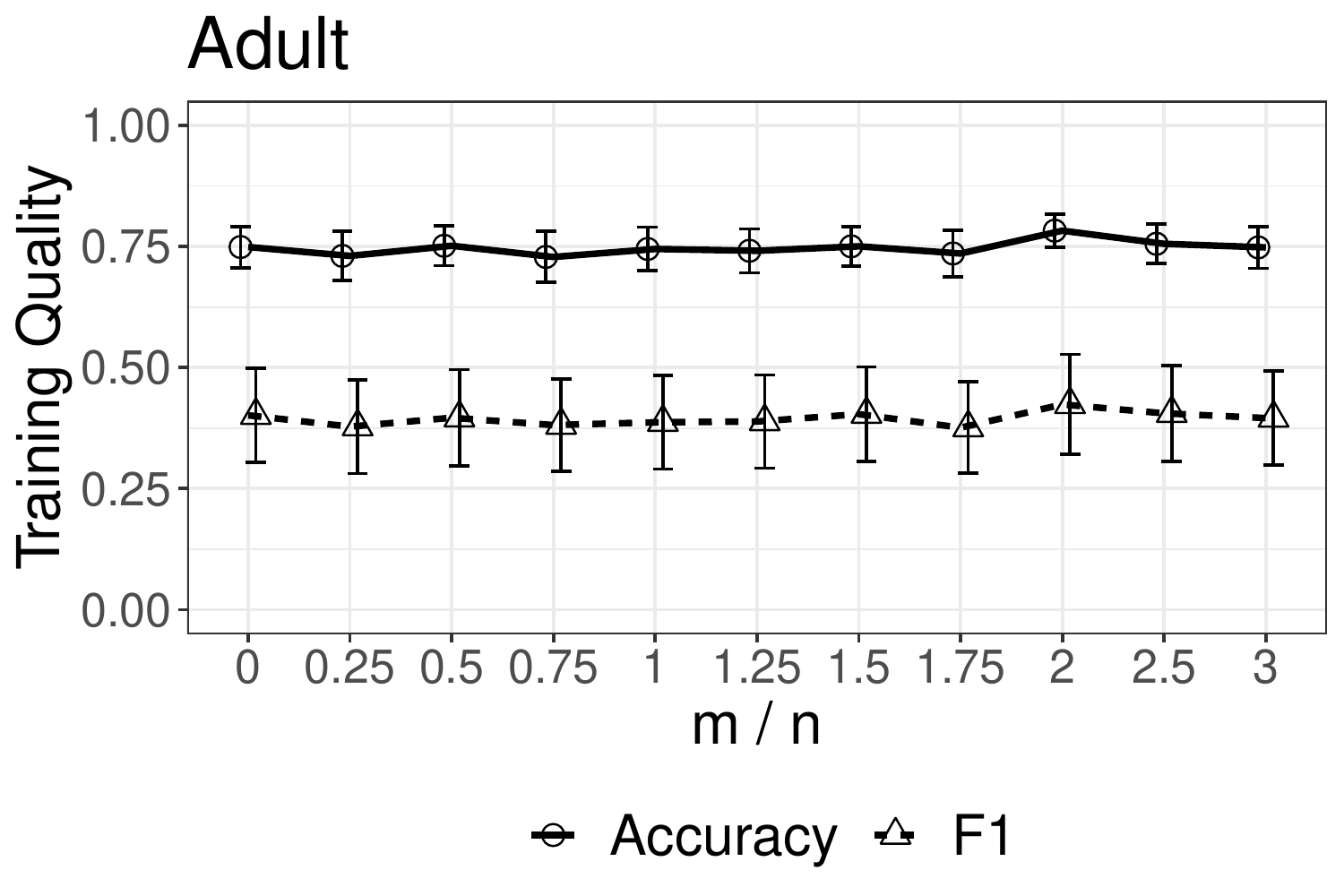}%
        \caption{Model Training}
    \end{subfigure}%
    \begin{subfigure}[h]{0.3\textwidth}%
        \includegraphics[width=\textwidth]{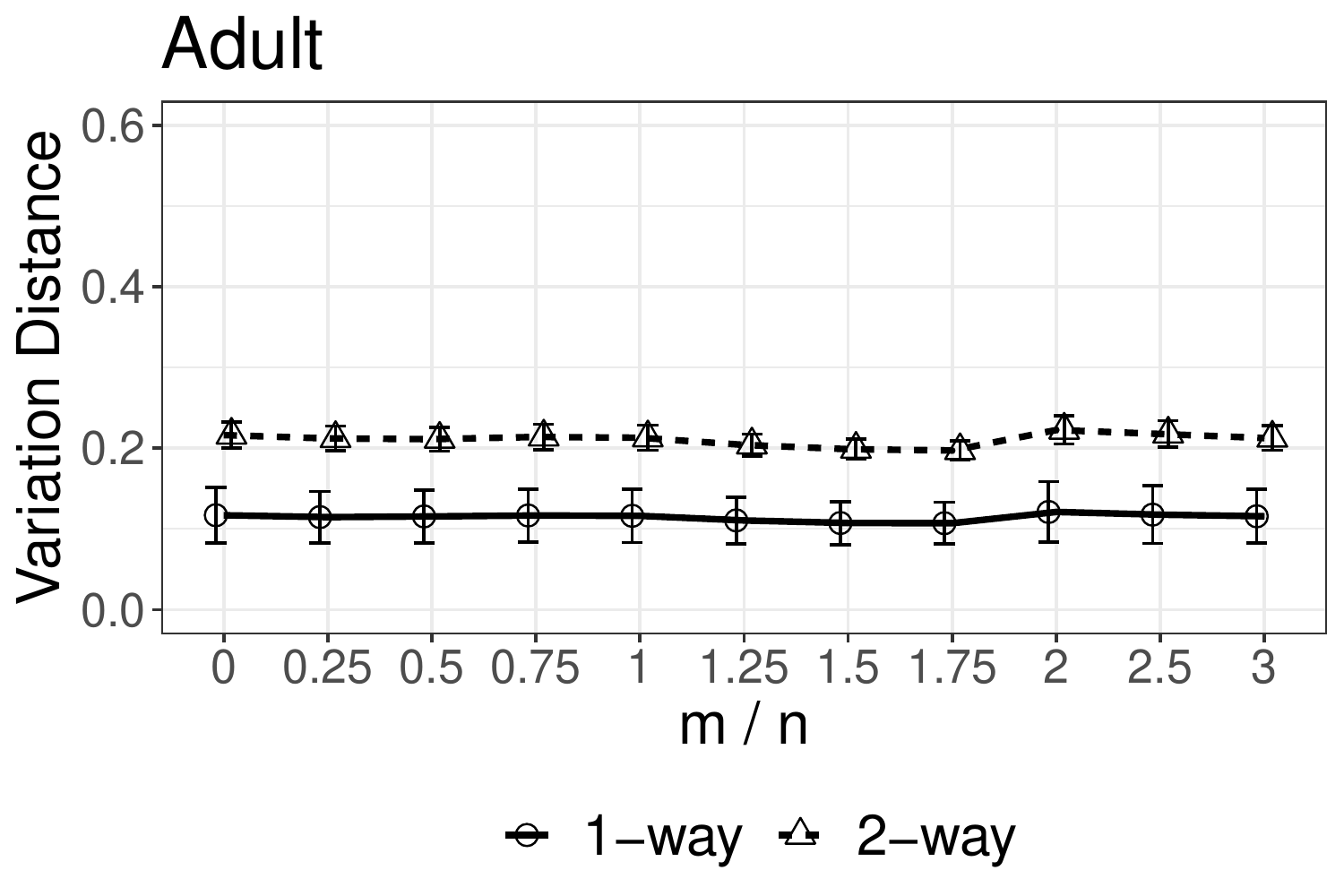}%
        \caption{Marginal Distance}
    \end{subfigure}%
    \begin{subfigure}[h]{0.3\textwidth}%
        \includegraphics[width=\textwidth]{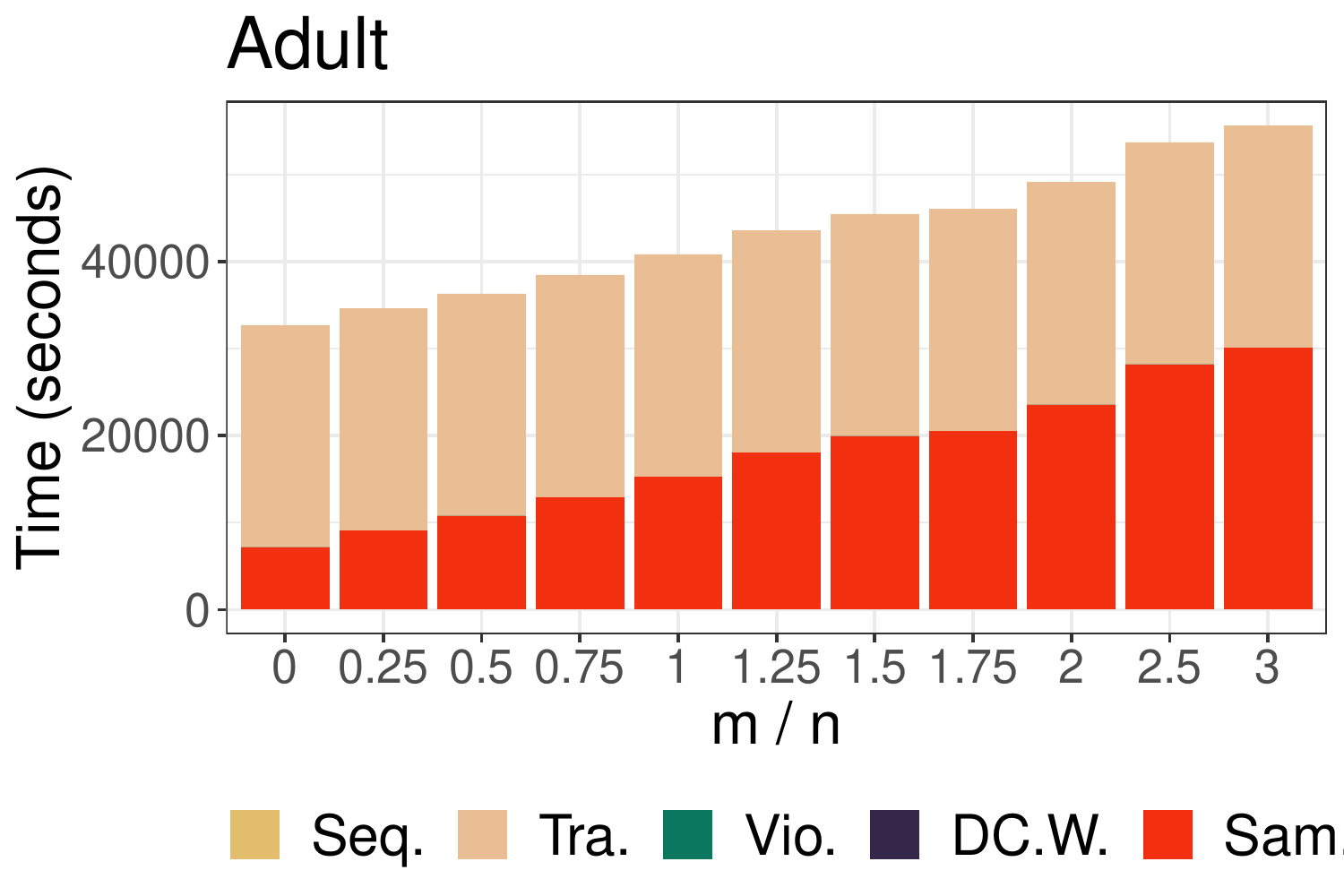}%
        \caption{Time Complexity}
    \end{subfigure}%
    \vspace{-.75em}
    \caption{Task quality and execution time by varying the number of resampling per attribute.}
	\label{fig:mcmc}
\end{figure*}

Recall that \system's sampling process adopts MCMC by re-sampling $m$ random cells after each column is synthesized (Algorithm~\ref{alg:syn}).
Figure~\ref{fig:mcmc} shows the effects of $m$, which is represented as a ratio over dataset cardinality $n$ on the x-axis.
Comparing to no re-sampling, the re-sampling up to $m=3n$ using the same probabilistic data model improves accuracy (by up to 0.03), F1 (by up to 0.02), and both 1-way and 2-way marginal distances (by 0.01 and 0.02, respectively).
Meanwhile, more re-sampling requires longer execution time (by up to 4$\times$).

\stepcounter{exp}
\subsubsection{Experiment \theexp: Efficiency Optimizations}\label{sec:exp:optimization}
This section presents two optimization techniques to speed up \system under certain conditions.
The first technique is to train \system's probabilistic data models (Algorithm~\ref{alg:model}) in parallel, where each model $M_{X, y}$ is trained on a separate machine and tuple embeddings are initialized randomly, instead of reusing previously trained ones.
Without reusing the tuple embeddings, we observe that although all task qualities drop 0.01 on the Adult dataset, the training time becomes 3.5$\times$ faster. 

The second optimization is to exploit the special property of DCs.
One special DC is the hard functional dependencies (e.g., keys), where the right-hand-side attribute has only one unique value given the left-hand-side values.
Instead of checking violations for the set of candidate values, we can find  the correct value from previously synthesized data. 
We scale up the TPC-H dataset to 1 million rows with the same set of DCs. 
\system can complete in 10 hours by leveraging the fact that all DCs are hard functional dependencies. 

\fi

}

\section{Related Work}
\label{sec:related}

There has been extensive literature on releasing differentially private synthetic data~\cite{nist, DBLP:journals/tkde/ZhuLZY17, fan_2020, Bowen_2020}. 
These approaches can be categorized into two classes: 
1) statistical approaches, which focus on synthesizing low-dimensional projections; and
2) deep learning approaches, which train a deep generative model to sample tuples.
Both classes assume tuples are i.i.d, and hence cannot preserve the structure of the data. 
Our approach is a combination of both, and 
more importantly, our method differentiates prior work in that we explicitly consider the denial constraints~\cite{DBLP:books/acm/IlyasC19} enforced among tuples, 
rather than simply assuming tuple independence.

Statistical approaches for generating synthetic data usually estimate low-dimensional marginal distributions~\cite{DBLP:journals/tkde/XiaoWG11, DBLP:conf/sigmod/QardajiYL14}, due to the hardness of privatizing high-dimensional data with differential privacy guarantee~\cite{DBLP:conf/stoc/BlumLR08, DBLP:conf/stoc/DworkNRRV09, DBLP:conf/tcc/UllmanV11, DBLP:conf/icml/GaboardiAHRW14}.
These low-dimensional distributions can be used to estimate the high-dimensional tuple distribution, based on the assumption of conditional independence among attributes, which can be modeled using probabilistic graphical models~\cite{DBLP:books/daglib/0023091}, such as using the Bayesian network~\cite{DBLP:conf/sigmod/ZhangCPSX14, DBLP:journals/pvldb/LiXZJ14, DBLP:conf/ssdbm/PingSH17} or undirected graphs~\cite{DBLP:conf/kdd/ChenXZX15, DBLP:conf/icml/McKennaSM19}.
Under this model, only correlations among dependent attributes are likely to be captured, but correlations that widely exist among conditional independent attributes and tuples are not captured in prior work. 

Deep learning models have been shown widely used in synthesizing unstructured data, such as images~\cite{DBLP:conf/cvpr/ShrivastavaPTSW17}, videos~\cite{Chawla2019DeepfakesH} and natural languages~\cite{DBLP:conf/icassp/Gupta19}.
Different from unstructured data, structured data is defined using relational schema and hence, stucture correlations naturally exist.
Na\"ivly applying deep learning models such as GAN~\cite{DBLP:journals/corr/GoodfellowPMXWOCB14} and auto-encoder~\cite{DBLP:journals/corr/KingmaW13} on structured data faces at least two challenges.
First, those models usually take numeric vectors as input,
and popular encoding schemes such as one-hot encoding or ordinal encoding do not work well on structured data~\cite{DBLP:journals/pvldb/FanLLCSD20},
Second, similar to statistical approaches, methods based on deep models (e.g.~\cite{DBLP:conf/sec/FrigerioOGD19, DBLP:conf/iclr/JordonYS19a, DBLP:journals/corr/abs-2001-09700, DBLP:journals/corr/abs-1802-06739}) suffer from missing structure correlations.

In general, generating differentially private synthetic data is hard, due to the tradeoff between accuracy and privacy~\cite{DBLP:conf/stoc/BlumLR08, DBLP:conf/stoc/DworkNRRV09, DBLP:conf/tcc/UllmanV11, DBLP:conf/icml/GaboardiAHRW14}.
On the other hand, an efficient private data generation algorithm fails to offer the same level of accuracy guarantees to all the queries.
Existing practical methods (e.g.,~\cite{DBLP:journals/corr/abs-1812-02274, DBLP:conf/kdd/ChenXZX15,DBLP:conf/iclr/JordonYS19a, DBLP:conf/sigmod/ZhangCPSX14,DBLP:conf/pods/BarakCDKMT07}) therefore choose to privately learn only a subset of correlations to model the true data.
However, the structure of the data is not explicitly captured by these methods and thus are poorly preserved in the outputs.

\section{Conclusion}
\label{sec:conclusion}

In this work, we are motivated to design a synthetic data generator that can preserve both the structure of the data, and the privacy of individual data records.
We present \system, an end-to-end data synthesis system for constraint-aware differentially private data synthesis. 
\system takes as input a database instance, along with its schema (including denial constraints), 
and produces a synthetic database instance.
Experimental results show that \system can preserve the structure of the data, while generating useful synthetic data for applications of training classification models and answering marginal queries, comparing to the state-of-the-art methods.

\balance
\bibliographystyle{ACM-Reference-Format}
\bibliography{sigproc}

\appendix

\ifpaper
\else
\section{DC Violation Analysis}\label{app:analysis}

\revise{
Consider in the non-private setting and assume that the true database has no violations for the given set of DCs.
Intuitively, we can learn an accurate probabilistic database model according to the learnability theorem (Theorem 14~\cite{DBLP:conf/icdt/SaIKRR19}).
When sampling from the learned probabilistic database model, the sampled synthetic database instance should have a small number of DC violations.
Formally, we state the following theorem.

\begin{theorem}\label{theorem:accuracy}
Given a true database instance $D^*$ with no constraint violations for a set of DCs $\Phi$, we learn the probabilistic database model $\mathbb{D}$ and sample a synthetic database instance $D^\prime \sim \mathbb{D}$ based on Eqn.~\eqref{equ:pdb2}.
The sampled $D^\prime$ incurring any DC violations has a low probability.  
\end{theorem}

\begin{proof}

Let $\Theta$ represent the parameter set for the tuple probability, and $W$ denote the weight vector of DCs.
Given the true database instance $D^*$ with no DC violations, the parameters can be found by maximizing the likelihood of $D^*$, i.e.,
\begin{equation}\label{equ:mle}
\Theta^*, W^* = \argmax_{\Theta,W} \frac{\Pi_{t\in D^*} \Pr(t; \Theta)}{Z}
\end{equation}
where 
\begin{align*}
Z =& \sum_{D\in \textbf{D}_{V=0}} \Pi_{t\in D} \Pr(t; \Theta) + \\
& \sum_{D\in \textbf{D}_{V\neq 0}} \Pi_{t\in D}\Pr(t; \Theta) \exp(-\sum_{\phi \in \Phi} w_\phi |V(\phi, D)|)
\end{align*} 
and $\textbf{D}_{V=0}$ represents the set of database instances with no DC violations.
To maximize the likelihood of $D^*$, the weight $w_\phi$ for each DC are set to be very large (i.e., $w_\phi \rightarrow \infty$).

Finally, suppose $\exists \phi \in \Phi$, such that $|V(\phi, D^\prime)|\neq 0$, and consider the probability of sampling $D^\prime$:
\begin{equation}\label{equ:sd}
\Pr(D^\prime; \Theta^*, W^*) = \frac{\Pi_{t\in D^\prime} \Pr(t)\times \exp(-\sum_{\phi \in \Phi} w_\phi |V(\phi, D^\prime)|)}{Z}
\end{equation}

Since $w_\phi \rightarrow \infty$ (Eqn.~\eqref{equ:mle}) and $|V(\phi, D^\prime)|\neq 0$,
then $\Pr(D^\prime; \Theta^*, W^*) \rightarrow 0$.
\end{proof}

According to Theorem~\ref{theorem:accuracy}, the synthetic database instance $D^\prime$ is most likely to have no violations.
In \S~\ref{sec:exp:dcs}, we empirically show that $|V(\phi, D^\prime)|=0$ for all hard DCs on all output instances.

More generally, even if the true database has a small number of DC violations (i.e., the low-noise condition~\cite{DBLP:conf/icdt/SaIKRR19}), we are still able to learn the parameters of the probabilistic database model $\mathbb{D}$ accurately. 
In addition, the sampling process follows the chain rule (Eqn.~\eqref{equ:pdb4}) for Eqn.~\eqref{equ:pdb2} and hence, allows an instance to be sampled correctly from $\mathbb{D}$. 
As a result, the true and the synthetic database instances come from the same distribution. The probability of a synthetic instance depends on the the softness of the DCs (Eqn.~\eqref{equ:sd}).

Finding the error bound of DC violations for general DCs with differential privacy guarantee is an exciting future work.

}

\fi

\end{document}